\documentclass[twocolumn,tighten]{aastex62}
\pdfoutput=1 
\usepackage{amsmath,amstext}
\usepackage[T1]{fontenc}
\usepackage{apjfonts} 
\usepackage[figure,figure*]{hypcap}
\usepackage{threeparttable}
\usepackage[htt]{hyphenat}

\newcommand{\medium}[1]{\textsc{#1}} 

\begin{document}
\shorttitle{Aemulus V: Galaxy Clustering}

\title{The Aemulus Project V: Cosmological constraint from small-scale clustering of BOSS galaxies}

\author{Zhongxu Zhai}
\affiliation{Department of Astronomy, School of Physics and Astronomy, Shanghai Jiao Tong University, Shanghai 200240, China}
\affiliation{Shanghai Key Laboratory for Particle Physics and Cosmology, Shanghai 200240, China}
\affiliation{Waterloo Center for Astrophysics, University of Waterloo, Waterloo, ON N2L 3G1, Canada}
\affiliation{Department of Physics and Astronomy, University of Waterloo, Waterloo, ON N2L 3G1, Canada}

\author{Jeremy L. Tinker}
\affiliation{Center for Cosmology and Particle Physics, Department of Physics, New York University, 726 Broadway, New York, NY 10003, USA}

\author[0000-0002-5209-1173]{Arka Banerjee}
\affiliation{Department of Physics, Indian Institute of Science Education and Research,
Homi Bhabha Road, Pashan, Pune 411008, India}
\affiliation{Fermi National Accelerator Laboratory, Cosmic Physics Center, Batavia, IL 60510, USA}

\author{Joseph DeRose}
\affiliation{Lawrence Berkeley National Laboratory, 1 Cyclotron Road, Berkeley, CA 93720, USA}
\affiliation{Berkeley Center for Cosmological Physics, Department of Physics, UC Berkeley, CA 94720,
USA}

\author{Hong Guo}
\affiliation{{Key Laboratory for Research in Galaxies and Cosmology, Shanghai Astronomical Observatory, Shanghai 200030, China}}

\author[0000-0002-1200-0820]{Yao-Yuan~Mao}
\altaffiliation{NASA Einstein Fellow}
\affiliation{Department of Physics and Astronomy, Rutgers, The State University of New Jersey, Piscataway, NJ 08854, USA}

\author{Sean McLaughlin}
\affiliation{Kavli Institute for Particle Astrophysics and Cosmology and Department of Physics, Stanford University, Stanford, CA 94305, USA}
\affiliation{Department of Particle Physics and Astrophysics, SLAC National Accelerator Laboratory, Stanford, CA 94305, USA}

\author{Kate Storey-Fisher}
\affiliation{Center for Cosmology and Particle Physics, Department of Physics, New York University, 726 Broadway, New York, NY 10003, USA}

\author{Risa H. Wechsler}
\affiliation{Kavli Institute for Particle Astrophysics and Cosmology and Department of Physics, Stanford University, Stanford, CA 94305, USA}
\affiliation{Department of Particle Physics and Astrophysics, SLAC National Accelerator Laboratory, Stanford, CA 94305, USA}

\correspondingauthor{Zhongxu~Zhai}
\email{zhongxuzhai@sjtu.edu.cn}

\begin{abstract}
We analyze clustering measurements of BOSS galaxies using a simulation-based emulator of two-point statistics. We focus on the monopole and quadrupole of the redshift-space correlation function, and the projected correlation function, at scales of $0.1\sim60~h^{-1}$Mpc. Although our simulations are based on $w$CDM with general relativity (GR), we include a scaling parameter of the halo velocity field, $\gamma_f$, defined as the amplitude of the halo velocity field relative to the GR prediction. We divide the BOSS data into three redshift bins. After marginalizing over other cosmological parameters, galaxy bias parameters, and the velocity scaling parameter, we find $f\sigma_{8}(z=0.25) = 0.413\pm0.031$, $f\sigma_{8}(z=0.4) = 0.470\pm0.026$ and $f\sigma_{8}(z=0.55) = 0.396\pm0.022$. Compared with Planck observations using a flat $\Lambda$CDM model, our results are lower by $1.9\sigma$, $0.3\sigma$ and $3.4\sigma$ respectively. These results are consistent with other recent simulation-based results at non-linear scales, including weak lensing measurements of BOSS LOWZ galaxies, two-point clustering of eBOSS LRGs, and an independent clustering analysis of BOSS LOWZ. All these results are generally consistent with a combination of $\gamma_f^{1/2}\sigma_8\approx 0.75$. We note, however, that the BOSS data is well fit assuming GR, i.e. $\gamma_f=1$. We cannot rule out an unknown systematic error in the galaxy bias model at non-linear scales, but near-future data and modeling will enhance our understanding of the galaxy--halo connection, and provide a strong test of new physics beyond the standard model. 
\end{abstract}


\section{Introduction}

Clustering analysis of galaxies provides important information for us to understand the spatial distribution and evolution of the underlying dark matter. The relationship between luminous galaxies and dark matter can serve as a constraint on galaxy formation physics, which is necessary for an unbiased determination of cosmological parameters. The observations from large-scale cosmological surveys, e.g., the Sloan Digital Sky Survey (SDSS-I/II, \citealt{SDSS_York, Abazajian_2009}), the Two Degree Field Galaxy Redshift Survey (2dFGRS, \citealt{Colless_2001, Cole_2005}), WiggleZ (\citealt{Drinkwater_2010}), BOSS (\citealt{Dawson_BOSS}), and eBOSS (\citealt{eBOSS_Dawson}) have provided spatial information for millions of galaxies and produced significant results that inform our understanding of the universe. Ongoing and future spectroscopic surveys such as the Dark Energy Spectroscopic Instrument (DESI, \citealt{DESI_2016}), 4MOST (\citealt{Jong_2016}), PFS (\citealt{Takada_2014}), Euclid (\citealt{Laureijs_2011, Laureijs_2012}), and NASA's Nancy Grace Roman Space Telescope (Roman, \citealt{Green_2012, Dressler_2012, Spergel_2015, Wang_2021}) will continue to map the structure of the Universe with unprecendented volumes and precision. The compilation and analysis of these current and future data will enhance our ability to measure the structure and evolution of the universe in the past billions of years. Using a method developed previously in this series, in this paper we analyze BOSS data to extract cosmological information from clustering on small scales and put constraints on cosmological parameters. 

Retrieval of cosmological information on small scales is challenging due to the lack of an accurate and convenient analytic model to describe the non-linear dynamics of dark matter, as well as our incomplete understanding of the baryonic processes that impact the spatial distribution of galaxies. In \cite{Zhai_2019}, we proposed an emulator approach for galaxy clustering based on $N$-body simulations to investigate this problem, targeting massive BOSS galaxies at $z=0.55$. The idea is to run a limited number of high-resolution cosmological simulations that can efficiently sample cosmological parameter space. We then augment these dark matter only (DMO) simulations with a galaxy bias model to compute galaxy correlation functions. Our galaxy bias model is based on the Halo Occupation Distribution (HOD; see \citealt{Wechsler_2018} and references therein). This theoretical template is combined with Gaussian Processes (GP) to model the dependence of galaxy statistics on both cosmology and galaxy bias. The resultant product is able to make predictions for an arbitrary set of parameters, both cosmological as well as those controlling the galaxy--halo connection, in an accurate and efficient manner. Therefore, it is possible to sample this enlarged parameter space to achieve the posterior distribution of unknown cosmological parameters and correctly marginalize over nuisance parameters. We demonstrate that there is more constraining power at non-linear scales than linear scales to constrain the growth rate of structure. In this paper, we apply this method to BOSS galaxies distributed within $0.18<z<0.62$ to constrain various parameters of our cosmological model and measure the growth rate of structure. Since the pioneering works of \citet{Heitmann_2009, Heitmann_2010}, \citet{Lawrence_2010}, and \citet{Heitmann_2014}, the emulator approach has been widely applied in literature to model the statistics of galaxy and dark matter on non-linear scales. Using the $N$-body simulation from the Aemulus Project (\citealt{DeRose_2018}), we have constructed emulators for the halo mass function (\citealt{McClintock_2018}), the halo bias function (\citealt{McClintock_2019}), the non-linear power spectra of biased tracers (\citealt{Kokron_2021}) and for investigating galaxy assembly bias (McLaughlin et. al. in preparation). This approach has also been used in studies such as the EuclidEmulator (\citealt{Knabenhans_2019}) to model the non-linear corrections to the dark matter power spectrum in preparation for the Euclid mission, the {\sc Dark Emulator} (\citealt{Nishimichi_2019, Kobayashi_2020, Miyatake_2021}) for dark matter halo statistics over a wide redshift range, extensions to $\Lambda$CDM cosmologies (\citealt{Giblin_2019, Ramachandra_2020}), baryonic effects in the matter power spectrum (\citealt{Arico_2020}), the modeling of the Lyman-$\alpha$ forest flux power spectrum (\citealt{Simeon_2019, Rogers_2019, Walther_2020, Pedersen_2020}), galaxy lensing and clustering for BOSS-LOWZ galaxies (\citealt{Wibking_2017, Wibking_2020}), weak lensing signal of galaxy clusters (\citealt{Salcedo_2020}), and so on.

Based on the Aemulus suite, \cite{Lange_2021} measured the linear growth rate for the BOSS-LOWZ galaxies using clustering measurement on small scales, while \cite{Chapman_2021} applied the emulator method of \cite{Zhai_2019} and performed a cosmological analysis with eBOSS LRG data using a similar range of scales. This work extends the previous works in three different ways. First, we use a different, and larger, redshift range. Second, we incorporate a model for the effects of galaxy assembly bias. Third, through a scaling parameter, we decouple the velocity field of dark matter halos from that predicted by general relativity (GR), enabling us to constrain departures from GR.

This paper is a direct application of \cite{Zhai_2019} to data from BOSS DR12. The method to construct the emulator for the galaxy correlation function remains the same but with two main differences in the analysis: (1) galaxy selection from BOSS DR12 and (2) incorporating assembly bias in the HOD model. \cite{Zhai_2019} modeled BOSS-CMASS galaxies at effective redshift $z\sim0.55$ with the number density of $n=4.2\times10^{-3}[h^{-1}\text{Mpc}]^{-3}$, corresponding to the peak value over the redshift range. However, the flux limit of the selection can introduce incompleteness in terms of luminosity or stellar mass. Thus, in this paper, we apply additional selection criteria to the BOSS galaxies to reduce the effect of sample incompleteness, although this effect is shown to be minor for the clustering analysis (\citealt{Tinker_2017, Zhai_2017}.

In \cite{Zhai_2019}, the connection between galaxies and dark matter halos is modeled by populating halos in the DMO simulations with a parameterized HOD. In particular, we assumed that the relationship can be simply modeled by $P(N|M)$, the probability distribution that a halo of mass $M$ contains $N$ galaxies of a given class, combined with parameters for spatial and velocity bias of galaxies {\it within} halos. This basic form assumes that the galaxy population, and therefore the clustering, is determined by halo mass only. However, result based on $N$-body simulations reveals that the halo clustering may depend on properties other than halo mass, which has been referred to variously as assembly bias or secondary bias (\citealt{Wechsler2001, Sheth_2004, Gao_2005, Harker_2006, Wechsler2006, Wechsler_2018}). The investigation of this assembly bias effect has been performed extensively with $N$-body simulations and semi-analytical models for galaxy formation and evolution. Depending on which secondary property is studied, the clustering of dark matter halos may have different levels of correlation. This effect can, theoretically, propagate into the distributions of galaxies that live in these halos and thus add additional complexity to the galaxy clustering analysis. Current studies have shown small but non-negligible correlation of galaxy and halo clustering with internal properties such as halo age, concentration, or spin, or external properties such as large scale environment (see e.g., \citealt{Mao_2015, Hearin_2016, Mao_2018, Villareal_2017, Lehmann_2017, Shi_2018, Salcedo_2018, Zentner_2019, Contreras_2020} and references therein). Although the precise details and physical reasons for the assembly bias are not fully understood, it is now clear that it is necessary to incorporate this effect into the standard HOD approach to create a robust model for the analysis of BOSS galaxy clustering, and to correctly marginalize over any such effect when obtaining constraints on cosmological parameters. Therefore, in this work, we extend the basic HOD model by adding more parameters to describe the clustering dependence on halo environment.

In addition to yielding unbiased cosmological constraints, searching for the galaxy assembly bias itself is also an active topic of research. However, recent attempts based on SDSS and BOSS galaxies present contradictory results on the significance of galaxy assembly bias (\citealt{Vakili_2019, Walsh_2019, Salcedo_2020, Yuan_2020}). Although the construction of the assembly bias models in these studies are different, it implies that our understanding of the galaxy assembly bias is not complete, necessitating a general and flexible parameterization within the galaxy bias model. Using the emulator approach and the extended HOD model, we will investigate this problem and examine any bias induced in the cosmological constraint in this paper.

The emulator approach applied in this paper enables constraints on fundamental cosmological parameters using small scale galaxy clustering, with an emphasis on the growth of the dark matter structure. Our pilot study (\citealt{Zhai_2019}) demonstrated that small scale clustering has more constraining power than large scales using perturbation theory to measure the parameter combination $f\sigma_{8}$. The result of this analysis can give accurate measurement over the entire redshift range of BOSS galaxies. In addition, our modeling of the dark matter halo velocity field also marginalizes over the modification of underlying gravity. We employ a phenomenological model by scaling the velocity field of dark matter halos with a free parameter. This extra degree of freedom can mimic, in a simplified manner, the effect of modified gravity. Therefore, clustering analyses in redshift space are able to probe deviations from GR. The result can help us better understand the families of cosmological models proposed to explain the cosmic acceleration.

Our paper is organized as follows: In Section \ref{sec:BOSS}, we introduce the BOSS galaxies and the sample selection. Section \ref{sec:simulations} describes the simulation suites used in the analysis. Section \ref{sec:HOD} lays out the galaxy--halo connection model. Section \ref{sec:galaxy_clustering} introduces galaxy statistics for the clustering measurement. Section \ref{sec:likelihood} describes the construction of the covariance matrix and prior for the likelihood analysis. Section \ref{sec:Results} presents our cosmological measurements and systematics analysis. We discuss and list our conclusions in Section \ref{sec:conclusion}.

\section{Observational Data} \label{sec:BOSS}

\begin{table}
\caption{Number of galaxies used in our analysis.}
\begin{center}
\begin{tabular}{rccc}
\cline{1-4}
 Redshift  & Type  & NGC & SGC \\ 
 \cline{1-4}
$0.18<z<0.32$  & DR12  &  115187   & 54399  \\
~             & Selected  &  82103  &  34249  \\
\cline{1-4}
$0.32<z<0.48$  &  DR12  &  230093  &  103861  \\
~             & Selected  & 158610  &  66509  \\
\cline{1-4}
$0.48<z<0.62$  &  DR12  &  342844  &  143024 \\
~             & Selected  &  209697  &  89950 \\
\cline{1-4}
\end{tabular}
\end{center}
\label{tab:galaxy_number}
\end{table}

In this paper, the analysis uses the large-scale structure catalog created from the BOSS observations as described in \cite{Reid_2016}, including the survey footprint, veto masks and observational systematics. We use both the galaxy and random catalogs created for clustering measurements. BOSS targets galaxies with two selection algorithms: the LOWZ sample at $z\sim0.3$ and the CMASS sample at $z\sim0.55$. We follow the strategy of \cite{Alam_2017} to use a combined LOWZ+CMASS sample covering the entire redshift range of $0.2<z<0.7$. We note that the color cuts and flux limits used in target selection introduce incompleteness in the BOSS galaxies, i.e., the BOSS galaxy sample is not a volume-limited sample. A typical halo occupation analysis assumes that the galaxy sample being modeled is volume limited. Although the incompleteness of BOSS galaxies has been quantified in \cite{leauthaud_etal:16} and \cite{Tinker_2017}, and the results show that the incompleteness does not have a significant impact on the clustering measurement or HOD analysis of the data (\citealt{Zhai_2017}), this incompleteness should be minimized in order to construct galaxy samples that are best suited for HOD analysis. Therefore we prepare the galaxy selection as follows. 

\begin{figure}[htbp]
\begin{center}
\includegraphics[width=9cm]{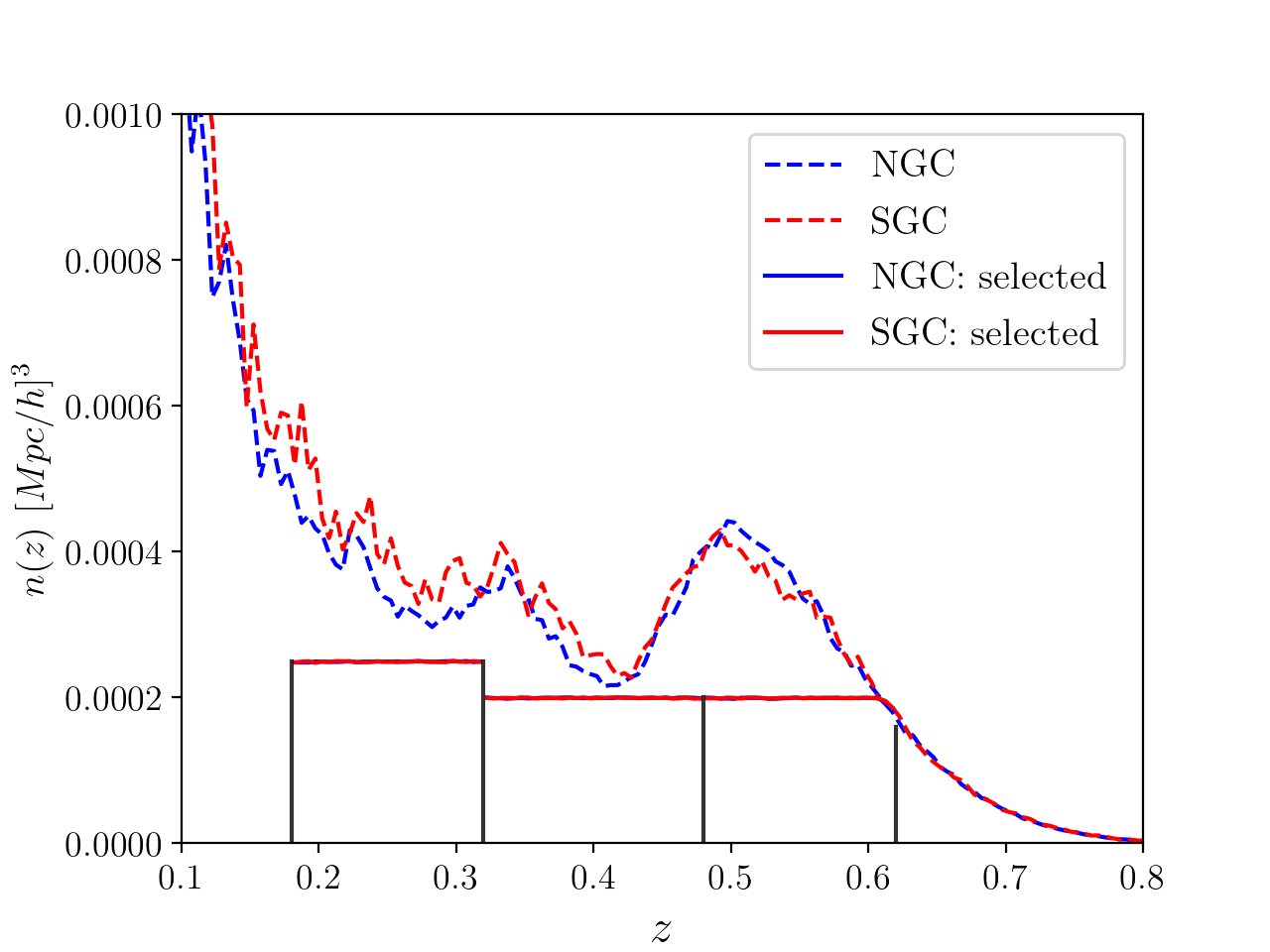}
\caption{The comoving number density of the BOSS DR12 galaxies as a function of redshift. The sample used in our analysis is defined by the galaxy brightness such that both NGC and SGC can reach a constant number density, as indicated by the horizontal lines. The grey vertical lines split the galaxies into low-$z$, med-$z$ and high-$z$ subsamples that will be analyzed individually.}
\label{fig:nz}
\end{center}
\end{figure}

\begin{figure*}[htbp]
\begin{center}
\includegraphics[width=19cm]{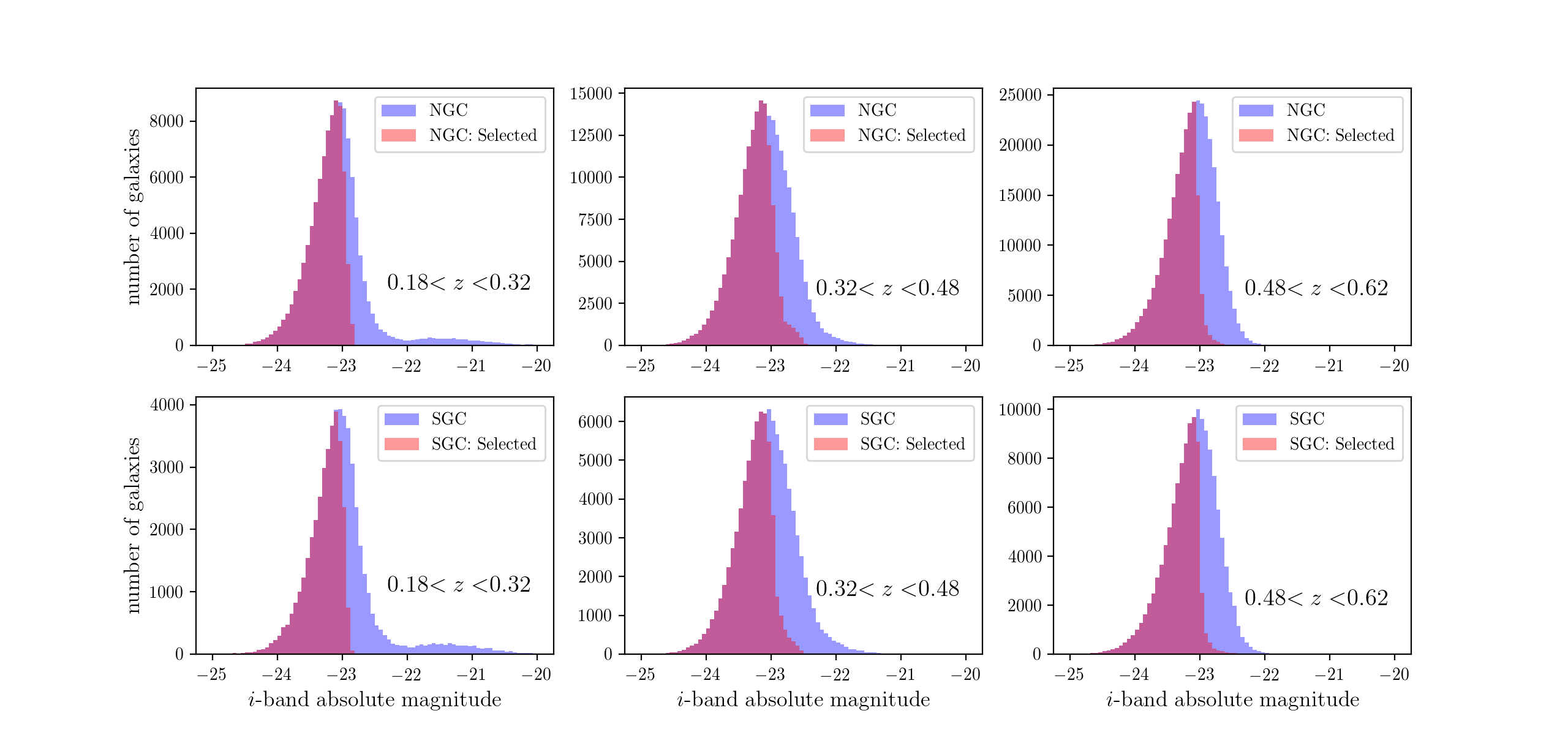}
\caption{Distribution of the $i-$band absolute magnitude of the BOSS galaxies. Our sample corresponds to the brighter end with a nearly hard cut at $-23$. This redshift-dependent selection can provide a galaxy sample with constant number density across redshift, which is close to a volume-limited sample.}
\label{fig:mag}
\end{center}
\end{figure*}

We first compute the galaxy number density, $n(z)$, of the BOSS sample assuming a spatially flat $\Lambda$CDM with $\Omega_{m}=0.31$ as a fiducial model to compute the cosmic volume. The result is shown in Figure \ref{fig:nz} for the North Galactic Cap (NGC) and South Galactic Cap (SGC) respectively. The local minimum at $z\sim0.4$ represents the transition between CMASS and LOWZ. We then split the galaxy sample into three redshift slices: $0.18<z<0.32$ (low-$z$), $0.32<z<0.48$ (med-$z$) and $0.48<z<0.62$ (high-$z$) and analyze the clustering measurements separately for each redshift slice. For each subsample, we apply a thin redshift binning with e.g., $\Delta z=0.005$. In each of these fine slices, we convert the $i-$band apparent magnitude of the galaxies to absolute magnitude and rank-order by the luminosity. Then we select the bright end of this galaxy subsample to have a number density that is constant across the redshift slice. For low-$z$, we use $2.5\times10^{-4}[h^{-1}\text{Mpc}]^{-3}$.  For both med-$z$ and high-$z$, we use $2.0\times10^{-4}[h^{-1}\text{Mpc}]^{-3}$. This results a galaxy sample with a constant number density across each of the redshift range of interest, as shown in the solid lines in Figure \ref{fig:nz}. We apply this selection separately for NGC and SGC and ensure they have the same number density by definition. Table \ref{tab:galaxy_number} summarizes the number of galaxies in the resultant samples. In Figure \ref{fig:mag}, we display the distribution of $i-$band absolute magnitude for each galaxy sample in NGC and SGC respectively. Although the luminosity threshold varies some across each bin in order to preserve the number density, this is significantly closer to a volume-limited sample than the original flux-limited target selection. Thus, these new galaxy samples are more appropriate for halo occupation analysis.

\section{Simulations} \label{sec:simulations}

The non-linear dynamics of dark matter can be well captured by $N$-body simulations (see, e.g., \citealt{Klypin_2011, Klypin_2016}). This also provides the theoretical framework for us to extract cosmological information from the BOSS galaxies. In this section we introduce the simulation suites employed in our work, as well as examinations of systematics. Table \ref{tab:simulations} summarizes the key information of these simulation suites and their functions in our analysis. Briefly, we use three different types of simulations: (1) The {\sc Aemulus} simulation suite (\citealt{DeRose_2018}), which is used to build the emulator, (2) high-resolution simulations that resolve substructure, which are used to test the emulator, and (3) lower-resolution PM simulations run with the GLAM code (\citealt{Klypin_2018}) to construct a covariance matrix for the BOSS clustering measurements.

\begin{table*}
\centering
\begin{tabular}{llllll}
\hline
Simulation  & Cosmology & Box size & Number of simulations & What it is used for & Reference \\
\hline
Aemulus & multiple    & $1.05h^{-1}$Gpc  & 75       & Building the emulator     & \cite{DeRose_2018}   \\
GLAM    & Planck-like & $1.0h^{-1}$Gpc   & 986      & Covariance matrix      & \cite{Klypin_2018}      \\
Uchuu   & Planck      & $2.0h^{-1}$Gpc   & 1        & External test of SHAM  & \cite{Ishiyama_2020} \\
UNIT    & Planck      & $1.0h^{-1}$Gpc   & 2 pairs  & External test of SHAM  &  \cite{Chuang_2019}  \\
\hline
\end{tabular}
\caption{Simulations used for the analysis in this paper.}
\label{tab:simulations}
\end{table*}

\subsection{The Aemulus suite}

We use the {\sc Aemulus}\footnote{\url{https://aemulusproject.github.io}} simulation presented in \cite{DeRose_2018} to build and test the emulator for galaxy clustering. The {\sc Aemulus} suite comprises 75 boxes of dark-matter-only $N$-body simulations, with 40 simulations at different cosmologies for training the emulator, and 35 additional simulations at 7 cosmologies for testing. The test set has 5 realizations at each cosmology for better statistics. The cosmologies of the training set are chosen based on $w$CDM model in an optimized Latin hyper-cube designed parameter space (\citealt{Heitmann_2009}). The cosmological parameters for the {\sc Aemulus} suite are the matter density $\Omega_{m}$, the baryon energy density $\Omega_{b}$, the amplitude of matter fluctuations $\sigma_{8}$, the dimensionless Hubble parameter $h$, the spectral index of the primordial power spectrum $n_{s}$, the equation of state of dark energy $w$ and the number of relativistic species $N_{\text{eff}}$. All the simulations have a volume of $1.05h^{-1}$Gpc with $1400^3$ dark matter particles, yielding a mass resolution appropriate for resolving halos that host massive galaxy populations like the BOSS samples. For $\Omega_m=0.3$, the particle mass is $\sim3.5\times10^{10}h^{-1}\text{M}_{\odot}$. We use the training set to build the emulator for the different galaxy clustering statistics, and we use the test set to evaluate the emulator performance and quantify the uncertainty, as done in \cite{Zhai_2019}. In Table \ref{tab:param}, the first section summarizes the cosmological parameters and the range relevant for the emulator construction in the following sections. We refer the readers to \cite{DeRose_2018} for more details on the {\sc Aemulus} simulations.

\subsection{Uchuu and UNIT}

We use the Uchuu\footnote{\url{http://www.skiesanduniverses.org/Simulations/Uchuu/}} \citep{Ishiyama_2020} and UNIT\footnote{\url{https://unitsims.ft.uam.es}} \citep{Chuang_2019} simulations to build galaxy mocks and do recovery tests of our emulator. The galaxy--halo connection in our emulator is based on an HOD in which the parameters of the mean occupation function vary with halo environment, thus mimicking the effect of galaxy assembly bias. In order to validate the robustness of this approach in modeling the galaxy clustering at non-linear scale and the inferred cosmological measurement, we test this approach against mock galaxy catalogs produced via the SubHalo Abundance Matching method (SHAM; i.e., \citealt{Kravtsov_2004, Vale_2004, Conroy_2006}; see \citealt{Wechsler_2018} for a review). This model assigns galaxies to dark matter halos based on the assumption that the stellar mass or luminosity of a galaxy is correlated with the properties of dark matter halo or subhalo hosting this galaxy. The SHAM model has far fewer parameters than HOD but it can well match observed galaxy statistics (e.g., \citealt{Lehmann_2017}). The standard mapping process between galaxies and dark matter halos in the SHAM approach  yields some amount of galaxy assembly bias. Therefore testing our HOD-based model with this SHAM model is able to tell us whether different models of assembly bias can bias the cosmological constraints.

We use two simulation suites to create SHAM mocks: Uchuu and UNIT; both adopt the Planck 2015 cosmology (\citealt{Planck_2015}). The Uchuu simulation has 2.1 trillion dark matter particles in a 8$[h^{-1}\text{Gpc}]^{3}$ box with a particle mass of $3.27\times10^{8}h^{-1}M_{\odot}$. This volume is $\sim 8$ times larger than the test boxes from {\sc Aemulus}, enabling a more precise measurement of clustering statistics. For the UNIT simulations, we use the $1h^{-1}$Gpc boxes with a particle number of $4096^{3}$. This simulation adopts the inverse phase technique (\citealt{Angulo_2016}) to reduce  cosmic variance. We use all the 4 boxes (2 pairs) in our analysis for galaxy clustering, implying that the effective volume is higher than 4$[h^{-1}\text{Gpc}]^{3}$. For both simulations, we use the method of \cite{Lehmann_2017} to assign galaxies to dark matter halos and subhalos. In this method, the property used to rank halos is a combination of the maximum circular velocity within the halo, $v_{\rm max}$, and the virial velocity, $v_{\rm vir}$. This combination allows the user to vary the amount of assembly bias the galaxies exhibit. More details on the parameters used and the results of the tests are in Appendix E.

\subsection{GLAM simulations}

We use the GLAM\footnote{\url{http://www.skiesanduniverses.org/Simulations/GLAM/}} simulations to construct the covariance matrix for the likelihood analysis. The GLAM simulations are run with the new Parallel Particle-Mesh (PM) $N$-body code (PPM-GLAM) with a box size of 1$h^{-1}$Gpc and 2000$^{3}$ particles. The high speed of this code has enabled nearly 1000 independent realizations at the Planck cosmology, a set of simulations large enough to construct a robust covariance matrix for all our clustering measurements. In our work, we use 986 boxes with redshift outputs equal to the mean redshifts of our BOSS samples, to estimate the covariance matrix for our BOSS correlation function measurements. The details of the GLAM simulation can be found in \cite{Klypin_2018}.

\section{galaxy--halo Connection Model} \label{sec:HOD}

In this paper, we adopt the HOD approach to model the galaxy--halo connection. The HOD describes the galaxy population within dark matter halos in a statistical manner. We start with the model of \cite{Zhai_2019} to define the mean occupation function for central and satellite galaxies, which is in turn based on the \cite{Zheng_2005} HOD model. Our implementation includes three additional parameters that control the concentration of the radial distribution of satellite galaxies and velocity biases of both centrals and satellites. The parameters are summarized in the second section of Table \ref{tab:param}. Note that the ranges of certain parameters are enlarged compared to \cite{Zhai_2019} to better fit the BOSS measurements. This is due to the fact that the galaxy number density in this paper is lower than that of the \cite{Zhai_2019} emulator, thus we are modeling the clustering of an intrinsically brighter galaxy population with a higher clustering amplitude. For reference, in Appendix \ref{appsec:boss_clustering}, we show the clustering measurements for the BOSS galaxies with and without our selection based on luminosity. Due to the correlation between galaxy luminosity and their host halo mass, these galaxies are likely living in more massive halos and are thus more clustered.

Compared with the previous model in \cite{Zhai_2019}, the critical change in this analysis is the addition of a galaxy assembly bias model, in which galaxy occupation depends on halo properties other than mass. With the lack of consensus on observational constraints on galaxy assembly bias, there is flexibility to choose secondary halo properties to investigate the assembly bias, including both internal and external properties. In this work, we focus on the bias introduced by the environment, i.e., an external property. In particular, we define the halo environment as the dark matter overdensity of dark matter halo. We measure the relative density $\delta$ of dark matter halos using a top-hat window function with radius of $10h^{-1}\text{Mpc}$. Then we modify the HOD model by scaling the parameter $M_{\text{min}}$ through
\begin{equation}\label{eq:Mmin_AB}
    \bar M_{\rm min} = M_{\text{min}}\left[1 + f_{\rm env}\text{erf}\left(\frac{\delta-\delta_{\rm env}}{\sigma_{\rm env}}\right)\right].
\end{equation}
This functional form modulates the dependence of $M_{\text{min}}$ as a function of local density and its value can be determined by other HOD parameters when galaxy number density is fixed (\citealt{Zhai_2019}). The amplitude parameter $f_{\text{env}}$ controls the overall strength of the dependence and the resultant level of assembly bias, the position parameter $\delta_{\text{env}}$ determines the threshold to split halos living in over- and under-dense regions, and the width parameter $\sigma_{\text{env}}$ controls the smoothness of the transition from under-density to over-density. This assembly bias model allows the minimum mass scale for dark matter halos to host a central galaxy to depend on halo environment, which has a direct impact on the occupancy of centrals and satellites and therefore can change the clustering signal compared with basic HOD model, see e.g., Figure 4 of \cite{Walsh_2019}. The degrees of freedom introduced through this parameterization can also enable the investigation of galaxy formation physics by looking at galaxies formed in halos of same mass but different environment. For the following analysis, setting $f_{\text{env}}=0$ can simply turn off the modeling of assembly bias and returns the basic HOD model. The parameterization chosen allows for significant flexibility in assembly bias; the impact on clustering can be negligible or very large. The change induced by the assembly bias can be to either increase or decrease the clustering amplitude, and the effect can occur at any density. 

The HOD model assumes the central occupancy $N_{\text{cen}}$ approaches 1 at the very massive end, i.e., the most massive halos must host a LRG at the center. However, this is not necessarily true given the target selection of BOSS galaxies and the mass incompleteness (\citealt{leauthaud_etal:16}). In order to investigate its impact on our measurement of linear growth rate from small scales, we introduce parameter $f_{\max}$ to scale the amplitude of $N_{\text{cen}}$ within the range [0.1, 1.0]. This parameter is equivalent to $f_{\Gamma}$ in \cite{Lange_2021} for a similar discussion. Other earlier works for a similar modeling can be found in \citet{Hoshino_2015} and \citet{Chapman_2021}.

The addition of these new parameters makes our previous HOD model more flexible. The construction of the emulator in this extended parameter space for galaxy correlation function follows the method developed in \cite{Zhai_2019}. More details, and the performance of the emulator, are described in Appendix \ref{appsec:emulator}. We note that the choice of the assembly bias model in this paper is not the only option. It is possible to incorporate galaxy assembly bias that depends on other properties of dark matter halos. However, recent studies based on hydrodynamic simulation and semi-analytic model show that the local environment of the halo at the present day is an excellent predictor of assembly bias (\citealt{Han_2019, Yuan_2020, Xu_2020}), as well as results from $N$-Body simulation (\citealt{Yuan_2021}). Therefore we apply this particular model of assembly bias throughout this work, but extensions to other model are also possible. In addition, this assembly bias model only applies to the host halos and not to subhalos. Correlations caused by satellites, for example introduced by relationships with concentration, halo formation time etc can be modeled using a more flexible method; but we do not address those here. We note that there are alternatives to extend the basic HOD model with assembly bias, e.g. the decorated HOD algorithm (\citealt{Hearin_2016}; see also McLaughlin et. al. (in preparation). This model redistributes galaxies within the same halo mass bin by secondary halo property, while preserving the original HOD after marginalization. Our implementation is less physically motivated, but has more flexibility to describe the types of galaxy assembly bias signals induced by various forms the assembly bias may take.

\begin{table*}
\centering
\begin{threeparttable}
\begin{tabular}{llcr}
\hline
& Parameter    & Meaning & Range \\
\hline
Cosmology & $\Omega_{m}$      & The matter energy density    & [0.255, 0.353]   \\
  & $\Omega_{b}$         & The baryon energy density        & [0.039, 0.062]      \\
  &  $\sigma_{8}$ & The amplitude of matter fluctuations on 8 $h^{-1}$Mpc scales.      & [0.575, 0.964]     \\
  &  $h$       & The dimensionless Hubble constant     &  [0.612, 0.748]  \\
  &  $n_{s}$       & The spectral index of the primordial power spectrum     & [0.928, 0.997]    \\
  &  $w$ & The dark energy equation of state     & [-1.40, -0.57]   \\
  &  $N_{\text{eff}}$ & The number of relativistic species  &  [2.62, 4.28] \\
  &  $\gamma_{f}$ & The amplitude of halo velocity field relative to $w$CDM+GR  & [0.5, 1.5] \\
\hline 
  HOD &  $\log{M_{\rm{sat}}}$  & The typical mass scale for halos to host one satellite  & [14.0, 15.5] \\
  &  $\alpha$  & The power-law index for the mass dependence of the number of satellites & [0.2, 2.0]\\
  &  $ \log{M_{\rm{cut}}}$ & The mass cut-off scale for the satellite occupation function  &  [10.0, 13.7]\\
  &  $\sigma_{\log{M}}$  & The scatter of halo mass at fixed galaxy luminosity  & [0.05, 1.0]\\
  &  $\eta_{\rm{con}}$ & The concentration of satellites relative the dark matter halo   &  [0.2, 2.0]\\
  &  $\eta_{\rm{vc}}$  &  The velocity bias for central galaxies  &   [0.0, 0.7]\\
  &  $\eta_{\rm{vs}}$  & The velocity bias for satellite galaxies  &   [0.2, 2.0]\\
  &  $f_{\text{max}}$ & The incompleteness parameter for central occupancy & [0.1, 1.0] \\ 
\hline
 Assembly Bias &  $f_{\rm{env}}$  & Amplitude parameter for assembly bias  &   [-0.3, 0.3]\\
  &  $\delta_{\rm{env}}$  & Position parameter for assembly bias  &   [0.5, 2.0]\\
  &  $\sigma_{\rm{env}}$  & Width parameter for assembly bias  &   [0.1, 1.0]\\
\hline
\end{tabular}
\end{threeparttable}
\caption{Parameters used in our emulator, their physical meaning, and the parameter space range for each parameter. For the HOD parameters, assembly bias parameters, and for $\gamma_f$, the range is used as a flat prior in analysis. For the cosmological parameters, we use a prior defined by the training cosmologies themselves. See text for further details. }
\label{tab:param}
\end{table*}

\section{Measuring Galaxy clustering} \label{sec:galaxy_clustering}

\subsection{Two-point Correlation Function (2PCF)}

We quantify the clustering for both BOSS and simulated data sets using the two-point correlation function $\xi(r)$, which measures the excess probability of finding two galaxies separated by a vector distance $\mathbf{r}$, relative to a random distribution, for all $|\mathbf r| = r$. In practical applications, the distance to galaxies is determined by redshift, which can be distorted by peculiar velocities, also known as the redshift-space distortion (RSD) effect. Therefore the measured galaxy distribution in redshift space is different than in real space, but these differences are driven by the amplitude of the peculiar velocity field and thus contain information about the growth rate of large-scale structure. We measure $\xi_{Z}(r_{p}, \pi)$ on a two-dimensional grid of separations perpendicular $r_{p}$ and parallel $(\pi)$ to the line of sight through 
\begin{equation}
\pi = \frac{\mathbf{s}\cdot\mathbf{l}}{|\mathbf{l}|}, \quad r_{p}=\mathbf{s}\cdot\mathbf{s}-\pi^2,
\end{equation}
with $\mathbf{l}=(\mathbf{s}_{1}+\mathbf{s}_{2})/2$ (\citealt{Davis_1983, Fisher_1994}), the subscript "Z" denotes redshift space. In order to reduce the effect of redshift-space distortions and extract information in real space (with subscript "R" in the following equation), we compute the projected correlation function (\citealt{Davis_1983})
\begin{equation}
w_{p}(r_{p})=2\int_{0}^{\infty}d\pi\xi_{Z}(r_{p}, \pi) = 2\int_{0}^{\infty}d\pi\xi_{R}(r=\sqrt{r_{p}^2+\pi^2}).
\end{equation}
This integral needs to be truncated at some scale in the measurement from the observational sample or mock catalog. We choose $\pi_{\text{max}}=80$ $h^{-1}$Mpc, which is large enough to give stable results. To encapsulate the clustering in redshift space, we measure the multipoles of the correlation function. With $\mu=r_{p}/s$, we use the standard decomposition with Legendre polynomial to obtain
\begin{equation}
\xi_{\ell}(s) =\frac{2\ell+1}{2}\int_{-1}^{1} L_{\ell}(\mu)\xi_{Z}(s, \mu)d\mu,
\end{equation}
where $L_{\ell}$ is the Legendre polynomial of order $\ell$. Most of the information in redshift space is contained in the first few even multipoles and in this work we use $\xi_0$ and $\xi_2$.

\subsection{Measurement from BOSS}

Using the galaxy samples defined in Section \ref{sec:BOSS}, we measure the 2PCF through the estimator (\citealt{LS_1993})
\begin{equation}\label{eq:LS}
\xi(r_{p}, \pi)=\frac{DD-2DR+RR}{RR},
\end{equation}
where $DD$, $DR$, and $RR$ are suitably normalized numbers of (weighted) data\textendash data, data\textendash random, and random\textendash random pairs in each separation bin. The positions of the BOSS galaxies are converted from RA, DEC and redshift to Cartesian coordinates assuming a Planck 2015 $\Lambda$CDM model with $\Omega_{m}=0.307$. The choice of cosmology for this transformation has a negligible impact on the final result (\citealt{Chapman_2021, Lange_2021}).

\begin{figure*}[htbp]
\begin{center}
\includegraphics[width=18cm]{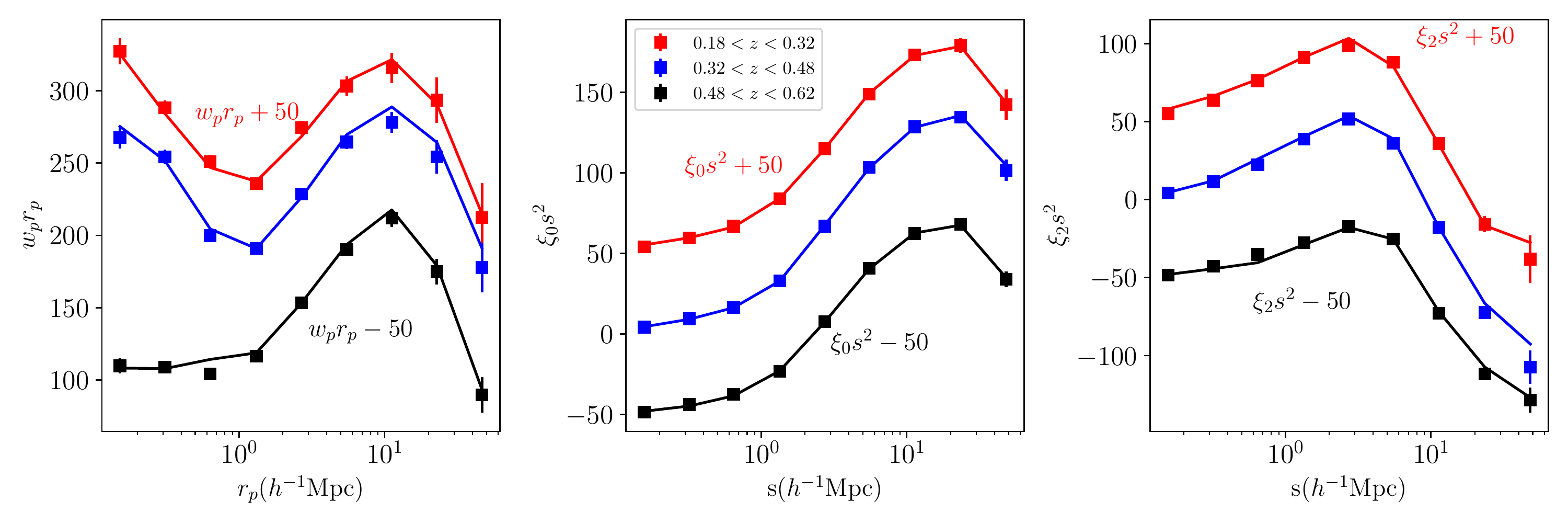}
\caption{The two-point correlation function of BOSS galaxies, including $w_{p}$(left), $\xi_{0}$(middle), and $\xi_{2}$ (right) for all three redshift bins. The lines are the prediction of the best-fit model with varying cosmology, HOD and assembly bias parameter, assuming the fiducial model for covariance matrix (see more details in Section \ref{sec:covariance}). The best-fit model uses data from $w_{p}+\xi_{0}+\xi_{2}$. The results for low-$z$ and high-$z$ are shifted slightly for plotting purpose.}
\label{fig:BOSS_2pcf}
\end{center}
\end{figure*}

In BOSS clustering measurements, there is an important systematic due to ``fiber collisions''. The spectroscopic redshift of BOSS galaxies is obtained by fibers, which have a physical scale of 62 arcsec on a given tile; any two fibers can not be placed closer than this scale. This leads to a fraction of galaxies without redshift determination. The loss of these galaxies has an impact on the clustering measurement at all scales, and is more significant on small scales. This fiber collision effect has been corrected using multiple methods in the clustering measurements, such as nearest neighbor or angular up-weighting methods. In our analysis, we adopt the method developed in \cite{Guo_2012}, which is based on the fact that the fiber collision can be resolved in areas that are observed within more than one tile. This method can recover the projected and redshift space two-point correlation function on scales below and above the collision scale. Our final measurements of the BOSS galaxies are shown in Figure \ref{fig:BOSS_2pcf} for all three redshift bins. For $w_{p}$ and $\xi_{\ell}$, we choose logarithmically-spaced bins for $r_{p}$ or $s$ from 0.1 to 60.2 $h^{-1}$Mpc, resulting in 9 data points for each statistic. For $\xi_{\ell}$, $\mu$ is binned linearly with 40 bins from 0 to 1. Both the measurement from BOSS galaxies and from the mocks use the same binning scheme.

\section{Likelihood Analysis} \label{sec:likelihood}

\subsection{Covariance matrix} \label{sec:covariance}

The covariance matrix is of critical importance in the likelihood analysis. Here we pursue several methods of quantifying the covariance matrix of the BOSS clustering measurements in order to determine the sensitivity of our results to the details of the matrix construction. It can be estimated using a large number of simulations, or through the data themselves, and the matrix can employ a combination of these two approached. 
Based on the observational data, we first measure the covariance matrix through jackknife re-sampling with 400 roughly equal sub-areas of the BOSS angular footprint. We note that this method may lead to a noise-dominated covariance matrix due to limited size of subsamples. Simulation-based methods have significantly less noise, but have inherent assumptions when constructing the mock galaxies used. 
To construct these mocks, we use the emulator of the 2PCF as constructed in the previous section to find an HOD model that can give consistent clustering measurements with the data. In particular, we fix the cosmological parameters to be those of the GLAM cosmology, and we also assume no assembly bias (i.e., $f_{\text{env}}=0$). The resulting HOD model is then used to populate the halos within the GLAM simulations to produce 986 galaxy mocks. We repeat this process for all three redshifts. 
The correlation matrix of these two approaches give quite consistent behavior of the galaxy 2PCF, with the fact that the mock-based method is substantially smoother and thus reduces the effect of noise in the calculation of the likelihood. 

With these two methods, we construct the following four covariance matrices:
\begin{itemize}
    \item Jackknife: Full covariance matrix from Jackknife re-sampling.
    \item GLAM: Full covariance matrix from the GLAM mocks.
    \item Fiducial: Diagonal elements from jackknife re-sampling combined with the correlation matrix from GLAM mocks.
    \item Modified Fiducial: Same as the fiducial, but using a different HOD model for the galaxy mocks.
\end{itemize}
As listed above, our fiducial matrix uses the data to set the amplitude of the errors, but we use the simulations for the shape of the covariance matrix. The off-diagonal elements are scaled by the corresponding diagonal elements. In the modified fiducial matrix, we choose a different HOD model from one of the training sets that is close to the 2PCF measurement from BOSS and use this HOD model to regenerate 986 GLAM mocks. This tests any sensitivity to the details of the galaxy bias model. We apply this test to the high redshift $z=0.55$ subsample. We find minimal dependence on the details of the HOD model. Further details on this test and the construction of the various covariance matrices can be found in Appendices \ref{appsec:covariance_matrix_construction} and \ref{appsec:covariance}.

The covariance matrices described above correspond to the contribution from sample variance in the data $C_{\text{sam}}$. When we perform the actual analysis, we also need to take into account the uncertainties from the emulator itself. Thus the final covariance matrix is
\begin{equation}
C = C_{\text{sam}}+C_{\text{emu}},    
\end{equation}
where $C_{\text{emu}}$ is the intrinsic error from the emulator prediction. To compute this, we adopt the same method as in \cite{Zhai_2019}. Simply put, the raw emulator performance show as the shaded area in Figure \ref{fig:emulator_error} has two contributions: intrinsic error of emulator and sample variance of the testing simulations. We assume these two terms are independent and therefore the intrinsic error of the emulator is the total error with the sample variance subtracted off in quadrature. In addition, we also assume the emulator error is independent among different $r_{p}$ and $s$ bins, thus $C_{\text{emu}}$ is diagonal only. This differs from \cite{Zhai_2019}, in which we assumed $C_{\text{emu}}$ had the same correlation matrix as $C_{\text{sam}}$, but our tests show this has a minor impact on the final constraints. A followup analysis is left for a future work (Storey-Fisher et al, in preparation).

\subsection{Sampling algorithm and priors} \label{sec:prior}

We perform the our analysis using the likelihood function
\begin{equation}\label{eq:likelihood}
\ln{\mathcal{L}} = -\frac{1}{2}(\xi_{\text{emu}}-\xi_{\text{obs}})C^{-1}(\xi_{\text{emu}}-\xi_{\text{obs}}),
\end{equation}
where $\xi_{\text{emu}}$ and $\xi_{\text{obs}}$ are the correlation function from the emulator and observational data respectively and $C$ is the covariance matrix as defined above. Depending on tests, $\xi_{\text{obs}}$ is from either BOSS measurements or galaxy mocks.

The likelihood analysis is done through Bayesian statistics. In particular, we explore the parameter space using the nested sampling algorithm (\citealt{Skilling_2004}) implemented in \medium{MultiNest} (\citealt{Feroz_2009, Buchner_2014}) package. This method can compute the Bayes Evidence through the integral
\begin{equation}
    \mathcal{Z}=\int\mathcal{L}(\mathbf{p})p({\mathbf{p}})d\mathbf{p},
\end{equation}
where $p(\mathbf{p})$ is the prior given parameter vector $\mathbf{p}$. The Bayes Evidence has been widely used to evaluate the model selection in different scientific fields. The output from \medium{MultiNest} also gives posterior distributions of the parameters in the model. Compared with traditional methods to obtain the posterior, such as Markov Chain Monte Carlo (MCMC)-like method as we used in \cite{Zhai_2019}, the nested sampling needs fewer evaluations to reach convergence. In our analysis, we choose 1000 live-point to sample the high-dimensional parameter space. A typical run for likelihood analysis using emulators of all three statistics $w_{p}+\xi_{0}+\xi_{2}$ takes roughly five thousand CPU hours to get converged results.

Another critical ingredient is the prior $p(\mathbf{p})$. For our HOD parameters, we choose a flat and un-informative prior defined by the range of the parameters (Table \ref{tab:param}). However, the cosmological parameter space in our analysis is restricted by the sampling of the training cosmologies, see for instance Figure 3 of \cite{DeRose_2018}. This means the that emulator is only guaranteed to produce reliable prediction of the 2PCF within this CMB+BAO+SNIa defined area. 
Therefore we define a prior space for the cosmological parameters based on the training cosmologies. In particular, this prior is defined by an ellipsoid in seven-dimensional space and we restrict the nested sampling to be within this prior range. We present more details of the training area in Appendix \ref{appsec:prior}.

\subsection{Recovery tests}

Before we apply this likelihood analysis to the BOSS galaxies, we first perform a recovery test using the SHAM galaxy mocks. The details and results are shown in Appendix \ref{appsec:sham}. It shows that with different details for the SHAM mock construction, our HOD-based emulator is able to recover the input cosmology successfully with the parameters constrained within $1\sigma$ level, and thus validate the emulator construction. In Appendix \ref{appsec:covariance}, we present the constraints using different setups for the covariance matrix. The consistency between different covariance matrices shows that the effect on the final cosmological measurement is not significant. Compared with the smoother correlation matrix from GLAM mock, the noise in the jackknife resampling method does not bias the cosmological constraints.

\section{Results from BOSS Galaxies} \label{sec:Results}

In this section, we present constraints using BOSS galaxies. We begin with a presentation of our results when implementing myriad priors and assumptions on the analysis. We then focus on the impact that galaxy assembly bias has on our results. Our results are summed up in our constraints on the growth rate of structure, both when assuming GR or when allowing $\gamma_f$, the scaling parameter of the velocity field, to be a free parameter. Last, we compare our results to others in the field that use simulation-based approaches to extract cosmological information from small scales.

\subsection{Constraints on key cosmological parameters}\label{sec:constraint_BOSS}

The fiducial covariance matrix is constructed using the jackknife resampling for the uncertainty and GLAM mocks for the off-diagonal elements of the correlation matrix as explained in the previous section. All results in this section will use this covariance mtarix. The observational data for galaxy statistics are $w_{p}+\xi_{0}+\xi_{2}$. In order to have a comprehensive investigation, we perform several tests as follows:
\begin{itemize}
    \item Fiducial: Varying cosmological parameters + HOD parameters + assembly bias parameters.
    \item Planck Prior: Gaussian prior on a subset of the cosmological parameters ($\Omega_{m}, \Omega_{b}$, $\sigma_{8}$, $h$, $n_{s}$) using the latest Planck 2018 measurements. In particular we use the result from the chain \texttt{plikHM\_TTTEEE\_lowl\_lowE\_lensing}. We use this particular chain as Planck 2018 measurements throughout the paper, unless it is described explicitly. The other cosmological parameters $w$, $N_{\text{eff}}$ and $\gamma_{r}$ have the same prior as the fiducial case. 
    \item Fixing $\gamma_{f}=1.0$: This forces the analysis to use General Relativity to describe gravity.
    \item Fixing $w=-1$: This forces the analysis assuming $\Lambda$CDM.
    \item No assembly bias: We set $f_{\text{env}}=0$ to turn off the assembly bias modeling.
\end{itemize}

\begin{figure*}[htbp]
\begin{center}
\includegraphics[width=5.96cm]{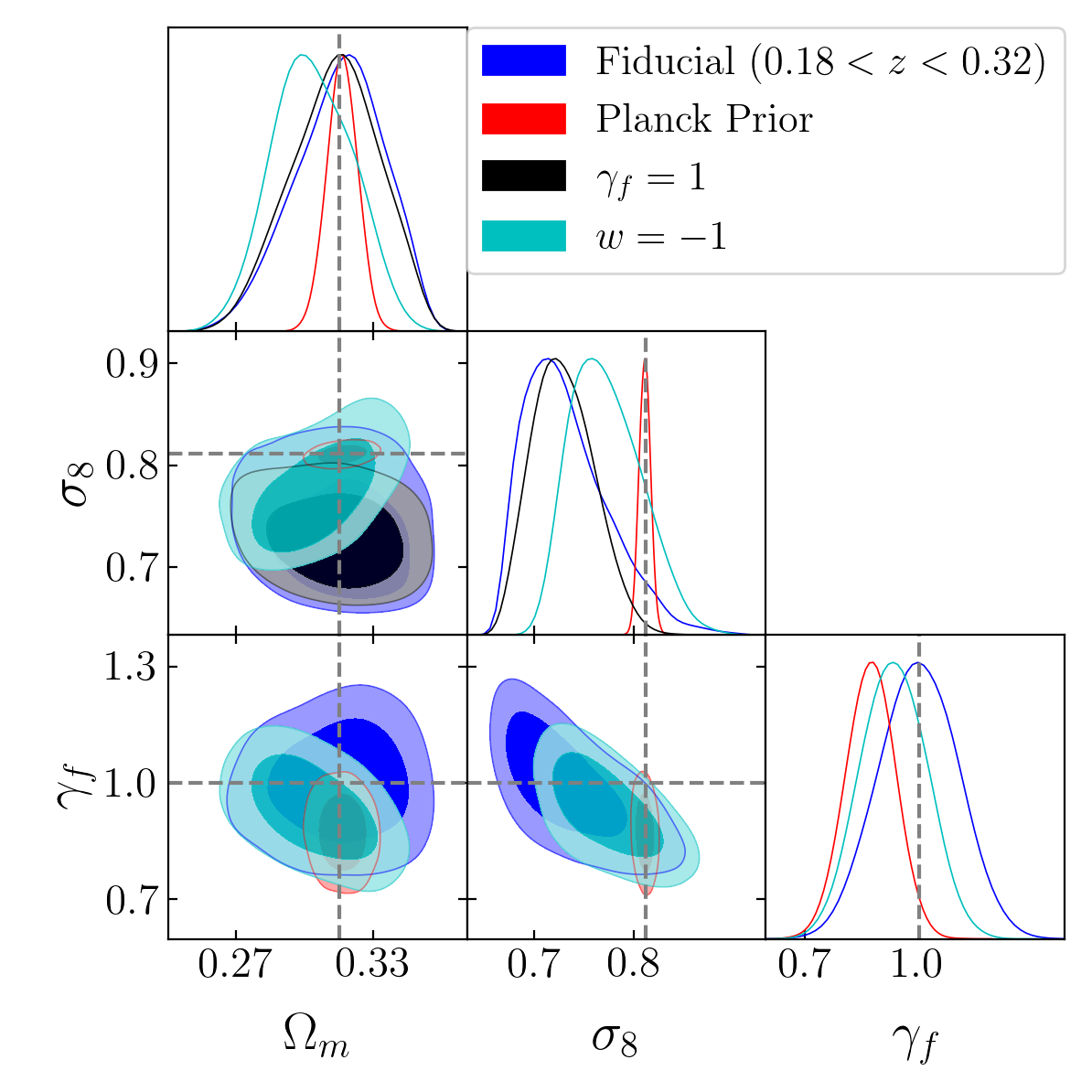}
\includegraphics[width=5.96cm]{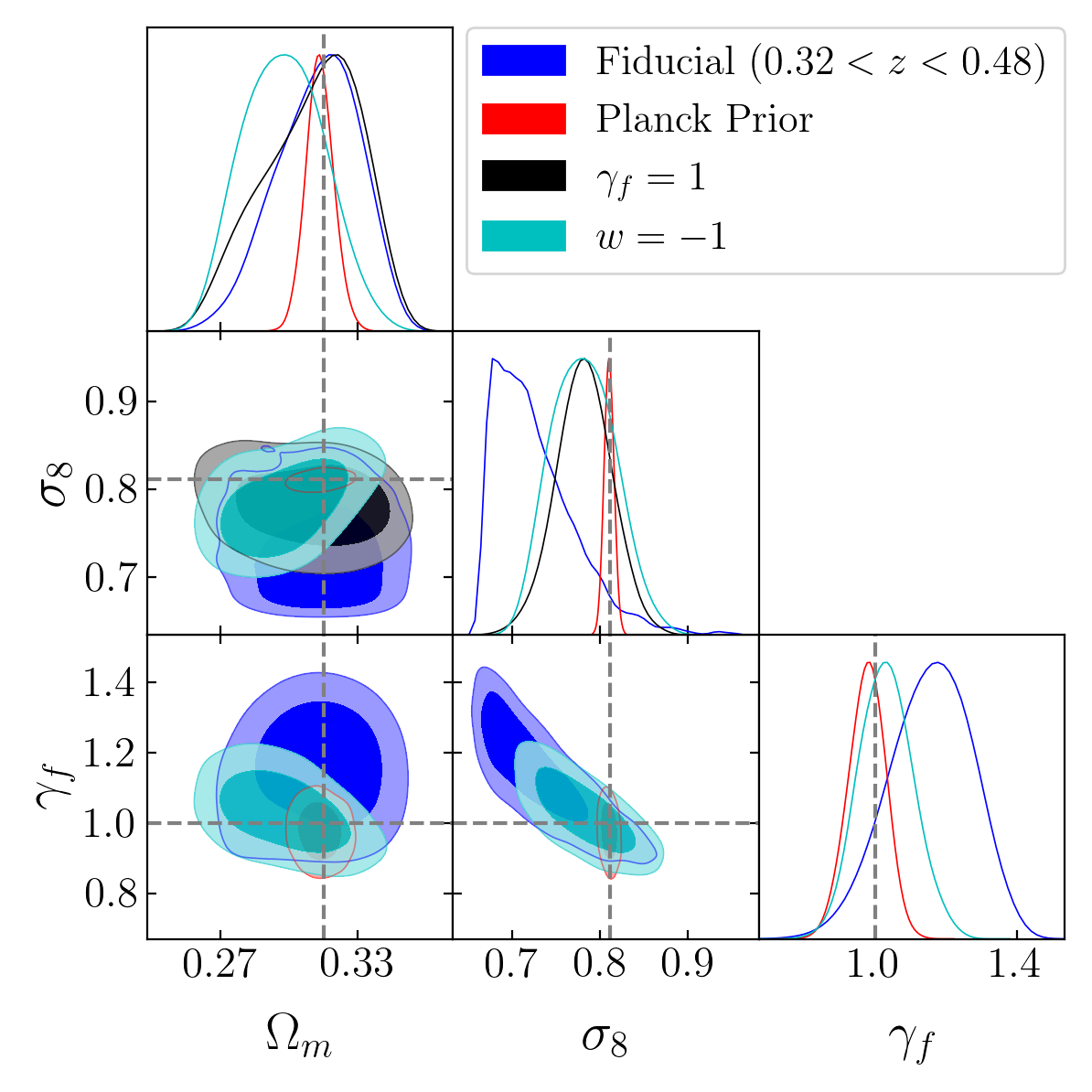}
\includegraphics[width=5.96cm]{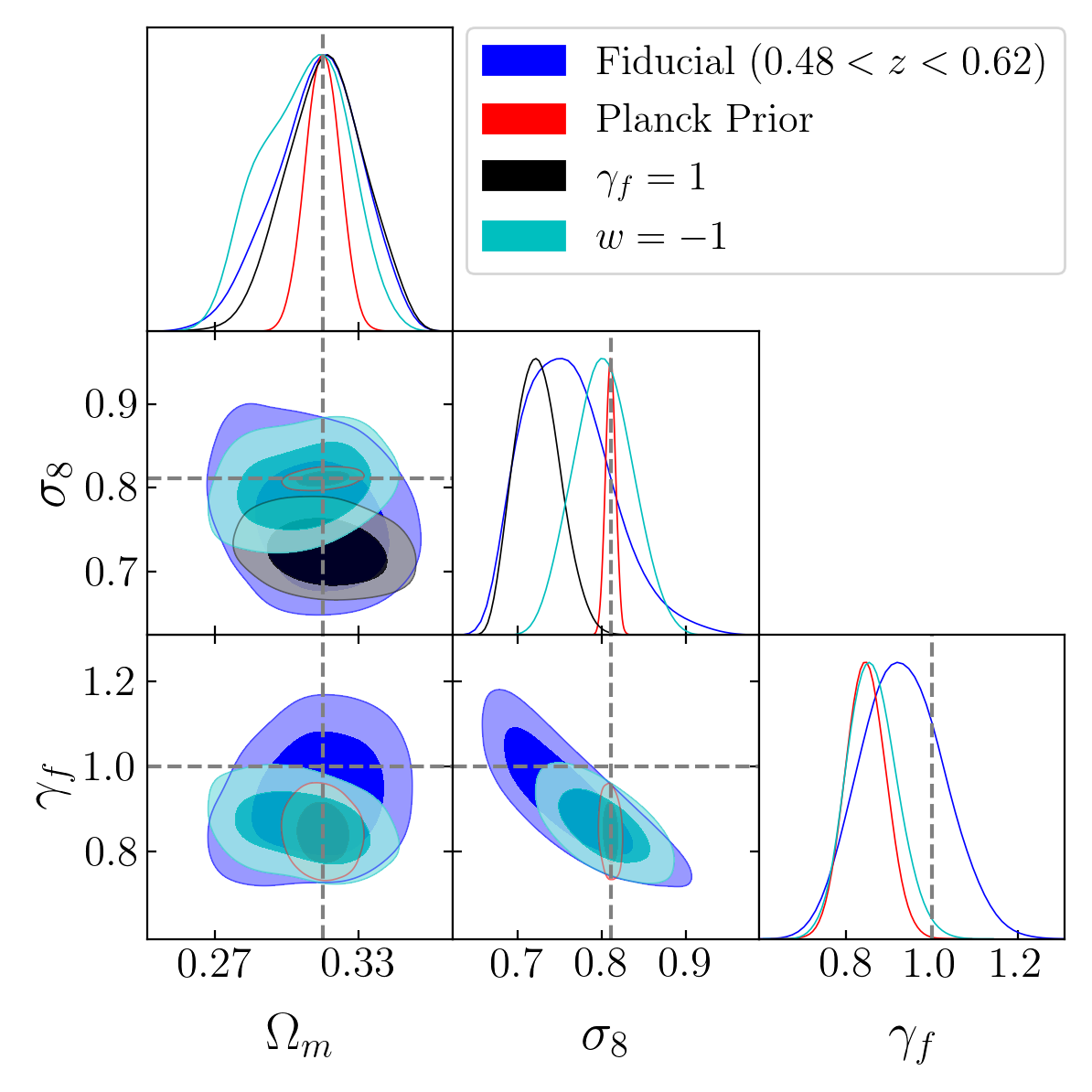}
\caption{Constraint on some of the key parameters of the low-$z$ (left), med-$z$ (middle), and high-z (right) subsamples, using $w_{p}+\xi_{0}+\xi_{2}$. The contours show 1- and 2-$\sigma$ confidence levels. The result shows a comparison of three different tests: (1) Fiducial test (blue): varying cosmology+HOD+assembly bias parameters with an uninformative prior; (2) adopting a Gaussian prior on a subset of the cosmological parameters $(\Omega_{m}, \Omega_{b}, \sigma_{8}, h, n_{s})$ using Planck 2018 observation (red); (3) fixing $\gamma_{f}=1$ with no deviation from GR (black); and (4) fixing $w=-1$ to be $\Lambda$CDM (cyan). The dashed lines for cosmological parameters indicate the best-fit model of Planck with $\gamma_{f}=1$ and $f_{\text{env}}=0$.}
\label{fig:lowz_constraint}
\end{center}
\end{figure*}

In Figure \ref{fig:lowz_constraint}, we present the constraints on the key parameters of interest ($\Omega_{m}, \sigma_{8}, \gamma_{f}$) for the three redshift bins respectively. The constraints on the model parameters are also summarized in Table \ref{tab:constraint}. Using our fiducial priors, all three redshift bins yield similar constraints on these parameters that influence the growth of structure. As expected, there is a degeneracy between $\sigma_8$ and $\gamma_f$, with higher values of $\sigma_8$ yielding lower values of $\gamma_f$. The amplitude of this degeneracy curve lies below the Planck+GR value of $(\sigma_8,\gamma_f)=(0.82,1.0)$. This implies that that the peculiar velocity field of BOSS galaxies is roughly 15\% lower than the Planck+GR prediction. Although we focus on the cosmological parameters that control the growth of structure, the constraints on all cosmological parameters, for each redshift bin, are shown in Figure \ref{fig:cosmology_summary} and presented in Appendix \ref{appsec:full_triangle}. 

Although the \medium{MultiNest} algorithm adopted in our likelihood analysis is not designed for goodness-of-fit analysis, the posterior produced from the computation allows the search of a minimum for $\chi^2$ that is close to the result from a minimization algorithm. In our fiducial analysis, the free parameters include 7 cosmological parameters, 1 parameter for dark matter halo velocity field $\gamma_{f}$, and 12 total HOD parameters. The data vector has 9 points for each of $w_{p}$, $\xi_{0}$ and $\xi_{2}$. Therefore the number of degrees of freedom is 8. We find $\chi^{2}=6.88$ for the low-$z$ subsample, $\chi^{2}=6.52$ for the med-$z$ subsample, and $\chi^{2}=16.8$ for the high-$z$ subsample. We note that the high-$z$ subsample gives a large $\chi^{2}$ relative to the degrees of freedom. However this result is dominated by the fit to $\xi_{0}$ at the smallest radial bin.  
We recompute the $\chi^{2}$ by excluding this single data point using the best-fit model. The resulting value is $\chi^{2}=12.32$, indicating a more reasonable result. Although these results imply a statistically good fit to the data, we note that $\chi^2$ per degree of freedom is only a rough indicator. The data points are correlated, reducing the number of degrees of freedom. However, for the clustering analysis in this paper and our particular attention to the measurement of growth rate of structure, most of the constraining power is only from a subset of parameters, implying that the ``effective" number of degrees of freedom is actually {\it higher} than the 8 listed above (i.e., see the discussion in \citealt{Lange_2021}). 

In addition to our fiducial constraints on the key cosmological parameters, Figure \ref{fig:lowz_constraint} also shows the results when adopting a Planck prior for all cosmological parameters. This then yields the value of the halo velocity field relative to GR, $\gamma_f$, that is required to match the data. As expected from the fiducial results, when enforcing this prior, the best-fit values of $\gamma_f$ are all below unity, with values of 0.87, 0.97, 0.84, all with errors of $\sim 0.05$. More notable, however, is that the $\chi^2$ values of the best-fit models are higher than in the fiducial analysis, with $\Delta\chi^2=5$, 4, and 5, for low-$z$, med-$z$, and high-$z$ respectively, indicating the mild difficulty of the Planck cosmology to fit the data, even with the added freedom of scaling the halo velocity field and an 12-parameter halo occupation model that includes assembly bias.

For comparison, these three figures also show results for an analysis in which the cosmological parameters adopt the fiducial prior, but with $\gamma_f=1$. This analysis is thus a fully $\Lambda$CDM+GR fit to the data. These results are indicated with the black contours in the $\Omega_m$--$\sigma_8$ panel. In all three redshift slices, the results are consistent with the fiducial analysis, and yield minimal changes to the best-fit $\chi^2$ values, with $\Delta\chi^2\approx 0$, 3, and 0 from low-$z$ to high-$z$, respectively. The constraints on $\Omega_m$ are consistent with the Planck results, but the constraints on $\sigma_8$ are significantly lower, with best-fit values of  $\sim 0.75$. This analysis demonstrates that a $\Lambda$CDM+GR model is sufficient to describe the data, but there is tension with the amplitude of clustering inferred from the CMB. The last test is to fix $w=-1$, i.e. force the model to be $\Lambda$CDM. Reducing this degree of freedom can change the best-fit $\chi^{2}$ values by $\sim1-2$ for all three redshift bins. The final constraint on the key cosmological parameters is quite consistent with the fiducial test but we note that fixing $w=-1$ can alter the degeneracies between some parameter pairs. The improvement on the measurement of linear growth rate is not surprising as in earlier work (\citealt{Chapman_2021}).

To visually demonstrate the impact of these prior assumptions on the fits to the data, in Figure \ref{fig:bestfit_BOSS_lowz} we present the best-fit emulator predictions for the galaxy correlation functions in all redshift bins, with residuals of the fits relative to the data. For $w_p(r_p)$, the fits are relatively consistent regardless of assumption. However, the adoption of the Planck prior impacts the amplitude of the monopole, especially for the low-$z$ and high-$z$ redshift slices. It is this change that drives the $\Delta\chi^2$ values. 

\begin{figure}[htbp]
\begin{center}
\includegraphics[width=8.5cm]{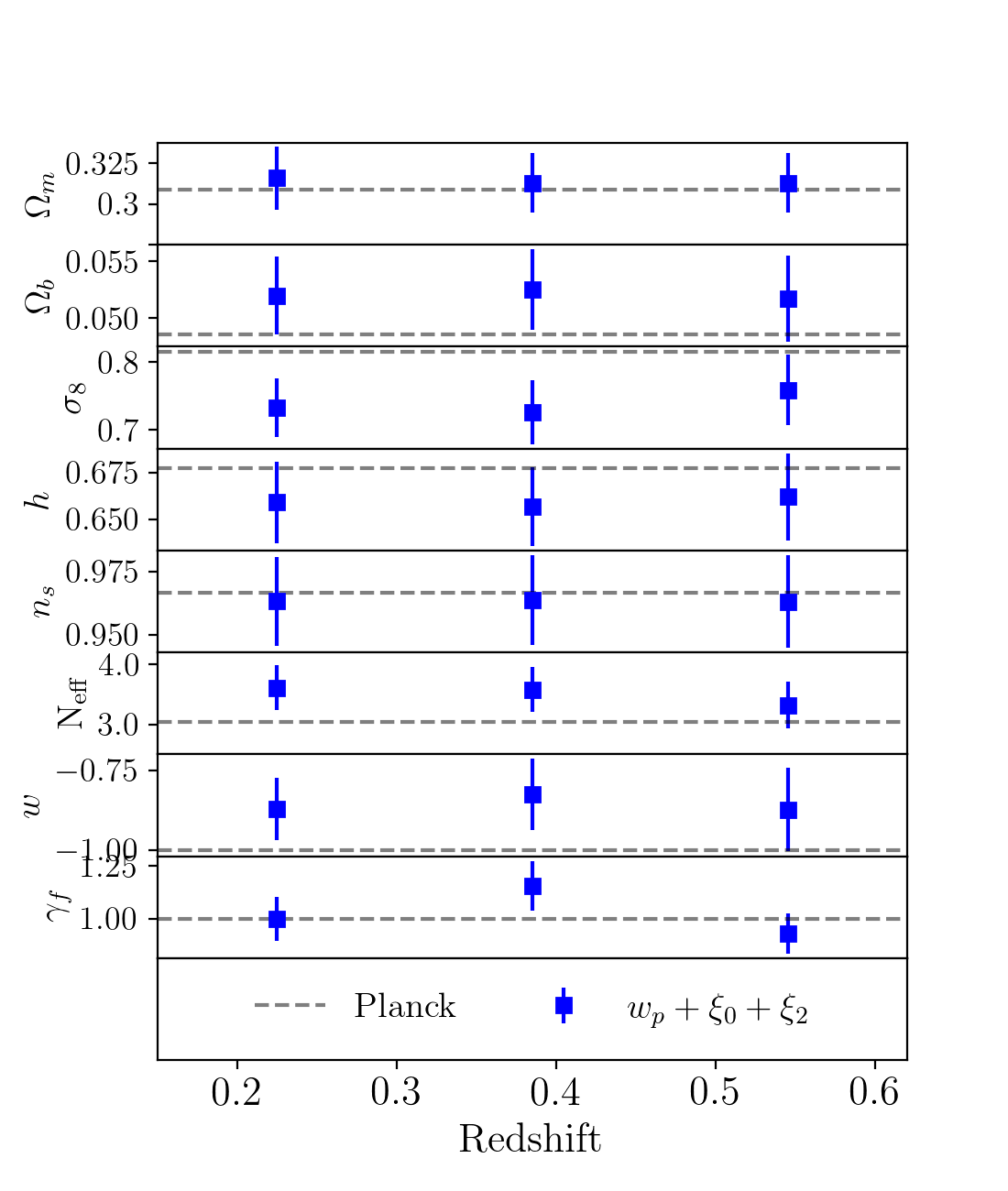}
\caption{Constraints on cosmological parameters from our analysis. The dashed lines correspond to the best-fit measurement from Planck using the baseline $\Lambda$CDM model.
}
\label{fig:cosmology_summary}
\end{center}
\end{figure}

\subsection{Halo occupation parameter constraints}

We note again that we include two parameters that encapsulate velocity bias between galaxies and dark matter. First, the orbits of satellite galaxies within their host halos may move faster or slower than expected from the virial velocity dispersion ($\eta_{vs}$). Second, central galaxies may have non-zero, random, velocities with respect to their host halos ($\eta_{vc})$. The velocity bias for BOSS central galaxies shows deviation from zero with significance of a few $\sigma$, and the deviation increases as we go to higher redshift. This indicates that centrals are not at rest with respect to their host halos, consistent with earlier findings in \cite{Guo_2015}, \cite{Yuan_2020}, and \cite{Lange_2021}. Note that we use a slightly different definition of the velocity bias for centrals than these works, but the results are consistent. 

The velocity bias for satellites is consistent with unity, indicating the velocity distribution of the satellites is well described by the virial dispersion of the dark matter halos. This is in agreement with \cite{Lange_2021}, who perform a similar clustering analysis using BOSS galaxies at $z=0.25$ and 0.4, as well as with the higher-redshift eBOSS LRG analysis of \cite{Chapman_2021}. This is in tension at some level with the result in \cite{Guo_2015} where the authors report a   $2\sigma$ constraint of $\alpha_{s}<1$ using BOSS DR11 galaxies, but we note their analysis is at fixed cosmology. Using CMASS galaxies, \cite{Yuan_2020} finds that satellite galaxies slightly prefer higher velocity than the dark matter particles but this result is still consistent with ours. However, their result has a large variation depending on the details of the fit. This implies the necessity of more detailed investigation of the velocity field traced by satellites. 

In our model, satellite galaxies follow an NFW density profile, with a concentration parameter that is proportional to that of the dark matter ($\eta_{\rm con}$). For the med-$z$ and high-$z$ bins, the satellite concentration parameter is roughly half that of the dark matter. For the low-$z$ bin, the best-fit value of $\eta_{\rm con}$ is higher, but the overall constraints on satellite concentrations are weak. Our constraints in the low-$z$ bin are consistent with those of \cite{Lange_2021} at the same redshift. It is important to note that $\eta_{\rm con}$ is not degenerate with any of our cosmological parameters.

In all three redshift bins, the data prefer a slight non-zero value of the assembly bias parameter $f_{\rm env}$, the parameter that governs the amplitude and sign of the bias. Positive $f_{\rm env}$ implies that the HOD halo mass scale increases at high densities by approximately 10\%, lowering the number of galaxies at high densities and reducing the overall amplitude of clustering. For reference, this value of $f_{\rm env}$ lowers the large-scale galaxy bias by $\sim 5$\% when all other HOD parameters are fixed. However, we note that the statistical significance of the result is $\lesssim 1\sigma$ in each redshift bin. This is in agreement with some earlier attempts, such as \cite{Walsh_2019}, \cite{Salcedo_2020}, and \cite{Lange_2021}, but in contrast to analyses by \cite{Zentner_2019} and \cite{Yuan_2020}. We note that these works employ different models for assembly bias with different data sets and simulations, thus a direct comparison of each result is not straightforward. Galaxy assembly bias may depend on multiple halo properties, such as concentration, age, and environment (\citealt{Han_2019}). Although the local density is an excellent indicator, it is possible that assembly bias due to other halo properties or their combinations is not fully accounted for. A detailed comparison of the assembly bias models and an evaluation of their impacts on the clustering analysis is necessary but beyond the scope of this paper. In McLaughlin et. al. (in preperation), we apply the similar emulator approach to model projected galaxy correlation function and galaxy lensing signal, and explicitly investigate the impact of assembly bias on the cosmological inference.

For reference, we present the full 1D and 2D constraints on all cosmological and halo occupation parameters in our analysis in Appendix \ref{appsec:full_triangle}.

\begin{table*}
\caption{Constraints on the cosmological parameters, HOD parameters and assembly bias parameters using clustering measurement of BOSS galaxies. The training priors for the cosmological parameters are adopted throughout the analysis and the numbers in the parenthesis represent the ratio of $68\%$ interval compared with the range of training space. }
\begin{center}
\begin{tabular}{rccc}
\cline{1-4}
  Parameter      & $0.18<z<0.32$ & $0.32<z<0.48$ & $0.48<z<0.62$  \\
\cline{1-4}
$\Omega_{m}$   & $0.315\pm0.019 ~(40\%)$ & $0.313\pm0.018 ~(37\%)$ & $0.313\pm0.018 ~(37\%)$\\
$\Omega_{b}$  & $0.052\pm0.0035 ~(30\%)$ & $0.053\pm0.0036 ~(31\%)$ & $0.052\pm0.0038 ~(33\%)$\\
$\sigma_{8}$ & $0.733\pm0.043 ~(22\%)$ & $0.726\pm0.047 ~(24\%)$  & $0.76\pm0.052 ~(27\%)$\\
$h$            & $0.659\pm0.022 ~(32\%)$  & $0.657\pm0.021 ~(31\%)$  & $0.663\pm0.024 ~(35\%)$ \\
$n_{s}$      & $0.963\pm0.017 ~(50\%)$  & $0.964\pm0.017 ~(50\%)$ & $0.963\pm0.018 ~(53\%)$\\
$N_{\text{eff}}$  & $3.61\pm0.378 ~(45\%)$ & $3.582\pm0.372 ~(45\%)$ & $3.319\pm0.387 ~(47\%)$\\
$w$          & $-0.87\pm0.105 ~(24\%)$ & $-0.825\pm0.112 ~(27\%)$  & $-0.874\pm0.131 ~(32\%)$\\
$\gamma_{f}$   & $0.998\pm0.104 (21\%)$ & $1.155\pm0.119 (24\%)$  & $0.931\pm0.095 (19\%)$\\
\cline{1-4}
$\log{M_{\text{sat}}}$    & $14.3\pm0.07 ~(9\%)$ & $14.44\pm0.127 ~(17\%)$ & $14.43\pm0.21 ~(28\%)$\\
$\alpha$      & $1.2\pm0.148 ~(16\%)$ & $0.978\pm0.154 ~(17\%)$ & $0.839\pm0.231 ~(26\%)$\\
$\log{M_{\text{cut}}}$ & $11.64\pm0.99 ~(53\%)$ & $11.78\pm0.96 ~(52\%)$ & $12.44\pm1.34 ~(73\%)$\\
$\sigma_{\log{M}}$  & $0.435\pm0.156 ~(35\%)$ & $0.336\pm0.163 ~(36\%)$ & $0.7\pm0.145 ~(32\%)$\\
$\eta_{\text{vc}}$   & $0.154\pm0.092 ~(26\%)$ & $0.135\pm0.083 ~(24\%)$ & $0.301\pm0.094 ~(27\%)$\\
$\eta_{\text{vs}}$   & $1.0\pm0.088 ~(10\%)$ & $1.131\pm0.107 ~(12\%)$ & $1.074\pm0.112 ~(12\%)$\\
$\eta_{\text{con}}$   & $1.227\pm0.342 ~(38\%)$ & $0.589\pm0.189 ~(21\%)$ & $0.606\pm0.24 ~(27\%)$\\
$f_{\text{max}}$  & $0.819\pm0.114 ~(25\%)$   & $0.738\pm0.13 ~(29\%)$    &  $0.82\pm0.124 ~(27\%)$  \\
\cline{1-4}
$f_{\text{env}}$   & $-0.054\pm0.091 ~(30\%)$ & $0.022\pm0.109 ~(36\%)$ & $-0.092\pm0.084 ~(28\%)$\\
$\delta_{\text{env}}$ & $1.261\pm0.481 ~(64\%)$  & $0.961\pm0.411 ~(55\%)$ & $1.155\pm0.45 ~(60\%)$\\
$\sigma_{\text{env}}$  & $0.551\pm0.257 ~(57\%)$ & $0.588\pm0.248 ~(55\%)$ & $0.577\pm0.25 ~(56\%)$\\
\cline{1-4}
\end{tabular}
\end{center}
\label{tab:constraint}
\end{table*}

\begin{figure*}[htbp]
\begin{center}
\includegraphics[width=18.5cm]{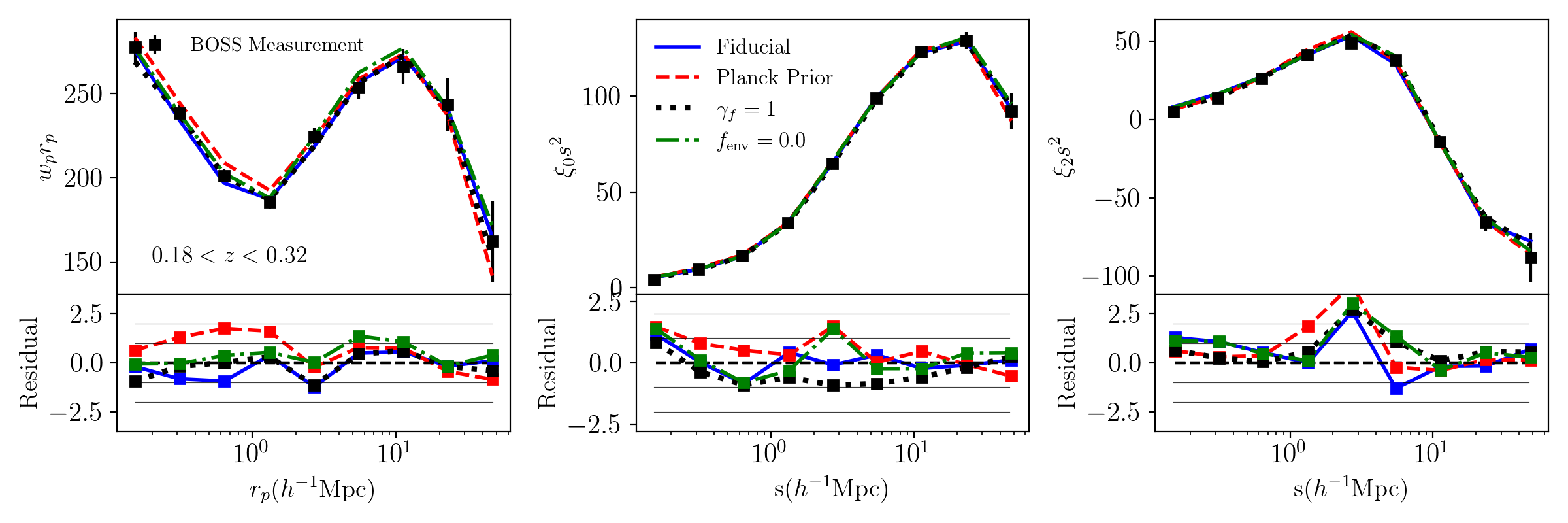}
\includegraphics[width=18.5cm]{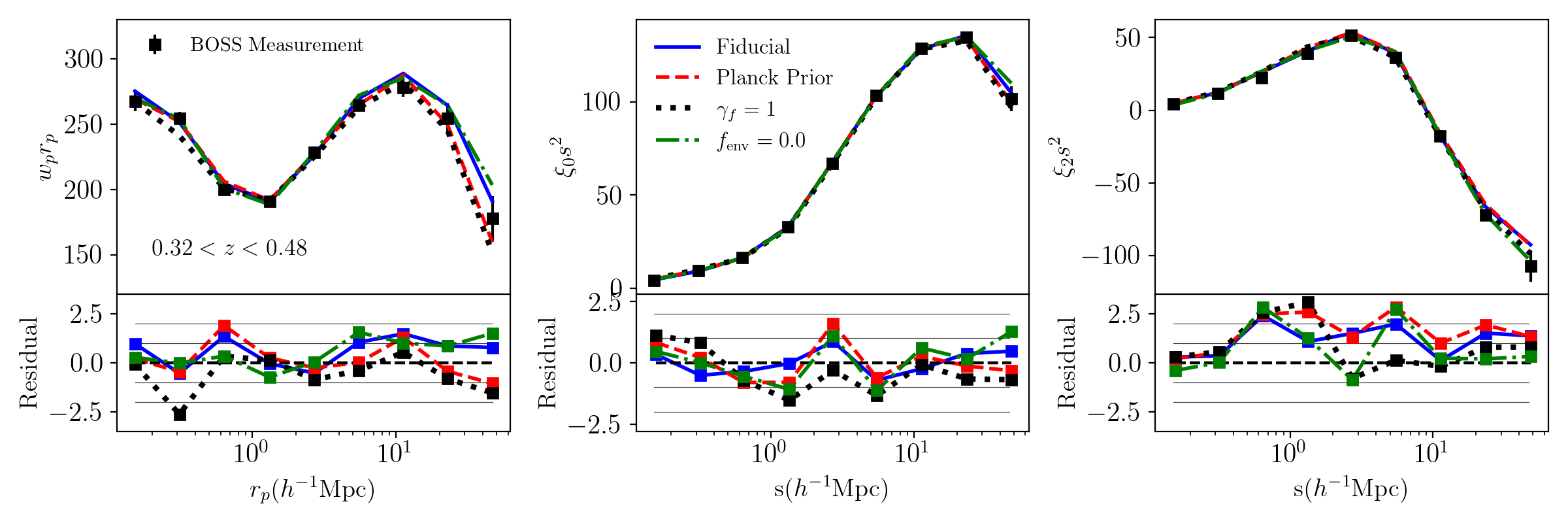}
\includegraphics[width=18.5cm]{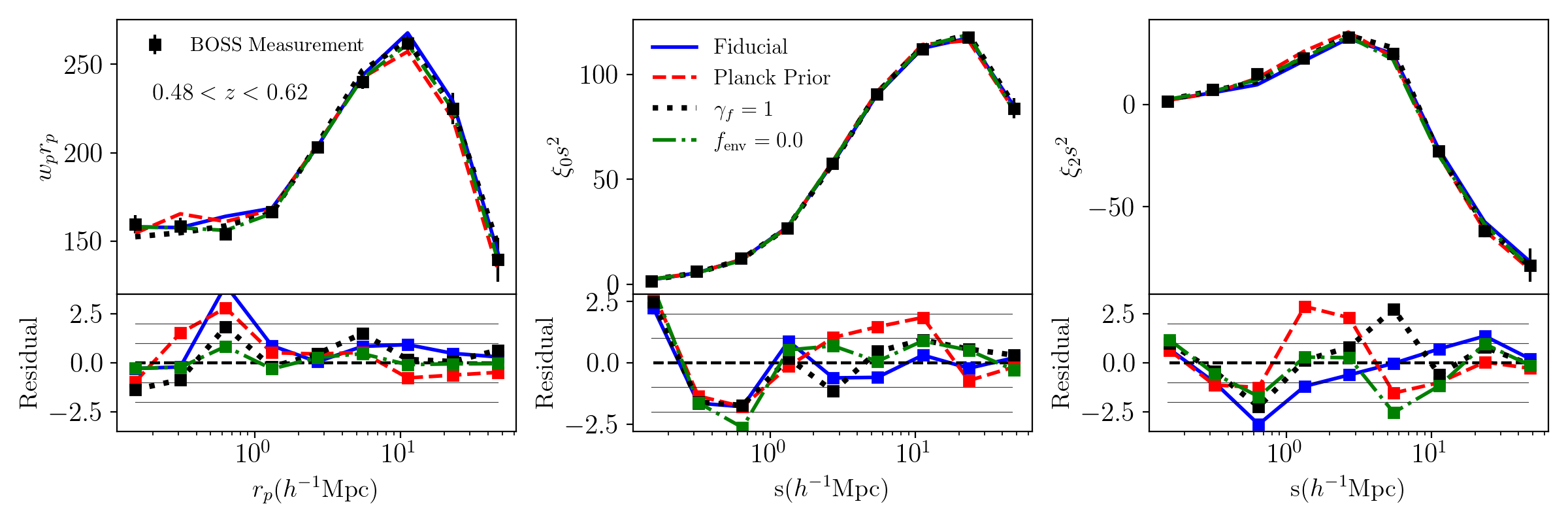}
\caption{Best-fit model of $w_{p}$ (left), $\xi_{0}$ (middle) and $\xi_{2}$ (right) for three subsamples, as well as the residuals with respect to the BOSS measurements. Different lines correspond to the tests as explained in the text. The bottom panels of each plot show the residual normalized by the observed uncertainty, and the horizontal thin lines correspond to 1 and 2$\sigma$ level.}
\label{fig:bestfit_BOSS_lowz}
\end{center}
\end{figure*}

\subsection{Effect of assembly bias}

\begin{figure*}[htbp]
\begin{center}
\includegraphics[width=5.5cm]{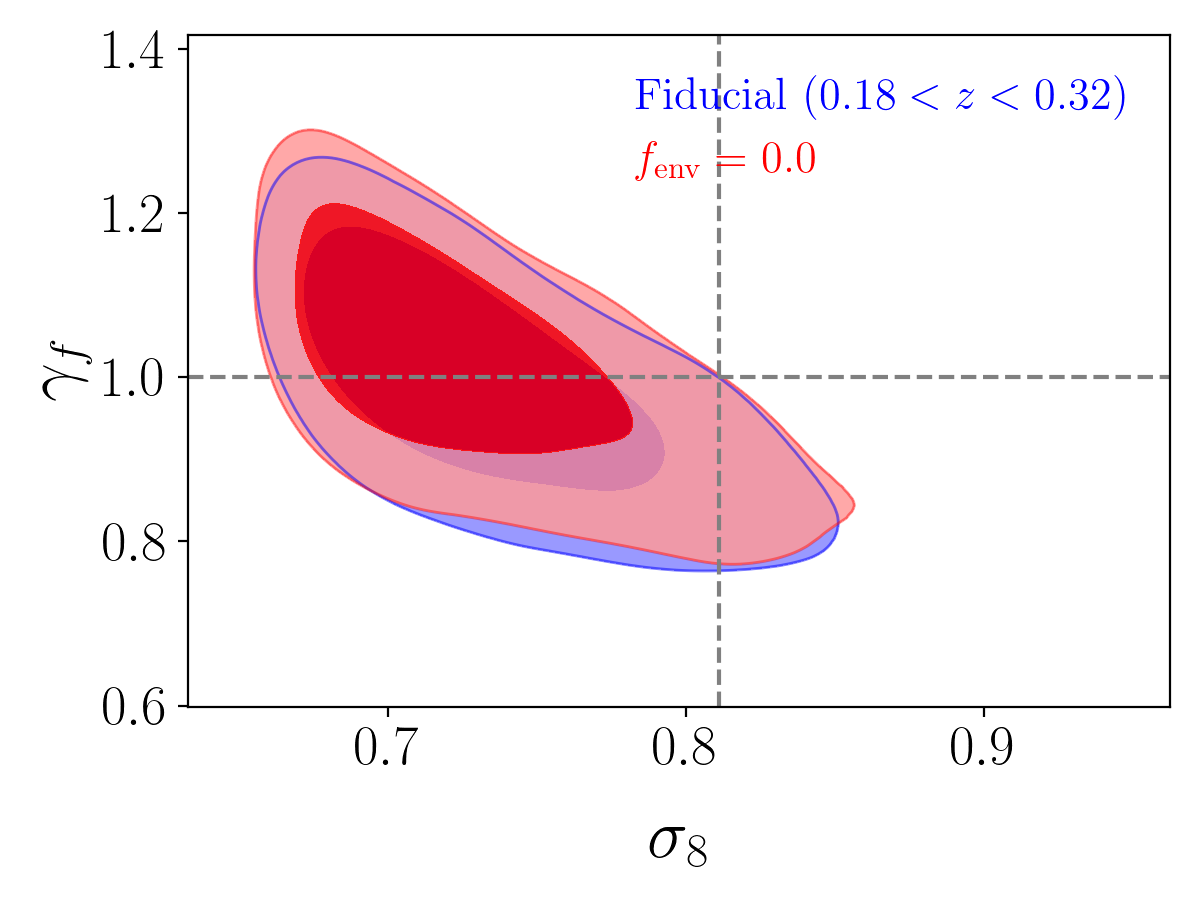}
\includegraphics[width=5.5cm]{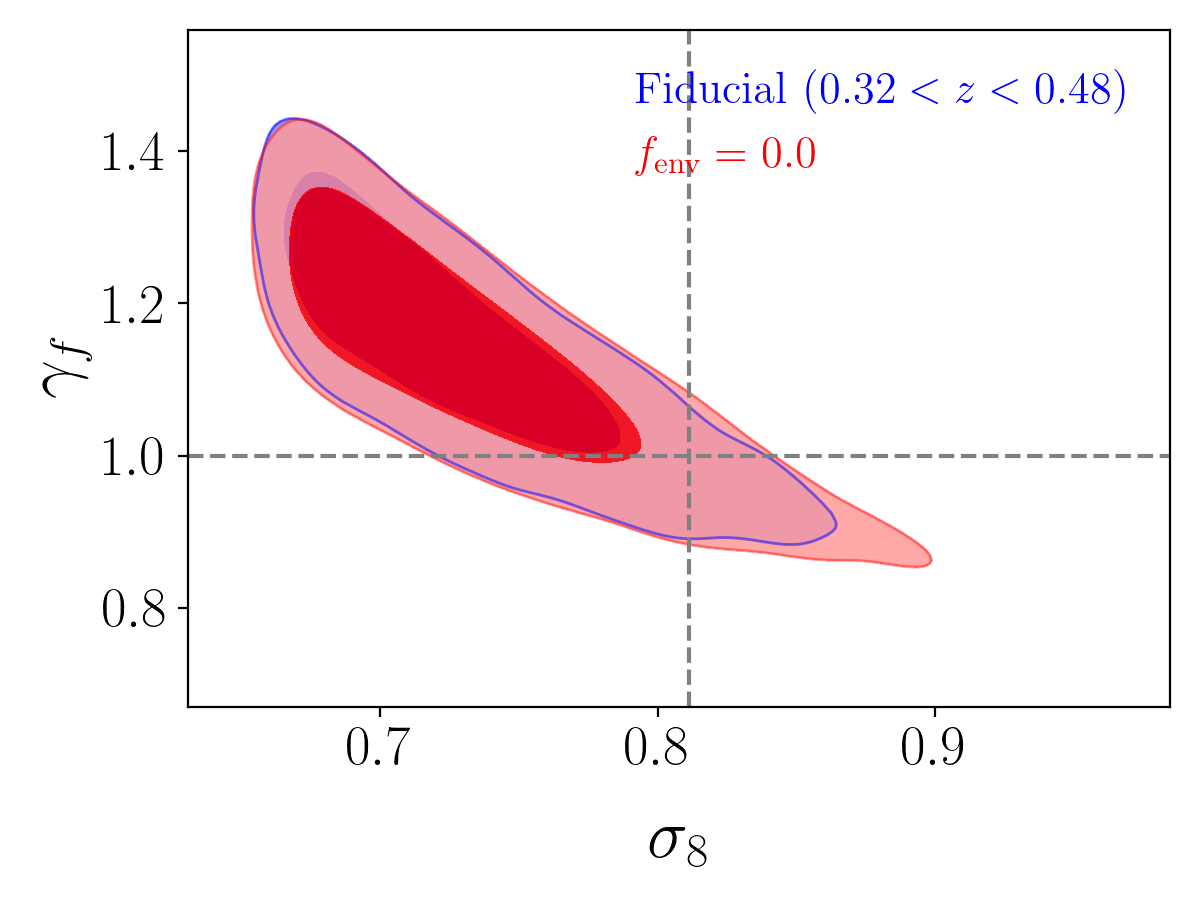}
\includegraphics[width=5.5cm]{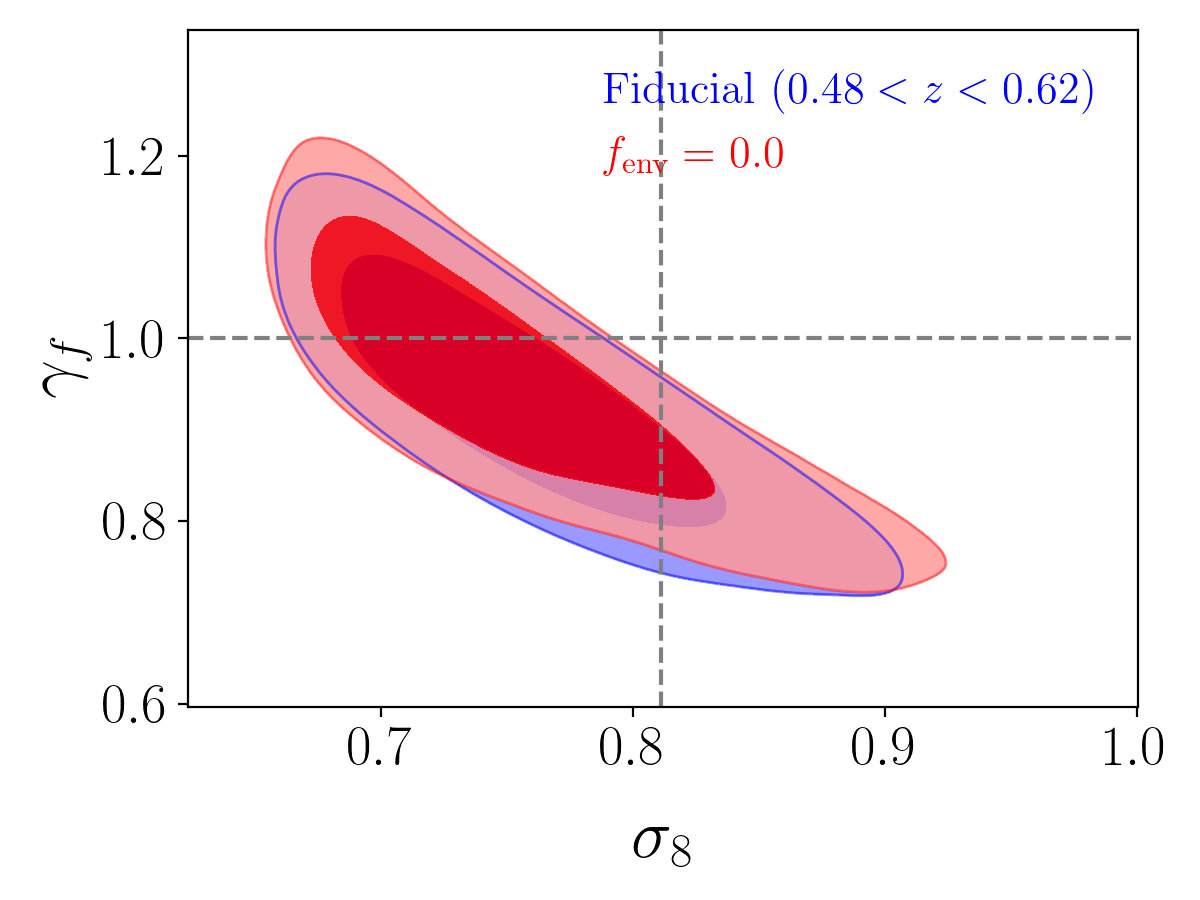}
\caption{Constraint on parameter $\sigma_{8}$ and $\gamma_{f}$ from the fiducial constraint (blue) and forcing $f_{\text{env}}=0$ (red), i.e., assuming a model with no assembly bias. The dashed line denotes the Planck observation for $\sigma_{8}$ and $\gamma_{f}=1$.}
\label{fig:constraint_noAB}
\end{center}
\end{figure*}

Incorporation of assembly bias in our emulator enables the investigation of its impact on the cosmological inference. Using our BOSS clustering data, we repeat the analysis by turning off the assembly bias, i.e., assuming a prior of $f_{\text{env}}=0$ in the likelihood analysis. The constraints on $\sigma_{8}$ and $\gamma_{f}$ for three redshift bins are shown in Figure \ref{fig:constraint_noAB}. Restricting the analysis in this way has a minimal impact on the key cosmological constraints. The largest shift on the contour plot is seen in the high-$z$ subsample, but it is still well within $1\sigma$ level. This result is consistent with expectation from the fiducial analysis that the constraint on $f_{\text{env}}$ shows only statistically weak deviations from 0. For comparison, the predictions of the emulator using the best-fit model without assembly bias is shown as the green dot--dashed curves in Figure \ref{fig:bestfit_BOSS_lowz}.

Although this analysis demonstrates that the inclusion of galaxy assembly bias is not required to achieve unbiased cosmological constraints from the BOSS sample, we still must note potential caveats. Although our model is designed with flexibility in mind, it may not mimic all potential impacts of galaxy assembly bias.
The assembly bias can be a combination of multiple secondary properties and the choice of the model in a particular analysis can be arbitrary. Our current model only applies to the host halos within the HOD formalism, and not to satellite galaxies separately. This requires a complicated model for the assembly bias (see e.g., \citealt{Xu_2020}). However, bringing in complementary observables can increase our ability to constrain more sophisticated models. With more statistics, such as higher order moments of the redshift space correlation function, void statistics, and galaxy lensing, we can take a greater leap in constraining both galaxy assembly bias and cosmological parameters. By constructing multiple GP-based emulators for galaxy correlation function and excess surface density of galaxy--galaxy lensing, McLaughlin et. al. (in preparation) explicitly investigate the impact on cosmological inference from galaxy assembly bias. Similar to the method used here, the local density of dark matter halo is used as the secondary halo property within the HOD framework. However that work extends the modeling to both centrals and satellites. The result based on CMASS and LOWZ-like mocks shows that the incorporation of assembly in the model can help reduce the bias for cosmological inference, and can become more important at small scales below 1 $h^{-1}$ Mpc.

\subsection{Measurement of structure growth rate}

\begin{figure*}[htbp]
\begin{center}
\includegraphics[width=7.0cm]{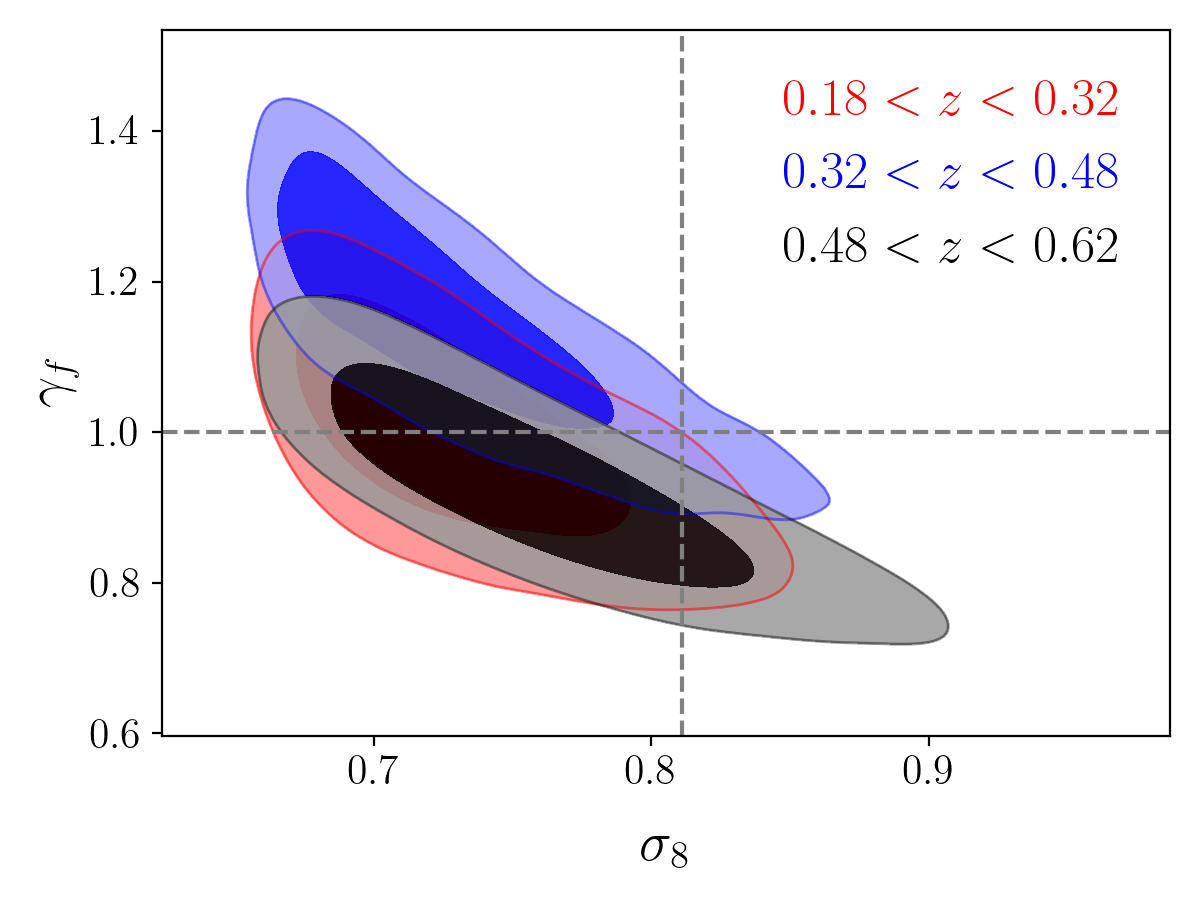}
\includegraphics[width=6.9cm]{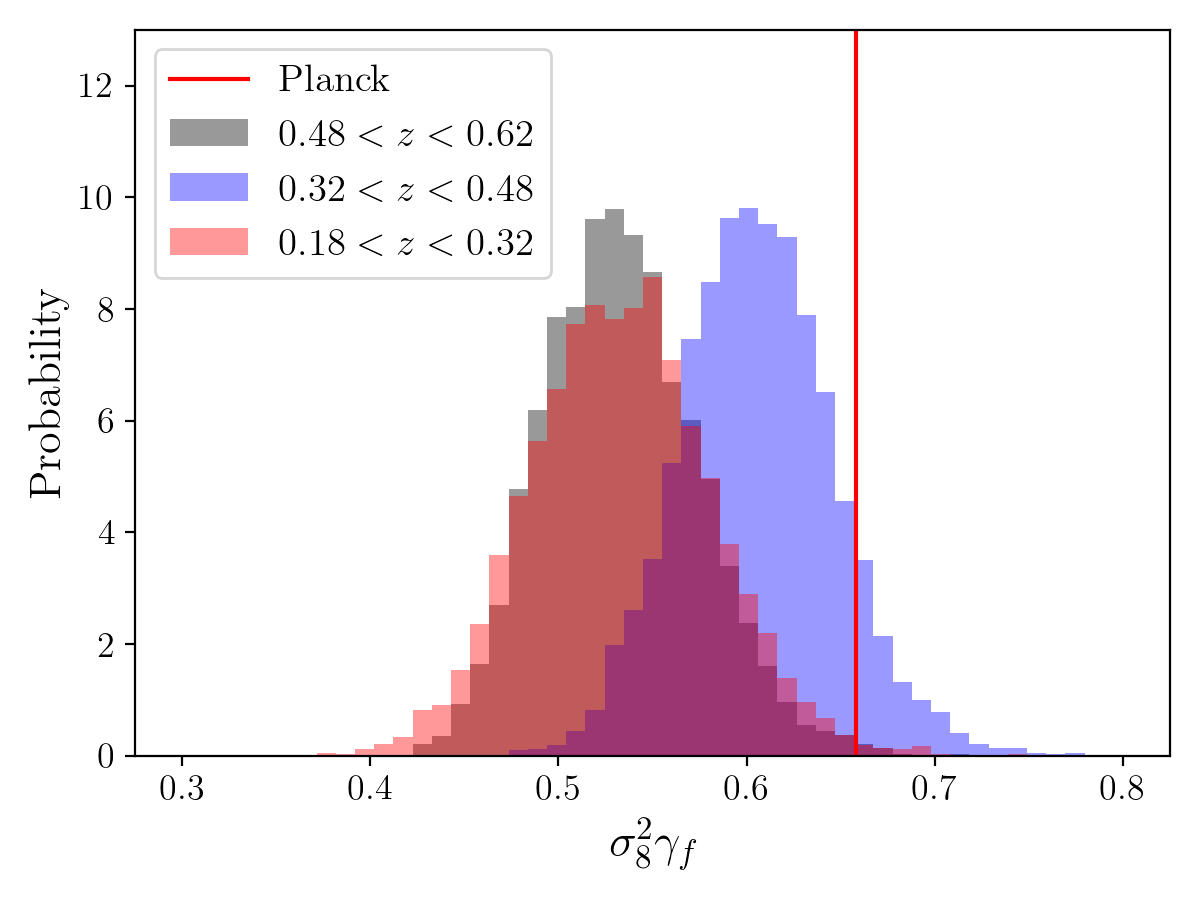}
\caption{\textbf{Left:} Constraint on parameters $\sigma_{8}$ and $\gamma_{f}$ for all three BOSS subsamples in the fiducial case. \textbf{Right:} One-dimensional distribution of the parameter combinations $\sigma_{8}^2\gamma_{f}$. The vertical red line shows the measurement from Planck 2018 observation with $\gamma_{f}=1$.}
\label{fig:1D}
\end{center}
\end{figure*}

One of the main goals of measuring galaxy clustering at non-linear scale is to precisely measure the growth rate of structure, quantified by the parameter combination $f\sigma_8$. In our model, the parameters that impact this quantity are $\Omega_m$, $\sigma_8$, and $\gamma_f$. Based on the fiducial analysis, our constraints on $\Omega_m$ are uncorrelated with other parameters and they are generally in agreement with the Planck results, regardless of prior assumptions. There is a clear degeneracy betwewn $\sigma_8$ and $\gamma_f$, seen in the previous figures, which we focus on in Figure \ref{fig:1D} by plotting the results from all three redshift slices together. The degeneracy curve traced out by these results is approximately $\gamma_f\sim \sigma_8^{-2}$. We note that the results approach the lower limit of $\sigma_8$ in the cosmological prior, which likely influences the exact shape of this degeneracy curve. The Planck $\Lambda$CDM+GR value lies just outside the $1-2\sigma$ constraints in this plane. The right panel shows the posterior probability of the parameter combination $\sigma_8^2\gamma_f$ for all three redshift bins, with the Planck value indicated with the vertical lines.  Expressed using this parameter combination, the results from the three different redshift bins are in good agreement with one another, and in tension with the Planck value. 

The growth of structure parameter, $f$, is determined by both the matter density and the amplitude of the halo velocity field. Thus, we one can think of our constraint on $f\sigma_8$ as $\gamma_f\Omega_{m}^{0.55}(z)\sigma_8(z)$. In detail, we use the CAMB software to compute the exact value of $f$ given the redshift and cosmology.
After marginalizing all the cosmological, HOD, and assembly bias parameters, are constraints are:
\begin{eqnarray}
    f\sigma_{8}(z=0.25) = 0.413\pm0.031 \\
    f\sigma_{8}(z=0.4) = 0.470\pm0.026 \\ 
    f\sigma_{8}(z=0.55) = 0.396\pm0.022
\end{eqnarray}
They correspond to fractional errors of $7.8\%$, $5.6\%$ and $5.5\%$, respectively, matching the precision expected from \cite{Zhai_2019}. The corresponding constraints from Planck 2018 are $0.473\pm0.006$, $0.478\pm0.005$, and $0.474\pm0.004$. Considering the uncertainties of both measurements, we find our measurements are lower than Planck by $1.9\sigma$, $0.3\sigma$ and $3.4\sigma$ for the three redshifts. We will discuss these differences in the following section. 

\begin{figure*}[htbp]
\begin{center}
\includegraphics[width=16cm]{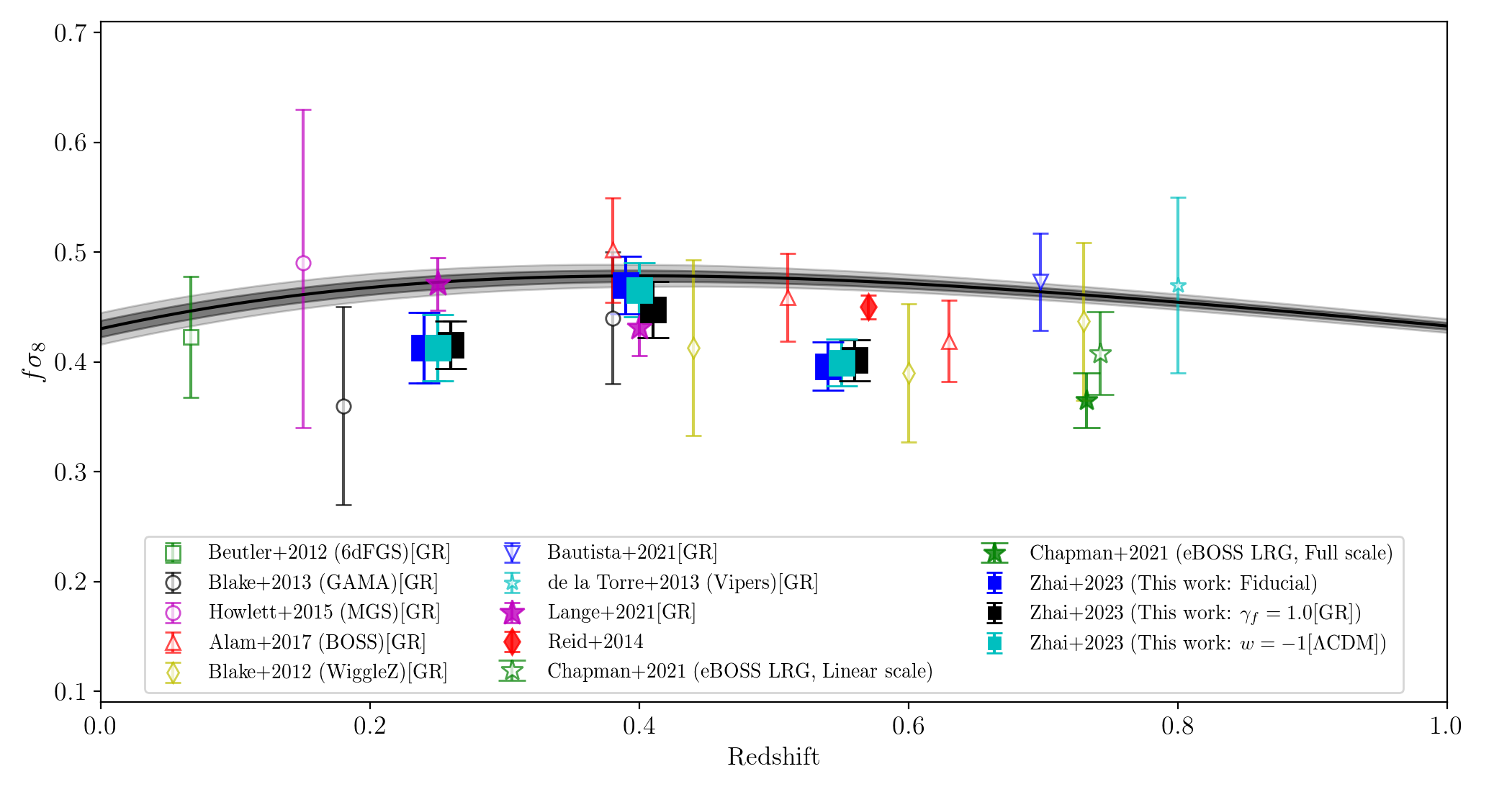}
\caption{Measurement of the growth rate of structure from our analysis using BOSS DR12 galaxies, as well as a compilation of the results in literature. The black line with shaded area is the prediction from the Planck 2018 release assuming a flat $\Lambda$CDM cosmology. The data include 6dFGS (\citealt{Beutler_2012}), GAMA (\citealt{Blake_2013}), SDSS-I/II main galaxy sample (\citealt{Howlett_2015}, MGS), WiggleZ (\citealt{Blake_2012}), Vipers (\citealt{de_la_Torre_2013}) and eBOSS-LRG (\citealt{Bautista_2021}). In addition, we also display measurements using BOSS galaxies similar to our work: DR12 final consensus results (\citealt{Alam_2017}), BOSS CMASS RSD analysis (\citealt{Reid_2014}) and BOSS LOWZ small scale analysis (\citealt{Lange_2021}). Note that our results of $f\sigma_{8}$ is quoted from measurement of $f\sigma_{8}\gamma_{f}$. We omit $\gamma_{f}$ in the y-axis to be in line with the measurements used for cosmological implications. In other words, except our analysis (including \citealt{Chapman_2021}) and \cite{Reid_2014}, all the other studies implicitly assume $\gamma_{f}=1$, i.e., the gravitational interaction is described by GR. The symbols are split into two categories: large scale (open) and small scale  (filled) measurements. Note that our measurements and eBOSS results at the same redshifts are shifted slightly for plotting purpose.}
\label{fig:fsigma8_z}
\end{center}
\end{figure*}

In Figure \ref{fig:fsigma8_z}, we display our measurements of $f\sigma_{8}$ for each of our three redshift bins, and we compare these values with other results in literature, as well as the prediction from a flat $\Lambda$CDM model using Planck 2018 results. The measurements are collected from clustering analyses of galaxies from surveys including 6dFGS (\citealt{Beutler_2012}), GAMA (\citealt{Blake_2013}), SDSS-I/II main galaxy sample (\citealt{Howlett_2015}, MGS), WiggleZ (\citealt{Blake_2012}), VIPERS (\citealt{de_la_Torre_2013}) and eBOSS-LRG (\citealt{Bautista_2021}). Note that all these analyses assume GR as the underlying gravity. We also include measurements using BOSS galaxies, using either large-scale or small-scale clustering data. On large scales, \cite{Alam_2017} gives the consensus constraints on $f\sigma_{8}$ and BAO distance scales over the redshift range of CMASS+LOWZ using RSD multipoles. Their measurement and uncertainty of $f\sigma_{8}$ shows a clear dependence on redshift. We linearly interpolate their results to our redshift and compare the constraints. At $z$=0.4 and 0.55, our fiducial result is consistent within 0.4$\sigma$ and 1.3$\sigma$, respectively, indicating the internal consistency of the BOSS analysis.\footnote{We note that the BOSS measurements are correlated due to overlapping redshift bins, thus the fact that our results are lower than the \cite{Alam_2017} cosntraints at both redshifts is not necessarily indicative of a systematic bias.}

On small scales, there are a number of other studies.  \cite{Lange_2021} perform the measurements using cosmological evidence modelling approach for BOSS galaxies at $z=0.25$ and 0.4, marginalized over {\sc Aemulus} cosmological models. \cite{Chapman_2021} applies the emulator approach to model the eBOSS LRG at $z=0.7$ and extract the measurement of linear growth rate that is close to our method. \cite{Reid_2014} present a 2.5\% measurement of $f\sigma_{8}$ for the CMASS galaxies, but we note that this analysis is at fixed cosmology and thus the error is likely underestimated (see the discussion in \citealt{Zhai_2019}).

We also extract constraints on structure growth assuming $\gamma_{f}=1$, which is consistent with most of the other analyses using GR as underlying gravity model. The results are
\begin{eqnarray}
    f\sigma_{8}(z=0.25) = 0.416\pm0.022 \\
    f\sigma_{8}(z=0.4) = 0.448\pm0.025 \\ 
    f\sigma_{8}(z=0.55) = 0.401\pm0.019
\end{eqnarray}
corresponding to a fractional error of $5.2\%$, $5.7\%$ and $4.7\%$ respectively. This result has similar accuracy as \cite{Lange_2021} where the authors claim a five percent measurement of $f\sigma_{8}$ at $z=0.25$ and 0.4. Our constraints for the samples at $z=0.25$ and $z=0.4$ are somewhat tighter because we include the projected correlation function, $w_{p}$, in the analysis to anchor the galaxy bias in real space. Although $w_p$ has limited cosmological sensitivity, it can strengthen the constraints on the HOD parameters themselves, which can yield tighter cosmological parameters by breaking degenercies between cosmology and bias that exist in the RSD data. The tension of our measurements with Planck is $2.5\sigma$, $1.3\sigma$ and $3.7\sigma$ for three redshifts. In general, the $\gamma_{f}=1$ prior raises the values of $f\sigma_{8}$, closer to the Planck values. However, the reduced uncertainties make the tension more significant. 
We also present the measurement of $f\sigma_{8}$ with $w=-1$ in Figure \ref{fig:fsigma8_z}. This prior yields quite consistent measurement as our fiducial analysis but can shrink the uncertainty slightly. It is known that a cosmology with $w\neq-1$ predicts different growth history of structure compared with $\Lambda$CDM cosmology (\citealt{Lue_04}). In our analysis, assuming this prior $w=-1$ mainly changes degeneracy for some of the parameters, for instance the contour plot for $\Omega_{m}$ (Figure \ref{fig:lowz_constraint}). However, it doesn't alter the estimate of $f\sigma_{8}$ significantly due to the velocity scaling parameter which is flexible to model a wide range of growth history.

\subsection{Scale Dependence}

\begin{figure*}[htbp]
\begin{center}
\includegraphics[width=5.5cm]{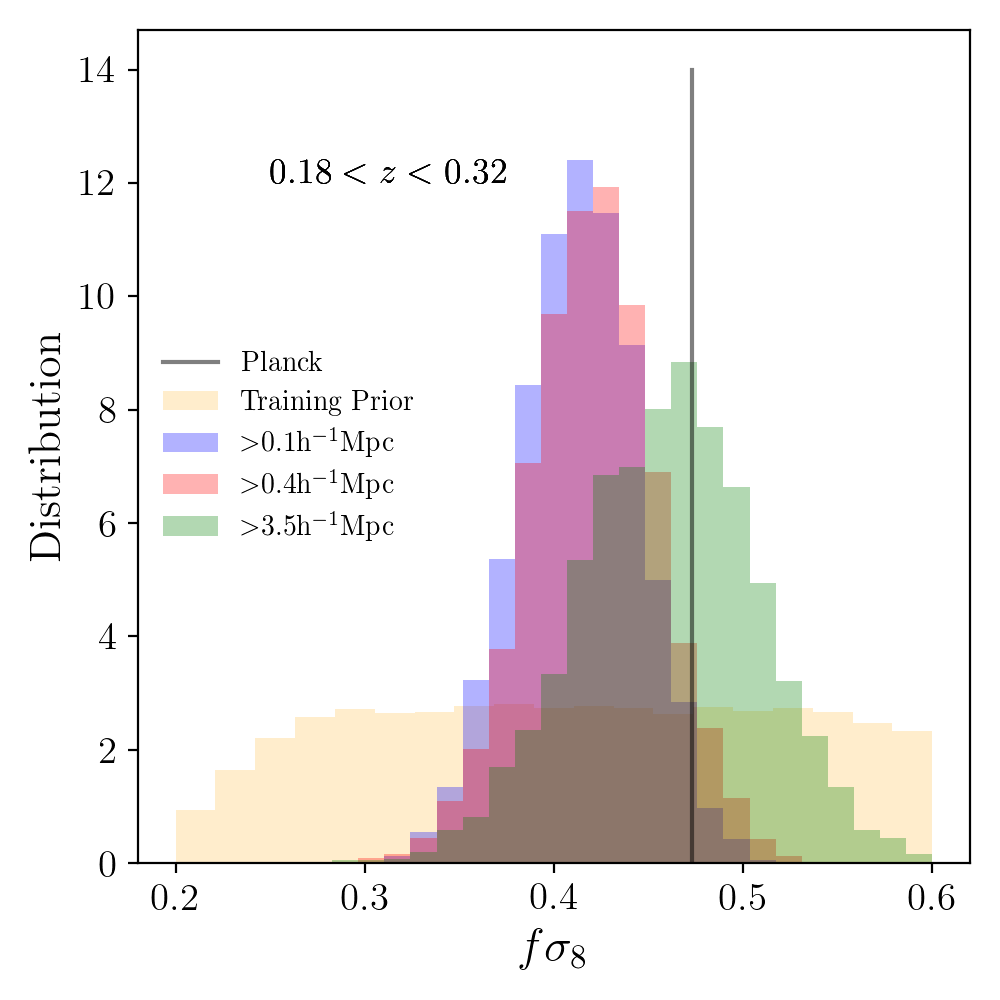}
\includegraphics[width=5.5cm]{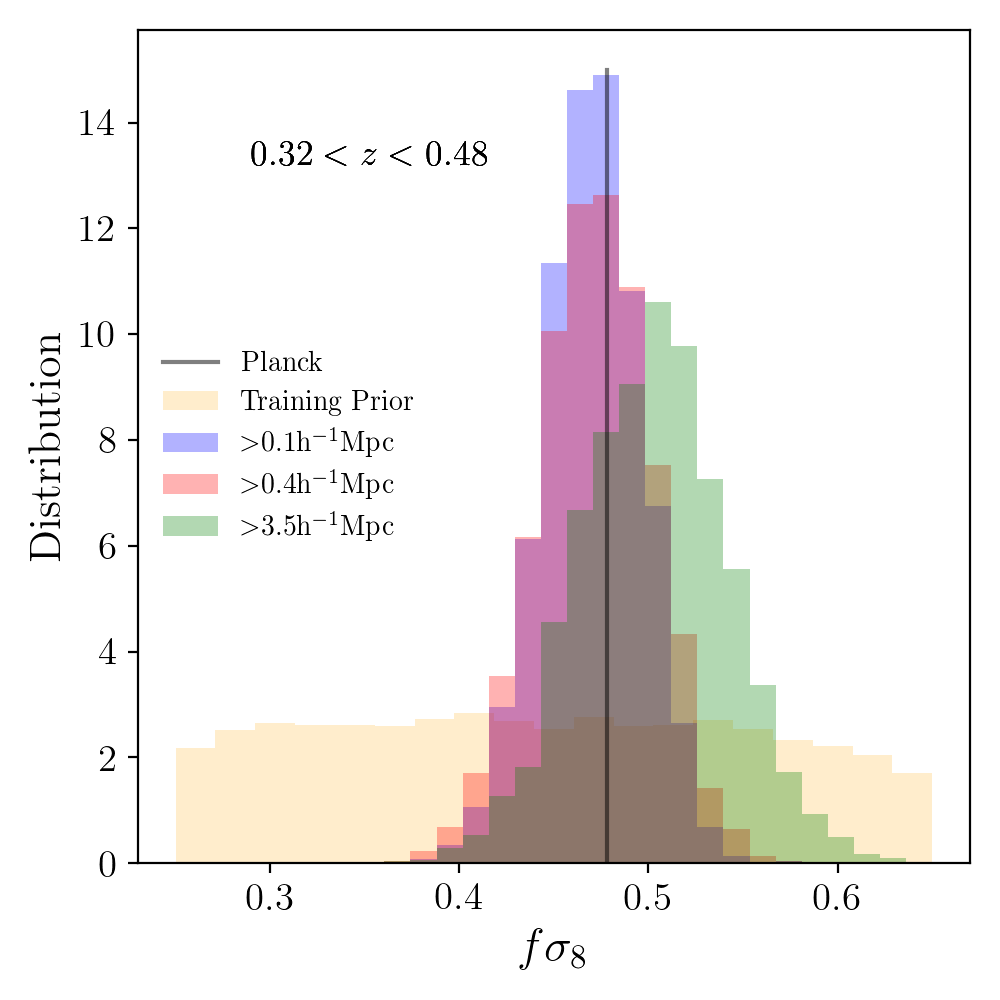}
\includegraphics[width=5.5cm]{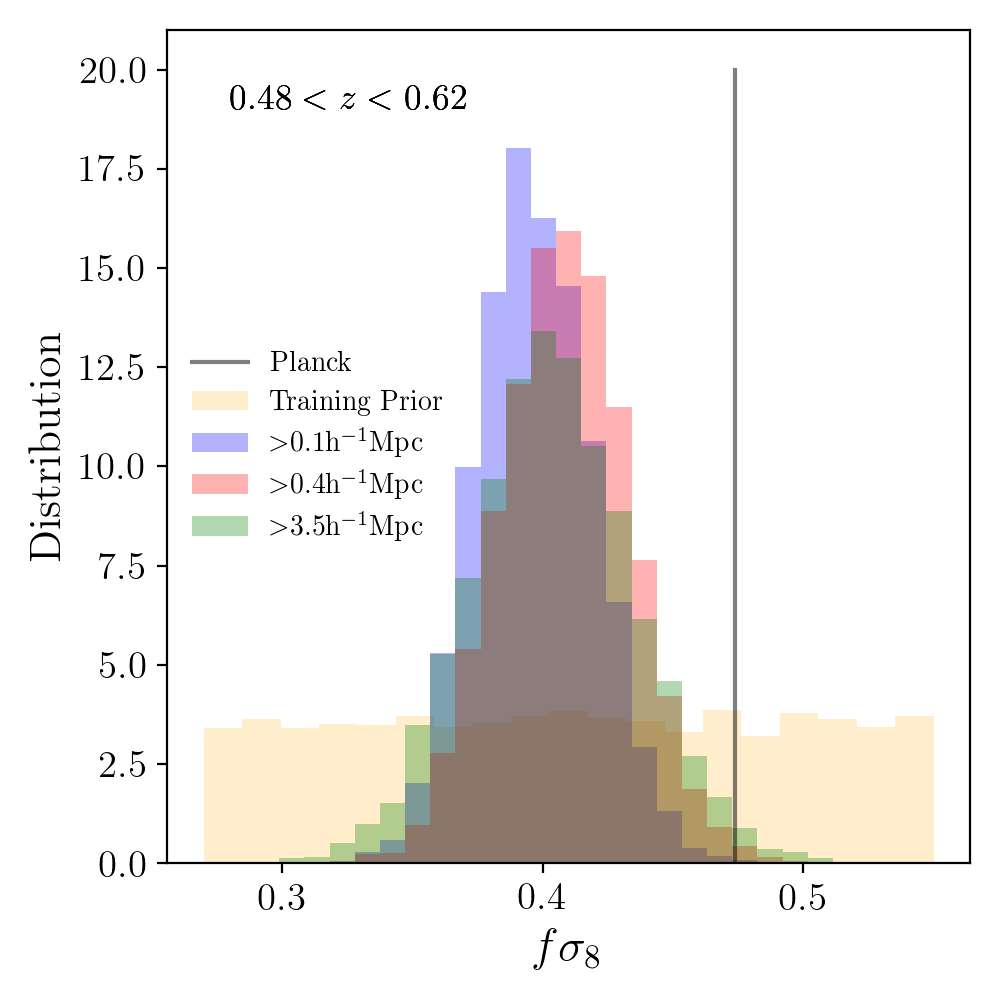}
\caption{Scale dependence of the constraint on $f\sigma_{8}$ from BOSS galaxies. Results are shown for three redshift ranges respectively as indicated in each panel. The range covered by the training prior is marked as yellow area.}
\label{fig:fsigma8_scale_cut}
\end{center}
\end{figure*}

In addition, we investigate the impact of scales considered in the clustering analysis. Figure \ref{fig:fsigma8_scale_cut} compares the posterior probabilities of $f\sigma_8$ for our fiducial analysis, which has a minimum scale of 0.1 $h^{-1}$Mpc, to two other analyses that increase the minimum scale used. Here we choose minimum scales of 0.4 $h^{-1}$Mpc and 3.5 $h^{-1}$Mpc. The former choice is motivated by the scale of the fiber collision effect; the latter choice is motivated by the transition between the one-halo and two-halo terms in the galaxy clustering signal. The results for a minimum scale of 0.4 $h^{-1}$Mpc are consistent with those of the fiducial scale, both in terms of the values of $f\sigma_8$ and their uncertainties. However, the results for 3.5 $h^{-1}$Mpc are substaintially different and the offset decreases with redshift. Although the new posteriors overlap with the fiducial, the best-fit values increase and the errors, as expected, widen. The tension with the Planck results is substantially alleviated, although the high-$z$ result is still low by $\sim 2\sigma$. These results agree with similar tests in \cite{Chapman_2021}, which imply that the lower values of $f\sigma_8$, and the tension with Planck, is driven by the fully non-linear regime. 

\subsection{Comparison with other studies using small-scale clustering}

As shown in Figure \ref{fig:fsigma8_z}, the small-scale, BOSS LOWZ anlaysis of \cite{Lange_2021} is in good agreement with our constraints on $f\sigma_8$ at $z=0.4$, but it is higher than ours at $z=0.25$. Even though these analysis use the same set of galaxies at this redshift, there are a number of differences in both the modeling and in the data that may drive this difference. \cite{Lange_2021} employs a novel statistical method using the {\sc Aemulus} simulations without explicitly constructing an emulator. Rather, they use the likelihood at each {\sc Aemulus} cosmology to fit the likelihood function of $f\sigma_8$. On the data side, the galaxy statistics used include anisotropic clustering in redshift space of moments up to hexadecapole, but not $w_p$. As shown in \cite{Chapman_2021}, including $w_p$ reduces $f\sigma_8$ relative to the mutipoles alone. In addition, their analysis is restricted to scales above $0.4h^{-1}$Mpc, which roughly corresponds to the fiber collision scale of BOSS LOWZ galaxies. Last, on the data side, is that \cite{Lange_2021} only uses the NGC, whereas our analysis uses both NGC and SGC data. On the modeling side, there are significant differences as well. \citealt{Lange_2021} assumes $\gamma_f=1$ (GR only), which we find also marginally changes the value of $f\sigma_8$. Although one single explanation does not seem able to explain the difference in these two analyses at $z=0.25$, the cumulative effect of all the differences can explain a significant amount.

As a direct application of the emulator approach, \cite{Chapman_2021} measure small-scale galaxy clustering of eBOSS LRGs at $z=0.7$ to constrain the growth rate of structure. This analysis is closest in spirit to our work, with the exception of our incorporation of assembly bias. They use the emulator of \cite{Zhai_2019}, updated to match the redshift and number density of eBOSS, and include the $f_{\rm max}$ parameter discussed above. \cite{Chapman_2021} present two constraints on $f\sigma_8$: one using clustering data down to 0.1 $h^{-1}$Mpc, and another restricted to scales about 7 $h^{-1}$Mpc. The full-scale anlaysis yields a value of $f\sigma_8$ that, like our BOSS measurements, is significantly below the Planck value. The larger-scale analysis lies in between, with errors such that it is consistent with both the small-scale result and the Planck value. \cite{Chapman_2021} also perform a number of notable tests that help to validate the emulator approach taken in this paper. 
Since measurements of the galaxy correlation function assume a cosmology in order to convert redshift to distance, this choice may lead to the so-called Alcock--Paczynski (AP, \citealt{Alcock_1979}) effect. \cite{Chapman_2021} examine this effect for small-scale clustering, finding that the impact on parameter constraints is negligible. Therefore we do not explicitly model this effect in our analysis. In addition, the eBOSS galaxies at higher redshift can experience significant errors in the galaxy redshifts, leading to a biased measurement of $f\sigma_{8}$ of up to $0.5\sigma$. Since BOSS galaxies are at lower redshift, the amplitude of redshift errors are significantly smaller. However, to be certain that there is no impact on our results, in Appendix \ref{appsec:z_error}, we examine this by running parameter recovery tests using test data that incorporate redshift uncertainty. Our results show that the impact of redshift uncertainty is negligible for the BOSS galaxies. 

\begin{figure}[htbp]
\begin{center}
\includegraphics[width=8.5cm]{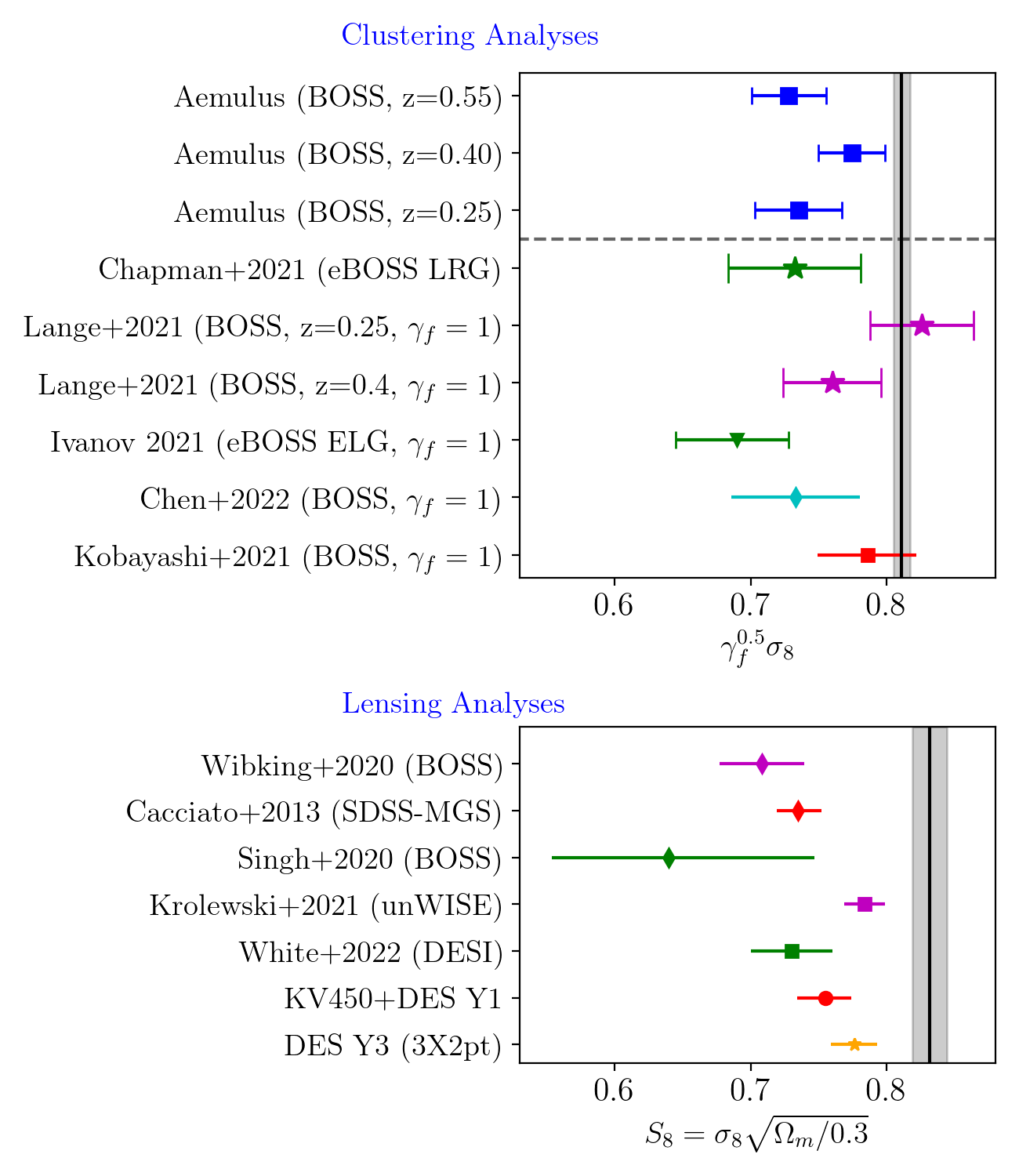}
\caption{\textbf{Top:} measurement of $\gamma_{f}^{0.5}\sigma_{8}$ from our BOSS galaxy analysis (blue square), compared with other works using similar galaxy statistics at both linear and non-linear scales. \textbf{Bottom:} latest measurements of $S_{8}=\sigma_{8}\sqrt{\Omega_{m}/0.3}$ using galaxy lensing statistics. The vertical lines in both panels correspond to the Planck measurements (\citealt{Planck_2020}).}
\label{fig:gammaf_sigma8}
\end{center}
\end{figure}

In Figure \ref{fig:gammaf_sigma8}, we project our measurement of $\gamma_{f}^{0.5}\sigma_{8}$ and compare with other analysis in literature, including the aforementioned works in \cite{Lange_2021}, \cite{Chapman_2021}, \cite{Ivanov_2021}, \cite{Kobayashi_2021} and \cite{Chen_2022} using redshift-space distortion, and galaxy lensing analyses from \cite{Singh_2020b}, \cite{Wibking_2020}, \cite{Krolewski_2021}, \cite{White_2022}, \cite{Asgari_2020}, and \cite{Abbott_2022}. In the top panel, we express the RSD measurement by parameter combination $\gamma_{f}^{0.5}\sigma_{8}$, which basically measures the amplitude of matter perturbation. Compared with the Planck result, all clustering analysis using low redshift galaxies, with the exception of the $z\sim0.25$ bin from \cite{Lange_2021}, give lower estimates of the perturbation amplitude than Planck (including \cite{Kobayashi_2021} which is slightly more consistent with Planck). Note that the eBOSS LRG (\citealt{Chapman_2021}) result with $\gamma_{f}=1$ has a consistent measurement with Planck, but with a high $\chi^{2}$. In the bottom panel, we compile some of the latest galaxy lensing measurements of $S_{8}=\sigma_{8}\sqrt{\Omega_{m}/0.3}$. The overall results of our clustering analysis are in agreement with these lensing analyses that the lensing amplitude is also lower than Planck prediction (\citealt{Leauthaud_2017}). A more robust analysis can combine the galaxy lensing measurement and RSD signals to improve the accuracy and the constraint on any deviation from GR or $\Lambda$CDM and also incorporate more flexible model for galaxy--halo connection (\citealt{Zu_2020}).

\section{Discussion and Conclusion} \label{sec:conclusion}

Galaxy clustering at small scales has been demonstrated to have a significant amount of cosmological constraining power, especially for the parameters that govern the growth and amplitude of structure. This paper extends the emulator approach for modeling galaxy clustering developed in \cite{Zhai_2019} to analyze the clustering of BOSS galaxies. In addition to the standard cosmological parameters of $\Omega_m$ and $\sigma_8$, we introduce the parameter $\gamma_{f}$ to scale the velocity field of dark matter halos to mimic the effect of modified gravity. At all redshifts covered by BOSS, our constraints on the parameter combination $f\sigma_8$ are below those predicted by the $\Lambda$CDM+GR model assuming the current Planck cosmology, with varying levels of statistical significance at each redshift bin. This result is similar to that of the ``lensing is low" phenomenon (\citealt{Leauthaud_2017}) in which the galaxy--galaxy lensing signal of the BOSS galaxies is lower than that predicted assuming a Planck cosmology by 30\% (\citealt{Leauthaud_2017, Wibking_2020}). As shown in Figure \ref{fig:gammaf_sigma8}, there are a growing number of results using small-scale clustering and lensing that are in tension with the current Planck cosmology.

There are three primary ways that this tension between our galaxy clustering results and the Planck constraints can be ameliorated. These include: (1) new physics that imparts deviations from general relativity, (2) cosmological solutions in the form of massive neutrinos, and (3) astrophysical solutions rooted in galaxy formation processes. 

The first of these solutions, in which gravity deviates from GR in order to change the prediction for the halo velocity field at a fixed matter density, is not favored. The first reason is that this class of explanation does not necessarily resolve the tension between the Planck cosmology and the aforementioned lensing results. For example, models with weaker gravity than GR boost the lensing signal in the two-halo term (\citealt{Leauthaud_2017}), but would reduce the value of $f\sigma_8$ (\citealt{Samushia_2014}). This resolves one tension at the cost of amplifying the other. But the second reason is that, by our own analysis, such a solution is not favored by the redshift-space distortions. In our fiducial analysis, $\gamma_f=1$ (i.e., gravity is GR) is within $\sim 1\sigma$. If we adopt a Planck prior for all cosmological parameters except $\gamma_f$, we do find that $\gamma_f<1$ to high significance, but this model is not a good fit to the data. However, when allowing the cosmology to be free but adopting a prior of $\gamma_f=1$, our model is a good descriptor of the BOSS clustering data.

The second solution addresses the tensions created by both lensing and clustering results simultaneously. From Figure \ref{fig:gammaf_sigma8}, the collected results from both of these probes indicate the need for a lower value of $\sigma_8$ than derived from CMB data. The values of these two constraints on the amplitude of matter fluctuations, one at $z<1$ and the other at $z\approx 1,100$, can be reconciled to some degree through the presence of massive neutrinos. The presence of massive neutrinos suppresses the growth of structure, which manifests in a scale-dependent manner with smaller scales being more highly affected. Given the current constraints on sum of neutrino masses to be $<0.12$ eV, it is unlikely that massive nuetrinos can fully resolve the tension between the results listed above and the Planck constraints on the clustering amplitude, but it goes in the right direction to resolve both lensing and RSD results, and it could alleviate a significant amount of the tension.

The last of the primary solutions involves the galaxy--halo connection. In this paper, we have attempted to make our galaxy bias model as flexible as possible, constructing a model with 11 free parameters. However, the tension between our results and the Planck cosmology is primarily driven by the clustering signal at $\lesssim 3$ $h^{-1}$Mpc, a scale that probes galaxy pairs within a single host halo as well as the transition between 1-halo and 2-halo galaxy pairs. Thus, a systematic error in the HOD approach cannot be ruled out. The tests presented in this paper use abundance matching techniques to create the test data. These test many of the assumptions in our model, such as spherical halos with isothermal velocity distributions and NFW density profiles for satellite galaxies. However, subhalo abundance matching does not incorporate baryonic effects that may influence the spatial distribution of galaxies. Currently, hydrodynamic simulations are not large enough to present statistically robust tests of the emulator. But it may be possible to incorporate the impact of baryons on the spatial distribution of galaxies in larger, dark matter only simulations, providing a more rigorous test (\citealt{Hearin_2016, Hearin_2021}). The effects of galaxy formation on the galaxy--halo connection, and thus galaxy clustering, may present itself either through assembly bias or in an occupation function that is not well-represented with our current model. 

The chief observational systematic, that of fiber collisions between BOSS spectra, is not likely to be a dominant source of bias. We note that the discrepancy with the Planck cosmology persists if we exclude data below 0.4 $h^{-1}$Mpc, the scale at which collisions become significant in the highest redshift bin. We also note that the method of correcting for collisions employed in this paper is distinct from the method used in the \cite{Chapman_2021} emulator analysis of eBOSS LRG clustering, but both studies are consistent in finding tension with Planck when including the smallest scales. On the other hand, we have incorporated the emulator uncertainty throughout the analysis which is non-negligible in the total covariance, implying that we can have another boost in the constraining power if the emulator error can be reduced significantly or totally removed. This can be non-trivial. In our pilot study (\citealt{Zhai_2019}), we use the analytical model of $w_{p}$ (\citealt{Tinker_analytical, Tinker_2012}) with full control of the uncertainty of the input training data and test data. With similar coverage of the parameter space as in this paper, the intrinsic emulator error can be reduced to sub-percent level across a wide range of scales, thus much lower than the data uncertainty. However the simulation-based emulator is worse and the performance varies with scale. This may indicate factors other than the GP algorithm that can weaken the emulator performance, including but not limited to the modeling of the training error and the specifics of the simulations such as volume, resolution and small scale density field. A more thorough exploration for the emulator error can  help to answer this question and will be useful for future studies.

For the last two of the proposed solutions listed above, the {\sc Aemulus} Project is in excellent position to make significant progress. The natural next step in development of simulations for emulator development is to create new simulations that incorporate neutrinos as an active particle species along with the dark matter. This will properly model the differential growth of structure induced by such particles, giving our next generation emulators the ability to incorporate and constrain the sum of the neutrino masses. These simulations are currently underway (DeRose et al, in preparation). Adding new parameters to the galaxy bias model, be they applicable to the mean occupation function or related to assembly bias, is a straightforward process. The addition of new parameters in a model necessitates increased constraining power from data. The {\sc Aemulus} approach, however, is open-ended. Non-standard galaxy clustering statistics, such as void statistics, marked statistics, and kNN statistics, can be emulated with equal efficacy to redshift-space distortions and weak lensing. These statistics bring in complementary information that can be used to constrain new freedom in HOD models \citep{Tinker_void_2008, Walsh_2019, Vakili_2019, Wang_2019, Szewciw_2021}. The next generation of {\sc Aemulus} emulators will include complementary, non-standard statistics (Storey-Fisher et al, in preparation). 

The measurement of the growth and amplitude of large-scale structure is a crucial test of our cosmological model. It provides complementary information to geometric probes, i.e., the cosmic distance measurements from observation of Type Ia supernovae and baryon acoustic oscillations. The measurement of $f\sigma_{8}$ as a function of redshift is able to constrain the growth history of the cosmic density field. For the large families of dark energy and modified gravity models that are proposed to explain cosmic acceleration, the growth measurement is able to distinguish them and pare down the viable parameter space. The accurate and model-independent measurement presented in this paper serves as the latest attempt. In future work, we will explore the cosmological implications of this latest measurement.

The {\sc Aemulus} Project aims at providing accurate and un-biased emulators for galaxy statistics at any redshift and number density. This paper represents the first application to the BOSS dataset. The overall performance is consistent with the estimates of our earlier work, and additional improvement is also possible. Ongoing surveys, most notably DESI, will increase the statistical constraining power of observational data and will thus require higher precision and accuracy from theoretical models. The next generation of Aemulus simulations and emulators will allow the analysis of small-scale clustering to continue down the path on which this paper lies.

\acknowledgments

ZZ thanks Michael Chapman, Johannes Lange, Surhud More, and Sukhdeep Singh for kindly providing their data.
This work received support from the U.S. Department of Energy under contract number DE-AC02-76SF00515 and from the Kavli Institute for Particle Astrophysics and Cosmology. ZZ is supported in part by NASA grant 15-WFIRST15-0008, Cosmology with the High Latitude Survey Roman Science Investigation Team (SIT). JLT and RHW acknowledge support of NSF grant AST-1211889. JLT also acknowledges support of NSF grant AST-2009291. YYM is supported by NASA through the NASA Hubble Fellowship grant no.\ HST-HF2-51441.001 awarded by the Space Telescope Science Institute, which is operated by the Association of Universities for Research in Astronomy, Incorporated, under NASA contract NAS5-26555. AB is supported by the Fermi Research Alliance, LLC under Contract No. DE-AC02-07CH11359 with
the U.S. Department of Energy, the U.S. Department of Energy (DOE) Office of Science Distinguished Scientist Fellow Program. This research made use of computational resources at SLAC National Accelerator Laboratory, and the authors thank the SLAC computational team for support. This research used resources of the National Energy Research scientific Computing Center, a DOE Office of Science User Facility supported by the Office of Science of the U.S. Department of Energy
under Contract No. DE-AC02-05CH11231.

\software{Python,
Matplotlib \citep{matplotlib},
NumPy \citep{numpy},
SciPy \citep{scipy},
George \citep{george_2014},
Corrunc \citep{Sinha_2020},
MultiNest (\citealt{Feroz_2009, Buchner_2014})
}

\appendix


\section{Luminosity selection on galaxy clustering} \label{appsec:boss_clustering}

\begin{figure*}[htbp]
\begin{center}
\includegraphics[width=18cm]{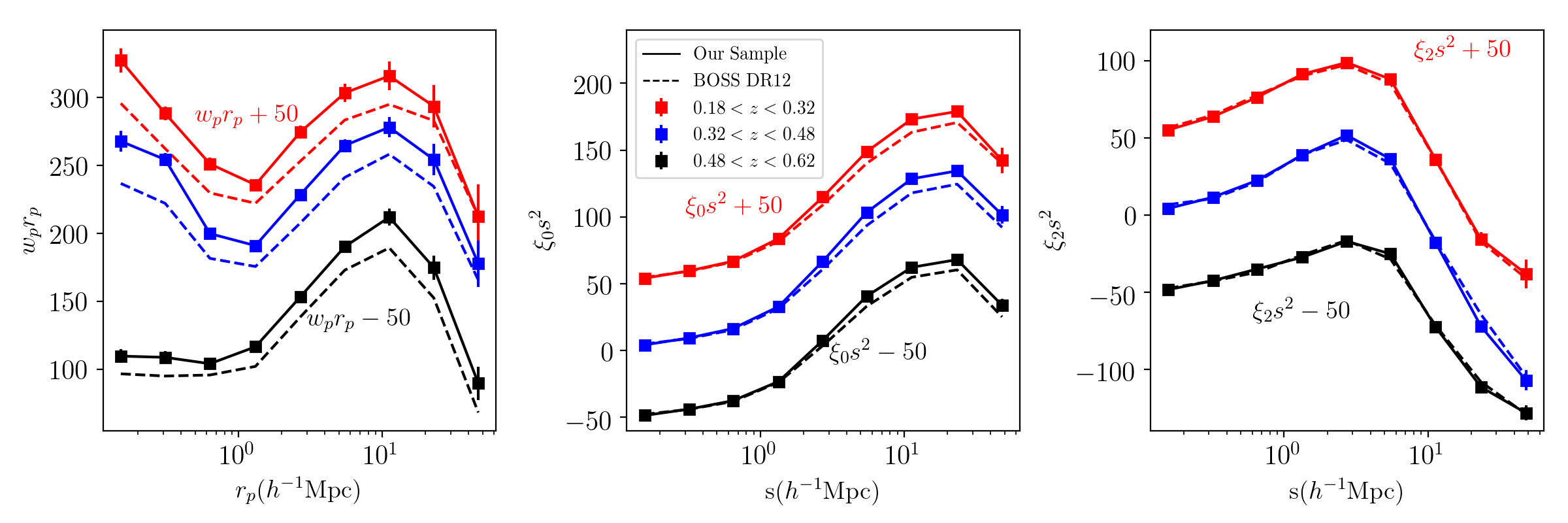}
\caption{The two-point correlation function $w_{p}$ (left), $\xi_{0}$ (middle) and $\xi_{2}$ (right) for the three subsamples of BOSS galaxies with and without luminosity selection. Dashed lines are measurements using all the galaxies, while the solid lines and squares denote our subsample selected by brightness. The low-$z$ and high-$z$ subsamples are shifted vertically for plotting purpose.}
\label{fig:BOSS_2pcf_z3}
\end{center}
\end{figure*}

The selection of galaxies based on their luminosity results in a brighter subsample. In Figure \ref{fig:BOSS_2pcf_z3}, we show the 2PCF of these galaxies and in comparison with the original BOSS galaxies. Due to the correlation between galaxy brightness and the host halo mass, our subsample reveals an increased amplitude of correlation function. The impact is significant for $w_{p}$ at all scales, and becomes weaker for RSD multipoles on small scales. This change requires the emulator of 2PCF from \cite{Zhai_2019} to be re-trained to explain the measurements as described in the text.

\section{Construction of covariance matrix} \label{appsec:covariance_matrix_construction}

\begin{figure*}[htbp]
\begin{center}
\includegraphics[width=18cm]{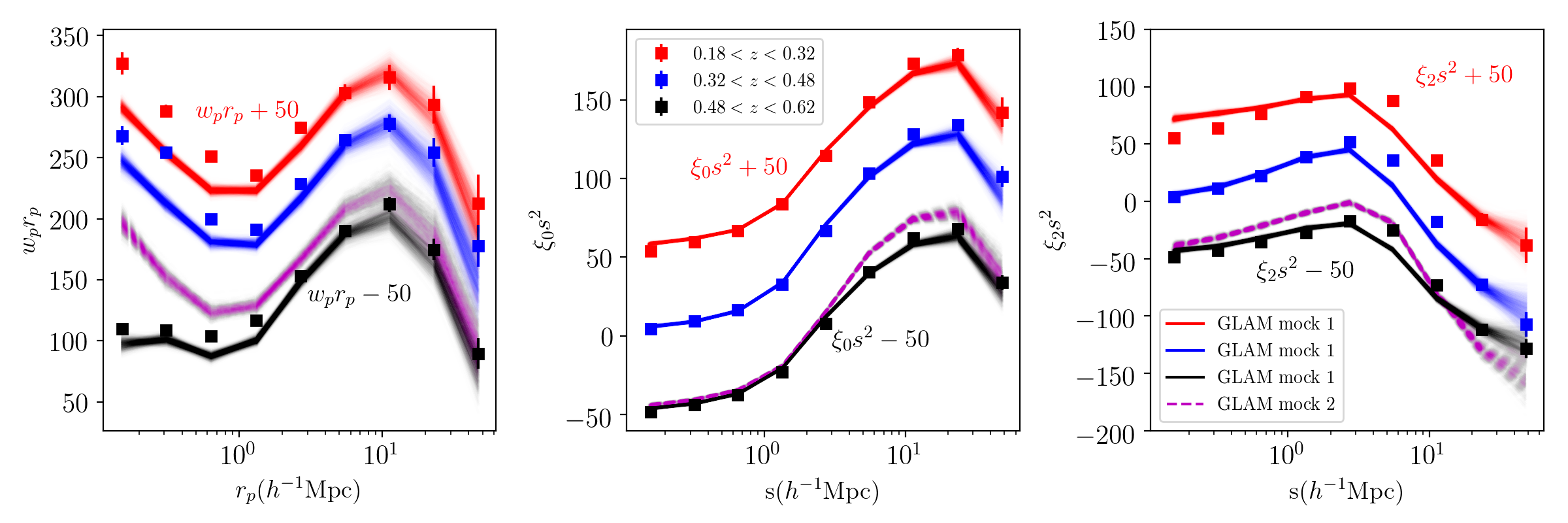}
\caption{Correlation function of the GLAM mocks to construct the covariance matrix. The squares with error bars are measurements from BOSS galaxies. The red, blue, and black lines are obtained by populating the GLAM mocks with an optimized HOD parameter set and the GLAM cosmology. We do not require the mocks to give consistent results as observation, since the GLAM cosmology may not be the true cosmology. In order to further test the covariance matrix, we ``randomly" pick a HOD model to populate the GLAM halo catalog for high-$z$ galaxies, as shown in the dashed purple lines. The clustering amplitude shows different behavior but the correlation matrix only shows mild variation.}
\label{fig:GLAM_2pcf}
\end{center}
\end{figure*}

\begin{figure*}[htbp]
\begin{center}
\includegraphics[width=19cm]{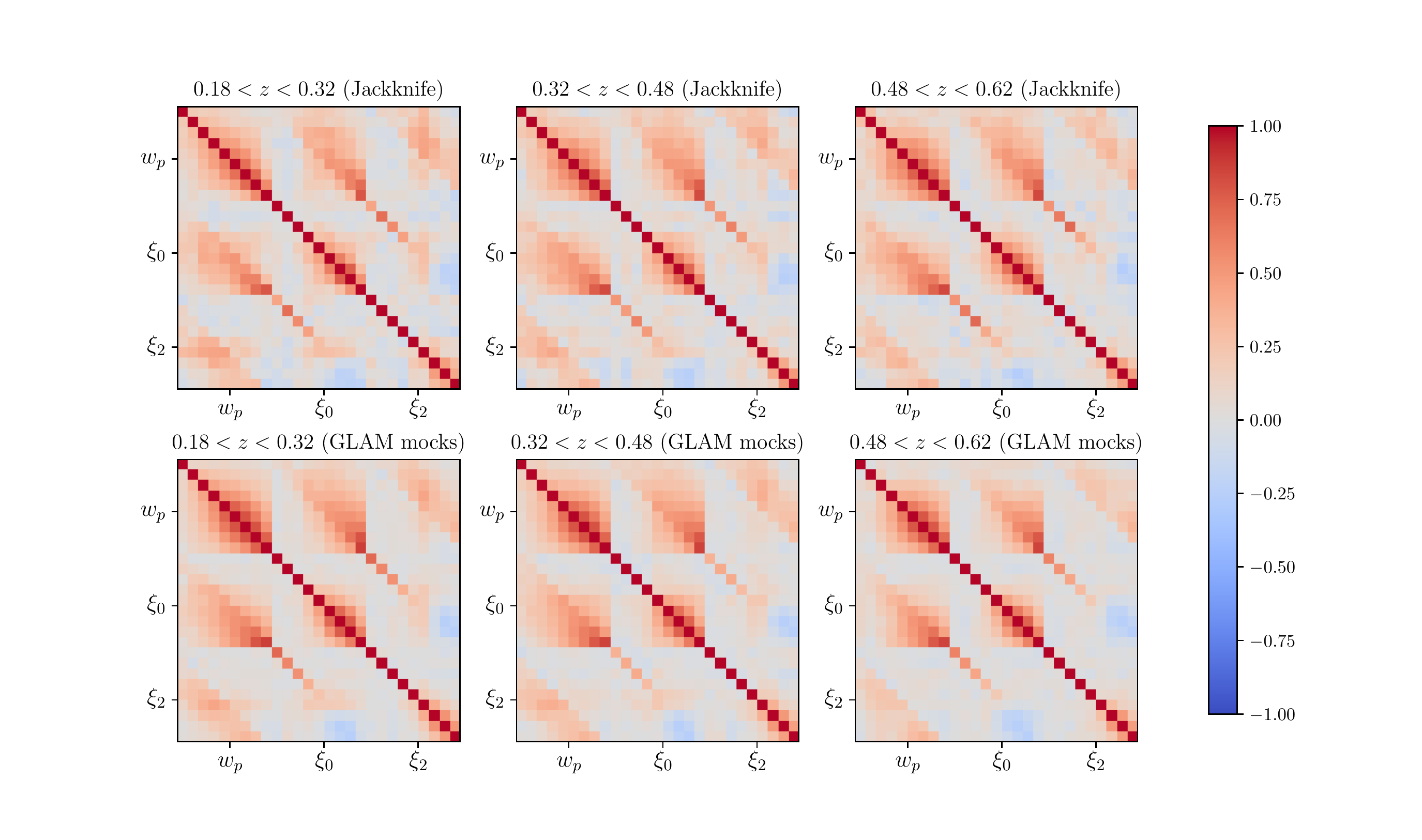}
\caption{\textbf{Top:} Correlation matrix for galaxy statistics $w_{p}$, $\xi_{0}$, and $\xi_{2}$ obtained by jackknife re-sampling of the BOSS galaxies. \textbf{Bottom:} The correlation matrix constructed using GLAM mocks.}
\label{fig:covariance}
\end{center}
\end{figure*}

The construction of the covariance matrix is described with details in Section \ref{sec:covariance}. In this appendix, we present the measurement of the 2PCF and the resultant correlation matrix in Figure \ref{fig:GLAM_2pcf} and \ref{fig:covariance} respectively. The top panel of Figure \ref{fig:covariance} shows the correlation matrix of $w_{p}$, $\xi_{0}$ and $\xi_{2}$ using jackknife resampling method of BOSS galaxies, and the bottom panel is from the GLAM mocks. For the high-$z$ subsample, we produce another set of GLAM mocks based on a random HOD. The correlation function is shown as the dashed purple line in Figure \ref{fig:GLAM_2pcf} with different clustering amplitude. However the resultant correlation matrix is similar to the first GLAM mocks (bottom panel of Figure \ref{fig:covariance}). Based on these galaxy mocks, we construct the covariance matrix in the likelihood analysis as explained in the text, and test the impact of different choices of covariance matrix in Appendix \ref{appsec:covariance}.

\section{Construction of the emulator}\label{appsec:emulator}

\begin{figure}[htbp]
\begin{center}
\includegraphics[width=16cm]{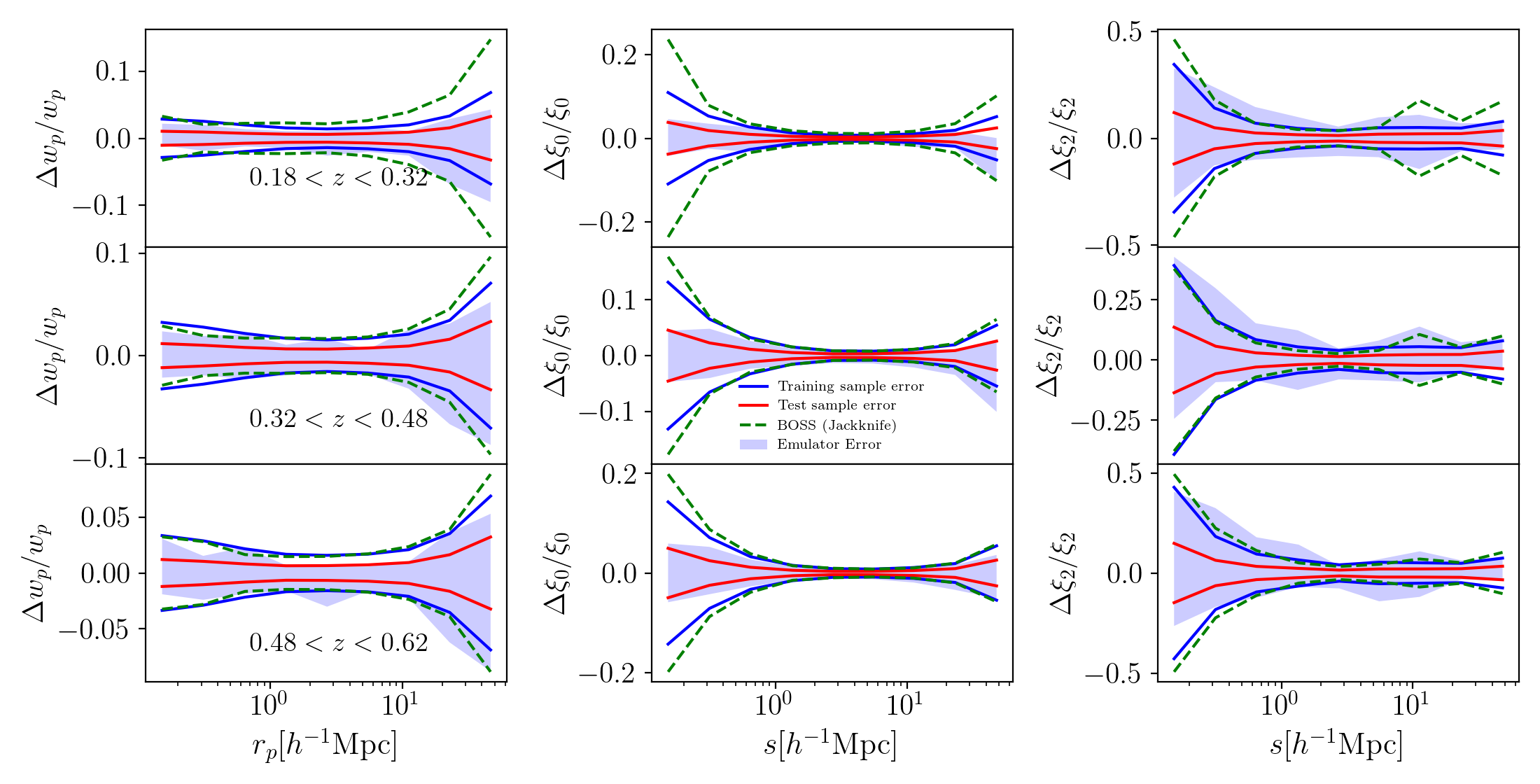}
\caption{Performance of the emulator for $w_{p}$ (left), $\xi_{0}$ (middle) and $\xi_{2}$ (right) for three redshifts. The solid blue line stands for the training error, while the red line is the error of the test samples. We also show the uncertainty from BOSS measurement as green dashed lines for reference. The test samples have multiple boxes and population to suppress sample variance and shot noise. The shaded area represents the inner 68\% distribution of the emulator performance.}
\label{fig:emulator_error}
\end{center}
\end{figure}

The construction of the emulator for $w_{p}$, $\xi_{0}$ and $\xi_{2}$ follows the same method as in \cite{Zhai_2019}, including the modeling of the training error, the choice of kernel function for the Gaussian-Process (GP) and the optimization method. In particular, we define the likelihood function with training sample as Eq (15) of \cite{Zhai_2019} for each statistics. The hyper-parameters in the kernel which define the distance metric are for individual model parameters. The training process is to optimize the above likelihood function and the resultant hyper-parameters form the final emulator to make prediction for new model parameter set. Since the HOD parameter space is extended with additional parameters for assembly bias and widened ranges for $M_{\text{sat}}$ and $\alpha$, the sampling of the emulator needs to be improved. Our test shows that using 80 HODs per cosmology is able to provide sufficient sampling for the emulator accuracy, which is a 60\% increase compared with \cite{Zhai_2019}. The 2PCF of the galaxy mocks from both the training sample and test sample is computed by the publicly available code \medium{Corrfunc} (\citealt{Sinha_2020}). Given the number density of our galaxy mocks, the average cost of one model evaluation using emulators for $w_{p}+\xi_{0}+\xi_{2}$ is at the level of 1 CPU-second. In Figure \ref{fig:emulator_error}, we present the overall accuracy of the emulator for $w_{p}$, $\xi_{0}$ and $\xi_{2}$ at three redshifts. The training error and test sample error are also shown for comparison. The emulator performance is quite similar as the pilot study in \cite{Zhai_2019}. For $w_{p}$ and $\xi_{0}$, the emulator accuracy is close or better than sample variance. At scales from 1 to 10 $h^{-1}$Mpc, the accuracy is better than 1 or 2\%, which enables the use of the clustering measurement at the most informative scales. The emulator performance is worse for $\xi_{2}$, but it is still able to contain useful information for cosmological constraint.

\section{Prior for the likelihood analysis}\label{appsec:prior}

\begin{figure}[htbp]
\begin{center}
\includegraphics[width=12cm]{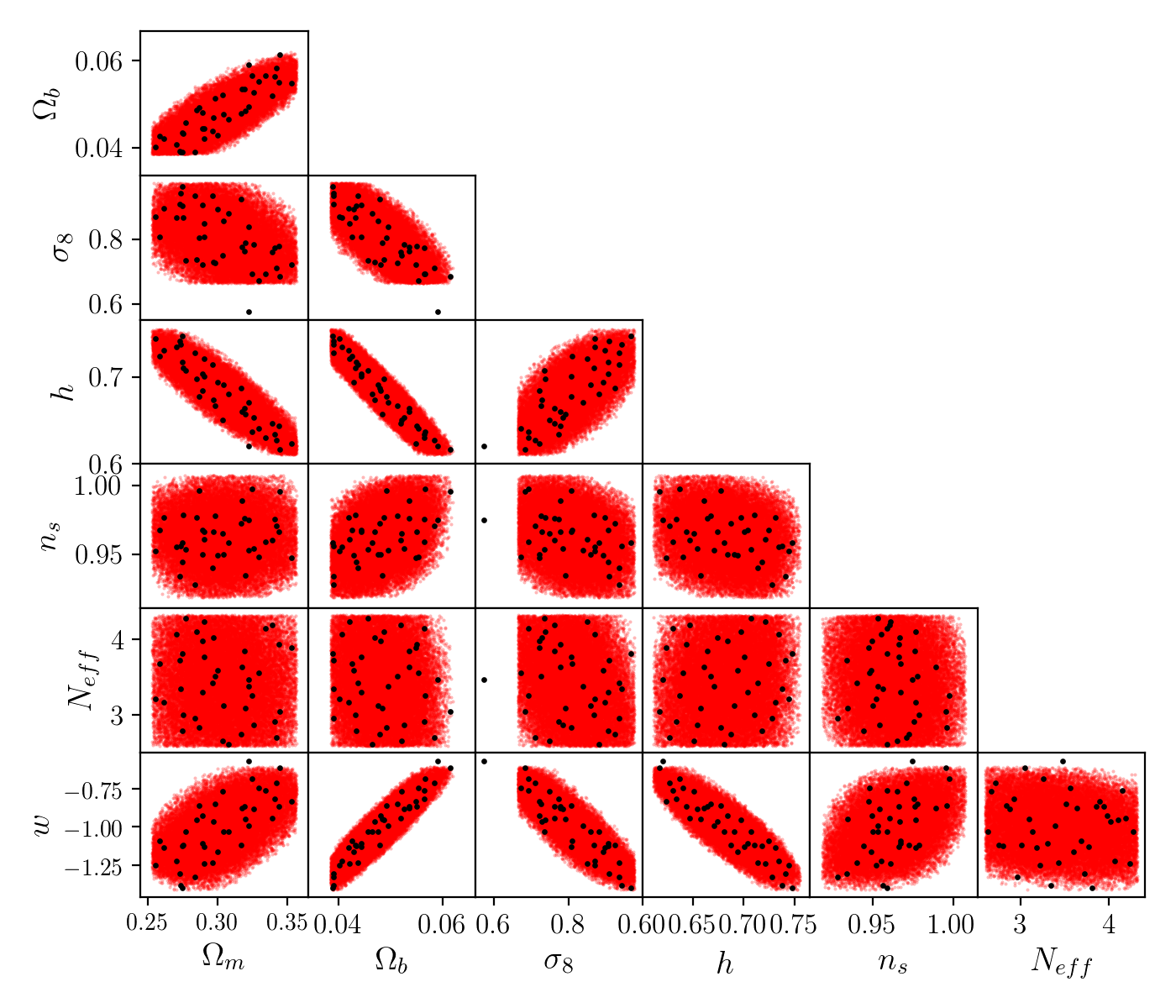}
\caption{Illustration of the prior space in the likelihood analysis for cosmological parameters. The black dots are the 40 training cosmologies from {\sc Aemulus} suite. Since Box023 has an outlier value of $\sigma_{8}$, we just use the other 39 models to get a covariance matrix of the cosmological parameters. Then we define a distance metric $\chi^{2}$ and only allow the sampling of the points within a threshold. This defines a 7D ellipsoid with uniform and un-informative distribution. The red dots show the resulting sampling in the prior space, i.e., the ellipsoid imposed with the range of each parameters as shown in the first section of Table \ref{tab:param}.}
\label{fig:training_prior}
\end{center}
\end{figure}

The sampling in the cosmological parameter space is restricted within the area covered by training cosmologies. We incorporate this restriction by defining a prior for cosmological parameters. We first compute the mean $\mu_{\text{cos}}$ and covariance matrix $C_{\text{cos}}$ of the seven parameters using {\sc Aemulus} training cosmologies. Then for an arbitrary point $\mu$ in the cosmological parameter space, we define a distance metric
\begin{equation}
    \chi^{2}_{\text{cos}}= [\mu-\mu_{\text{cos}}]C_{\text{cos}}^{-1} [\mu-\mu_{\text{cos}}].
\end{equation}
We choose a threshold value for $\chi^{2}_{\text{cos}}<12$ and only allows sampling that can satisfy this condition. This roughly corresponds to 3$\sigma$ level of this multivariate Gaussian distribution. The resulting seven-dimensional ellipsoid is able to enclose the training cosmlogies. In Figure \ref{fig:training_prior}, we display the projection of the training cosmologies, as well as the prior space.

\section{Test on SHAM catalogs}\label{appsec:sham}

We construct the SHAM catalog following the model proposed in \cite{Lehmann_2017}. This model adopts a velocity proxy to allow continuous transition between two halo properties 
\begin{equation}
    v_{\alpha} = v_{\text{vir}}\left(\frac{v_{\text{max}}}{v_{\text{vir}}}\right)^{\alpha_{\text{sham}}},
\end{equation}
where $v_{\text{max}}$ is the maximal circular velocity and 
\begin{equation}
    v_{\text{vir}} = \left(\frac{GM_{\text{vir}}}{R_{\text{vir}}}\right)^{1/2}.
\end{equation}
Note that we add subscript ``sham" to distinguish $\alpha_{\text{sham}}$ from the HOD parameter $\alpha$. When $\alpha_{\text{sham}}=0$, $v_{\alpha}=v_{\text{vir}}$, equivalent to matching galaxy by halo mass. When $\alpha_{\text{sham}}=1$, $v_{\alpha}=v_{\text{max}}$, the matching is based on maximal circular velocity. The typical value of $\alpha_{\text{sham}}$ is restricted within $[0,1]$, however we can artificially increase the value to increase the dependence of clustering on $v_{\text{max}}$ and therefore boost the level of assembly bias. 

\begin{figure}[htbp]
\begin{center}
\includegraphics[width=17cm]{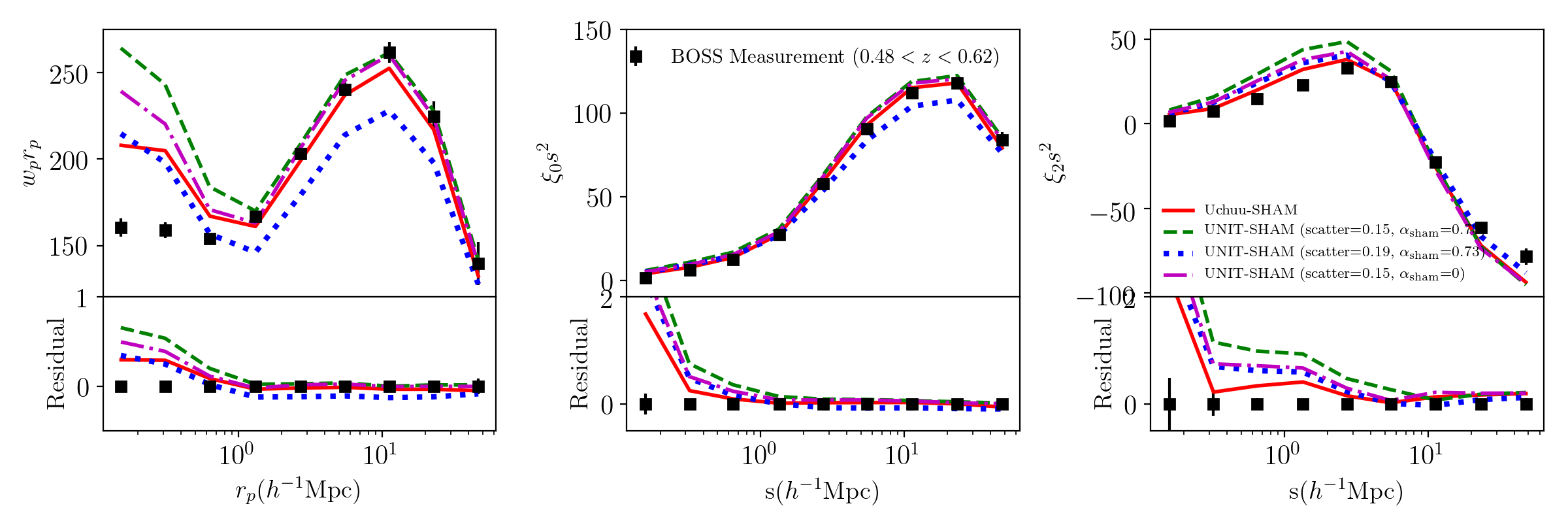}
\caption{Galaxy 2PCF of the SHAM catalogs using Uchuu and UNIT simulations at $z\sim0.55$, in comparison with high-$z$ BOSS galaxies.}
\label{fig:CF_SHAM}
\end{center}
\end{figure}

\begin{figure}[htbp]
\begin{center}
\includegraphics[width=8cm]{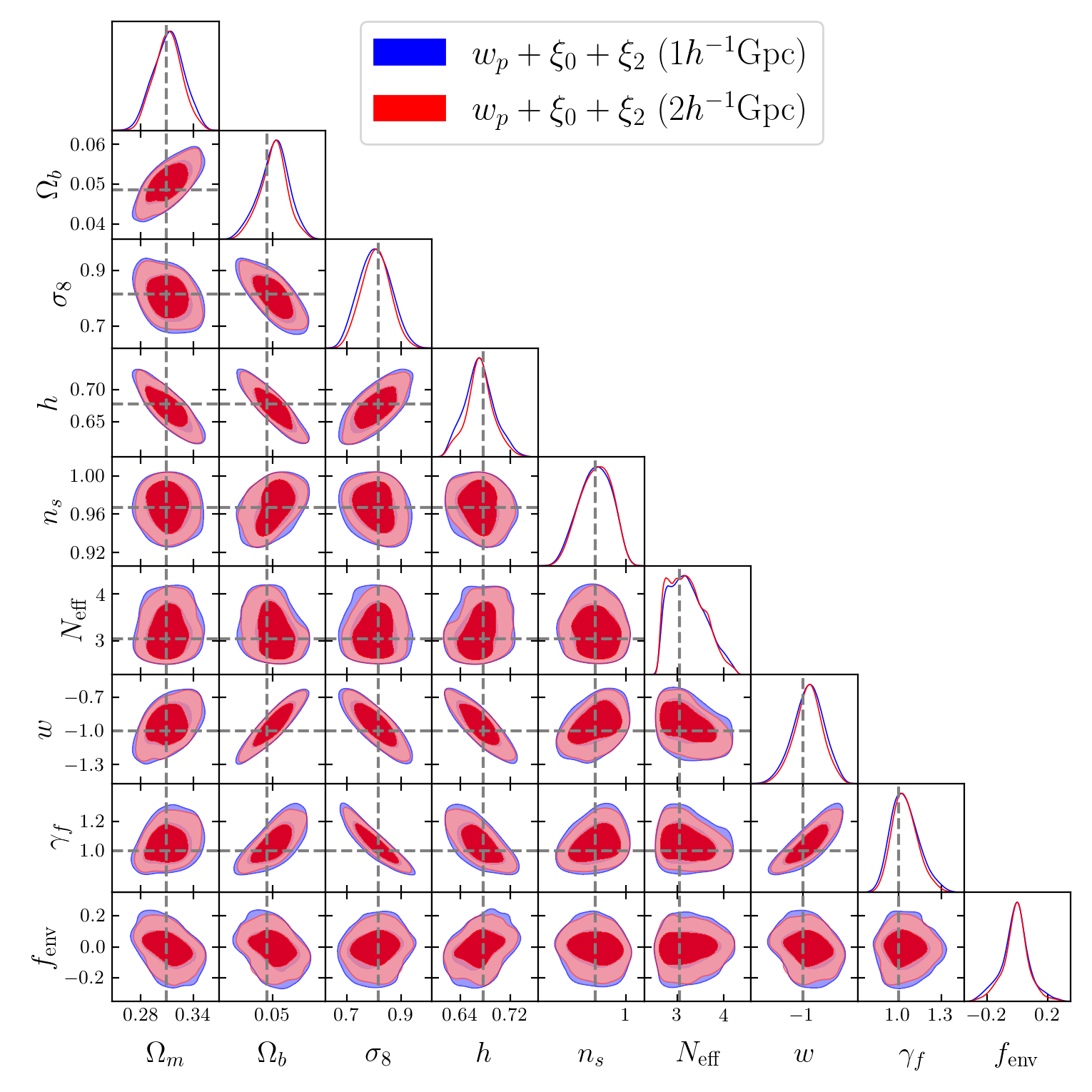}
\caption{Recovery test of the Uchuu SHAM catalog with $\alpha_{\text{sham}}=1.0$ and scatter=0.15. The constraint uses $w_{p}+\xi_{0}+\xi_{2}$ (right). The covariance matrix corresponds to the fiducial choice. Since the Uchuu simulation has larger volume than BOSS, we scale the uncertainties of the sample variance to match a $1h^{-1}$Gpc and $2h^{-1}$Gpc box respectively and compare the constraints. Our result shows that this difference is diluted to some extent by the contribution from the emulator error, however the sample variance corresponding to a larger box still gives a tighter constraint, consistent with expectation. The result shows that the input cosmology can be recovered within $1\sigma$ for all the cosmological parameters. This demonstrates the robustness of our assembly bias augmented HOD model and the construction of the emulator.}
\label{fig:constraint_Uchuu}
\end{center}
\end{figure}

In Figure \ref{fig:CF_SHAM}, we present the measurements of $w_{p}$, $\xi_{0}$ and $\xi_{2}$ for the SHAM catalogs using different values of scatter and $\alpha_{\text{sham}}$. The scatter parameter denotes the standard deviation of the stellar mass of galaxies at a fixed value of halo mass. We first perform the recovery test on the Uchuu SHAM catalog. The covariance matrix in the likelihood analysis is similar to the fiducial model as introduced in Section \ref{sec:covariance}: the correlation matrix is from the GLAM mocks and the diagonal elements correspond to the sample variance of the {\sc Aemulus} training box. Since the Uchuu simulation has a box size of $2h^{-1}$Gpc, we also use the ratio of the volume to scale the diagonal elements to match the Uchuu volume. These two tests are distinguished as ($1^{-1}$Gpc) and ($2h^{-1}$Gpc) respectively. Since this volume factor only applies to the sample variance $C_{\text{sam}}$, while the emulator error remains the same, the difference in the finalized covariance matrix is diluted. Figure \ref{fig:constraint_Uchuu} displays the final constraints on the cosmological parameters and the key assembly bias parameter $f_{\text{env}}$. We can see that the input cosmology of the Uchuu simulation is well recovered within $1\sigma$ using galaxy correlation function at non-linear scale. It shows that the assembly bias induced in the sub-halo matching process doesn't bias the cosmological constraint, and supports the robustness of our emulator. The two tests with different diagonal elements of the covariance matrix ($1^{-1}$Gpc) and ($2h^{-1}$Gpc) show quite similar results, with the ($2h^{-1}$Gpc) case presenting slightly tighter constraint on the contour plot. It is consistent with expectation that the larger volume can suppress the sample variance, but with a smaller difference due to the intrinsic emulator error since the volume scaling is only applied to the sample variance and the contribution from emulator error remains unchanged. This can make the (volume scaled) sample variance subdominant in the final covariance matrix.

\begin{figure}[htbp]
\begin{center}
\includegraphics[width=8cm]{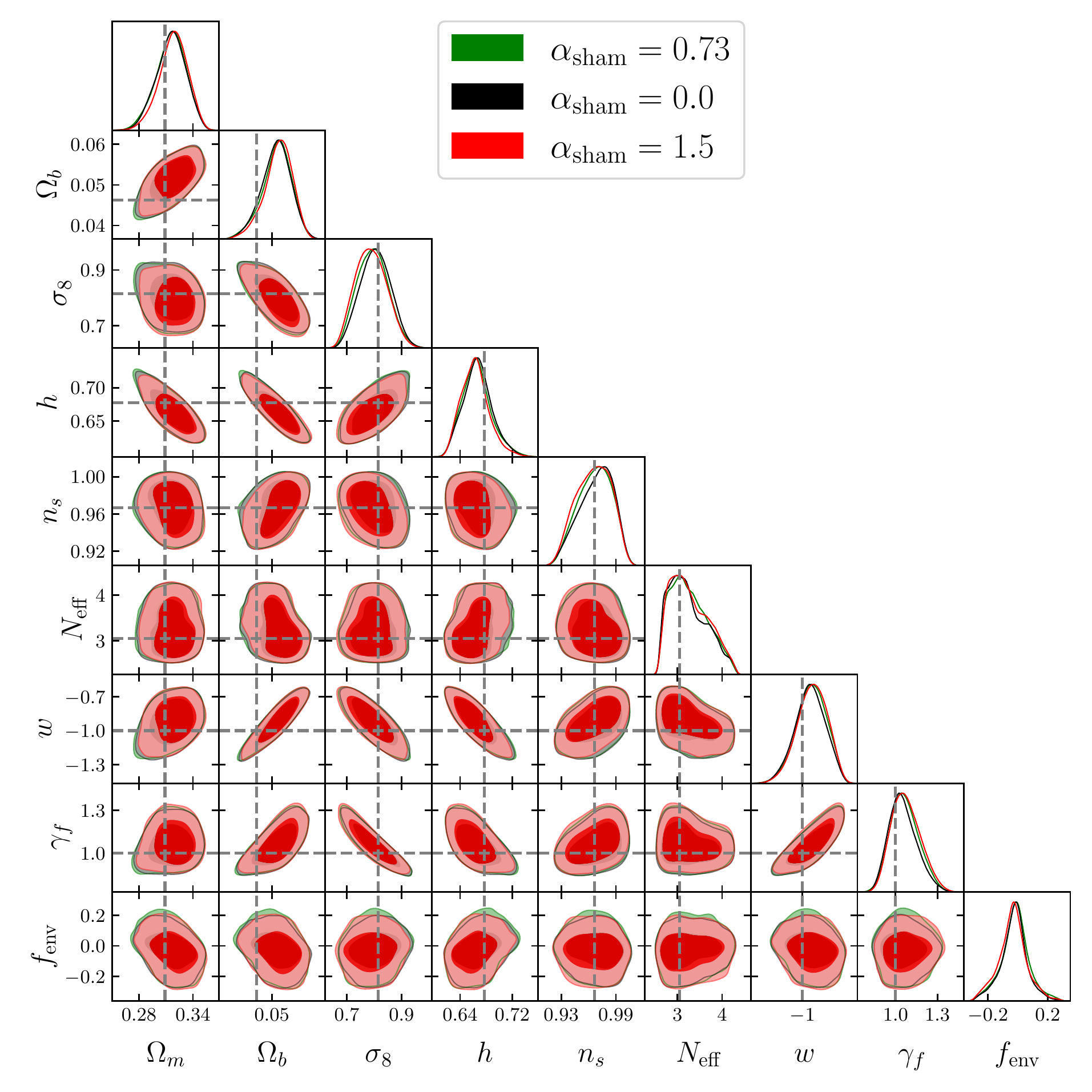}
\includegraphics[width=8cm]{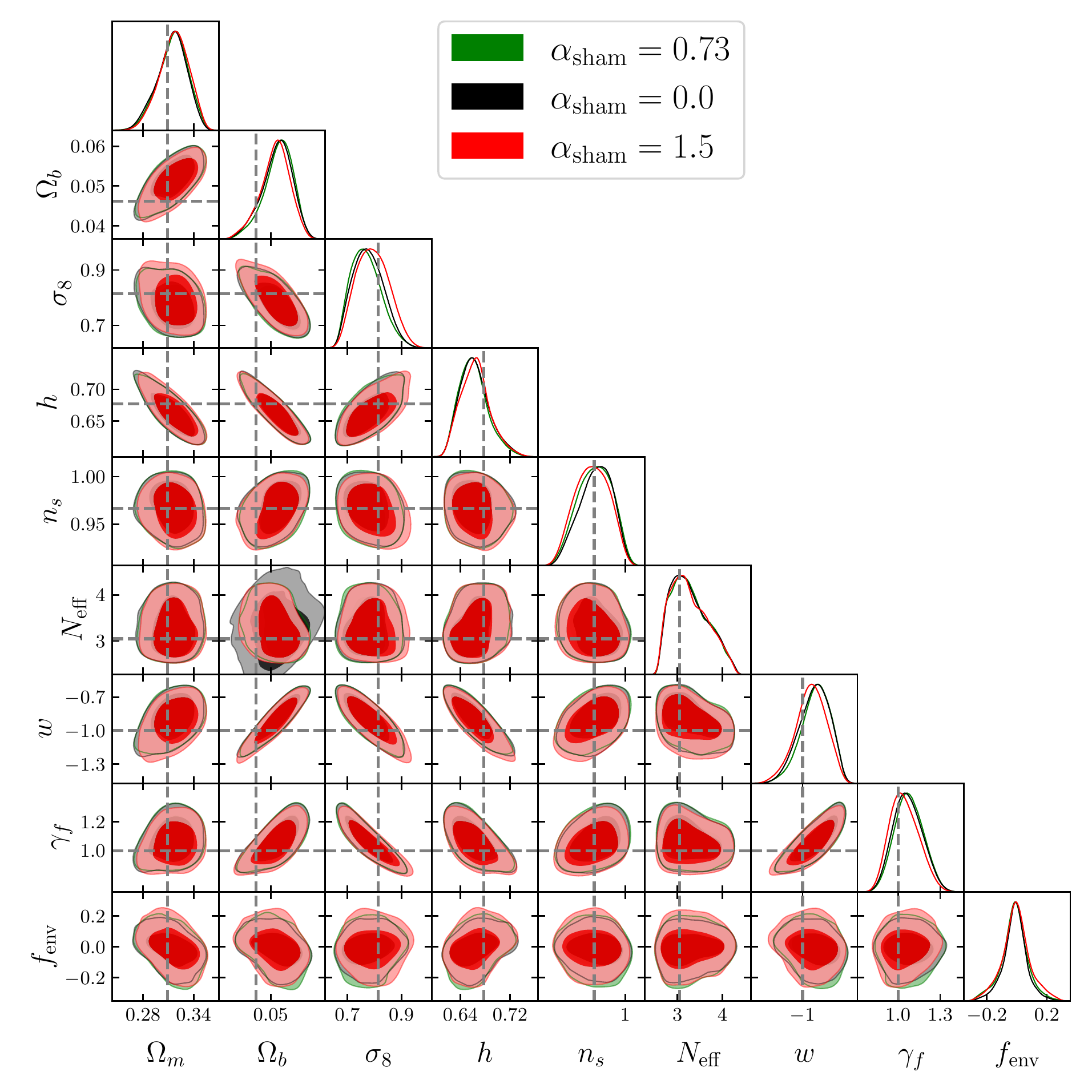}
\caption{Test on the SHAM catalog constructed using UNIT simulation with $\alpha_{sham}=1$ and scatter=0.08. The galaxy statistics employed in the analysis is $w_{p}+\xi_{0}+\xi_{2}$ for high-$z$ subsample. The left hand panel shows SHAM catalog constructed with scatter$=0.15$, while the right hand panel is for scatter$=0.19$. The test is done on SHAM catalogs with different values of $\alpha_{\text{sham}}$, which is a measure of the significance of assembly bias. The result shows that our HOD based model can give unbiased recovery of the input cosmology.}
\label{fig:constraint_UNIT}
\end{center}
\end{figure}

In Figure \ref{fig:constraint_UNIT}, we present the same recovery tests on the UNIT SHAM catalog. For simplicity, we assume the sample variance of the measurement of correlation function is equivalent to the training box of the {\sc Aemulus} suite. The actual ``effective volume" of the UNIT boxes is non-trivial to estimate due to the inverse phase technique to reduce cosmic variance. However, the overall findings based on this recovery test is likely to hold. For different values of the scatter and $\alpha_{\text{sham}}$, our HOD-based model is also able to recover the input cosmology. The result is quite similar to Uchuu simulation.

\section{Constraint on all the parameters} \label{appsec:full_triangle}

\begin{figure*}[htbp]
\begin{center}
\includegraphics[width=17.5cm]{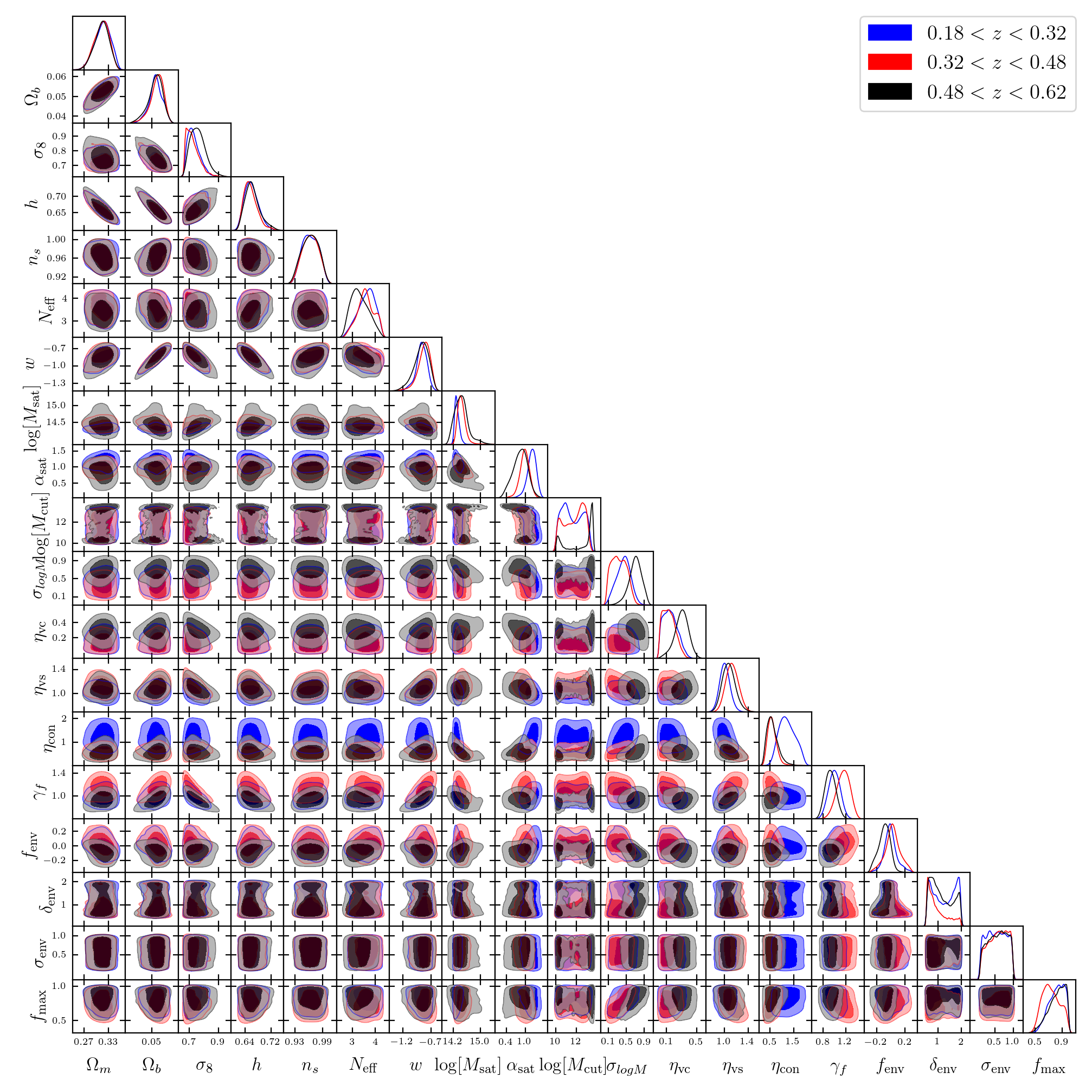}
\caption{1D and 2D contours of the parameters for our fiducial constraint using $w_{p}+\xi_{0}+\xi_{2}$.}
\label{fig:full_triangle}
\end{center}
\end{figure*}

Figure \ref{fig:full_triangle} displays the full triangle plot for the constraint on all the parameters in our model using measurements of $w_{p}+\xi_{0}+\xi_{2}$. The results of the three subsamples are shown on top of each other.

\section{Likelihood test of covariance matrix}\label{appsec:covariance}

\begin{figure*}[htbp]
\begin{center}
\includegraphics[width=18cm]{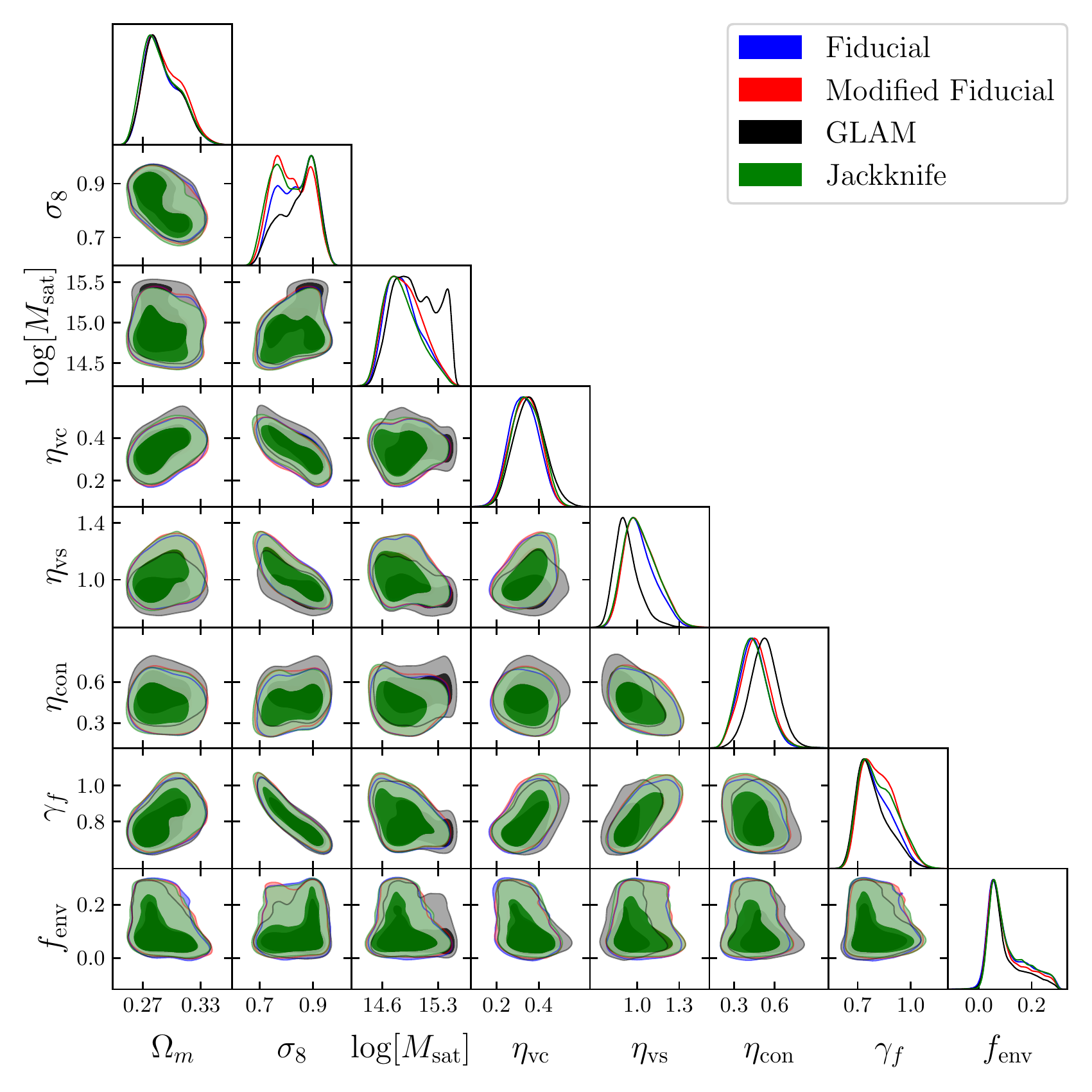}
\caption{Test of the different constructions of covariance matrix. The result shows constraint on a subset of the cosmological, HOD and assembly bias parameters using $w_{p}+\xi_{0}+\xi_{2}$ for high-$z$ ($z\sim0.55$) subsample. The consistency of these results demonstrate that the construction of the covariance matrix for sample variance doesn't have a significant impact on the cosmological inference.}
\label{fig:constraint_covariance_matrix_test}
\end{center}
\end{figure*}

\begin{figure*}[htbp]
\begin{center}
\includegraphics[width=7cm]{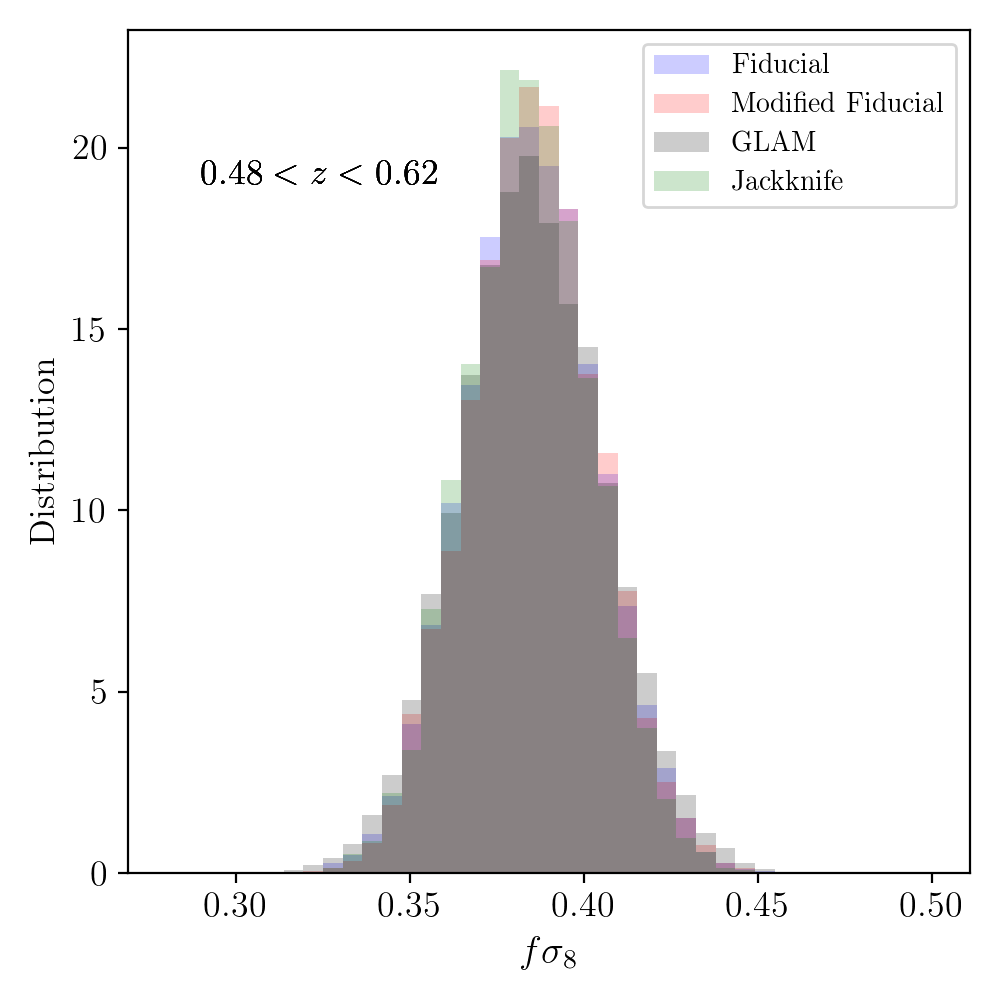}
\caption{Projection of $f\sigma_{8}$ using different constructions of the covariance matrix, labeled as Figure \ref{fig:constraint_covariance_matrix_test}. The results shows that the different constructions of the covariance matrix give quite consistent measurement of $f\sigma_{8}$.} 
\label{fig:fsigma8_covariance_matrix_test}
\end{center}
\end{figure*}

In this section, we show the constraint using different covariance matrices as constructed in Section \ref{sec:covariance}. We use the high-$z$ ($0.48<z<0.62$) subsample to present our results, but the low-$z$ and med-$z$ subsamples give similar results. For the high-$z$ subsample, we remind the readers that we create another set of GLAM mocks using a ``randomly" chosen HOD. Figure \ref{fig:constraint_covariance_matrix_test} presents the finalized constraint using $w_{p}+\xi_{0}+\xi_{2}$ for all these different covariance matrices. We find that the overall constraint is stable against difference choices of covariance matrix with some offset for a subset of the parameters. In order to explicitly investigate the impact on resultant measurement of structure growth rate, we extract $f\sigma_{8}$ from these analysis and present the distribution in Figure \ref{fig:fsigma8_covariance_matrix_test}. The results shows that different constructions of the covariance matrix give quite consistent measurement of linear growth rate of structure.
 
\section{Redshift uncertainty} \label{appsec:z_error}

\begin{figure*}[htbp]
\begin{center}
\includegraphics[width=8cm]{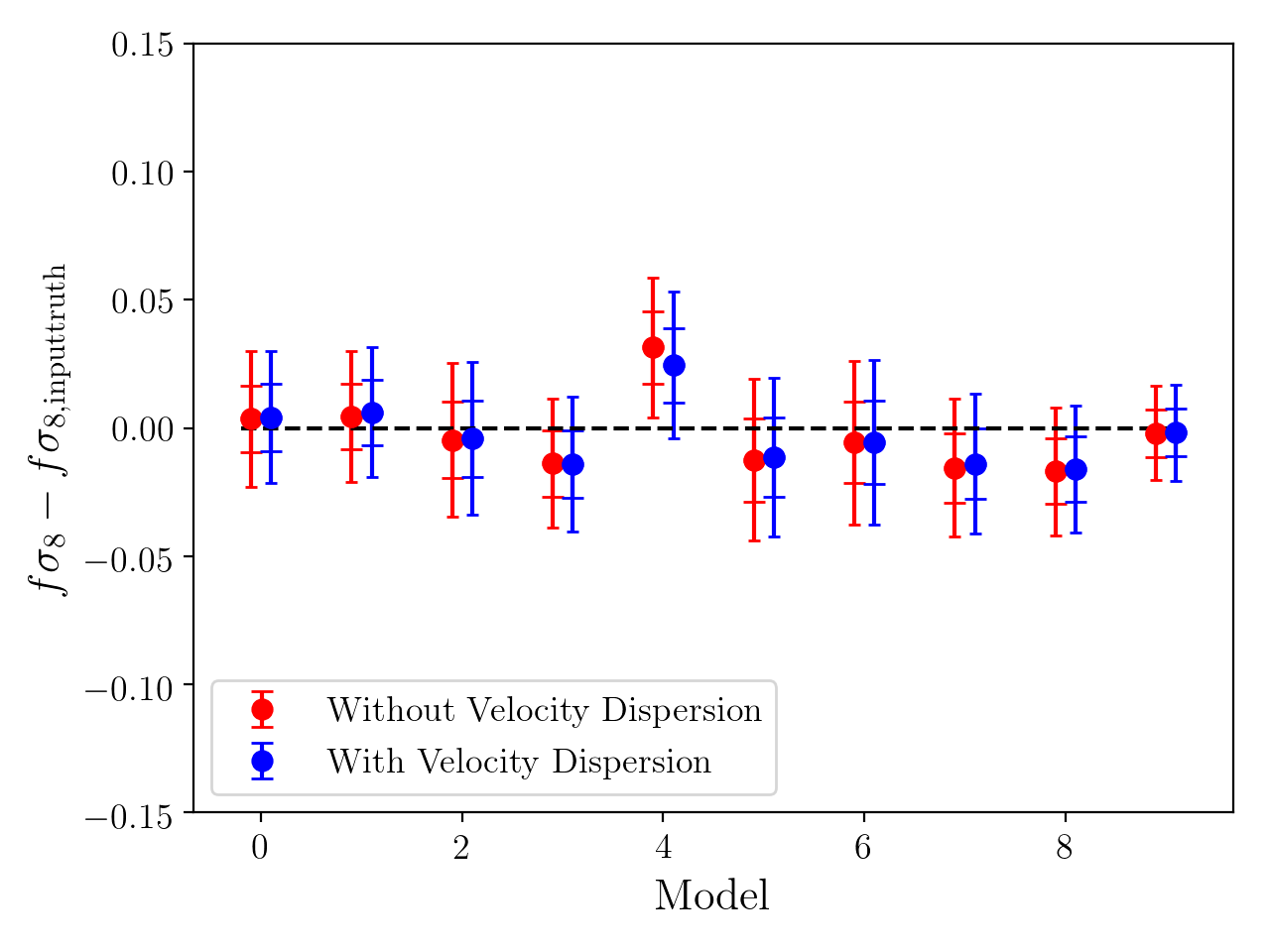}
\caption{Impact of velocity dispersion on the measurement of structure growth rate. The tests are done with ten randomly chosen models. The result shows the difference of $f\sigma_{8}$ recovered from the mock with respect to the input truth. The red dots with errorbars are result without additional velocity dispersion, while the blue dots are from mocks with additional velocity component. For plotting purpose, the points for the same model are shifted slightly along the x-axis and the errorbars represent 1 and 2 $\sigma$.}
\label{fig:constraint_velocity_dispersion}
\end{center}
\end{figure*}

One of the systematics in the clustering analysis is redshift uncertainty, or velocity dispersion, which can be estimated by repeat observations. The examination of BOSS galaxies shows that this uncertainty can be described by a Gaussian distribution with some value of standard deviation. The result based on both LOWZ and CMASS reveals a clear dependence of the dispersion on redshift, see \cite{Bolton_2012} for a detailed analysis. From a theoretical point of view, this velocity dispersion can be captured through our HOD modeling by the parameters $\gamma_{f}$ and velocity bias for central and satellites. In order to explicitly investigate the impact of this systematics on our cosmological measurements. We randomly pick 10 HOD models from our test suite, and add this velocity dispersion to the velocity of the galaxies in the mock and do the recovery test using the same set of emulator. In particular, we choose the high-$z$ subsample ($0.48<z<0.62$) in this analysis, add an additional velocity component randomly draw from a Gaussian distribution with a standard deviation of $30$ km/s corresponding to the maximum estimate in \cite{Bolton_2012} within our redshift range. We perform the recovery tests using galaxy statistics $w_{p}+\xi_{0}+\xi_{2}$ for models with and without this additional velocity component, and present the inferred measurement of structure growth rate in Figure \ref{fig:constraint_velocity_dispersion}. The dots with error bars represent the $1\sigma$ and $2\sigma$ uncertainties of $f\sigma_{8}$ recovered from the mocks compared with the input truth. This result shows that this velocity dispersion has a negligible impact on the final constraint. Since the velocity dispersion is increasing with redshift, the low-$z$ and med-$z$ subsamples can experience a less significant effect. Therefore this result validates the robustness of our measurement. 

In \cite{Chapman_2021}, the velocity dispersion is examined for the eBOSS LRG sample. Since these high-redshift galaxies have an average velocity dispersion of $91.8~\text{km}~\text{s}^{-1}$, a factor of three higher than our high-$z$ BOSS galaxies, it can impact the cosmological constraint to a higher extent. For full scale measurement, the parameter $\gamma_{f}$ can be biased with an offset of $0.5\sigma$. Although this component of systematics doesn't bias the BOSS measurements as in this paper, it imposes an additional concern for high-redshift galaxies in the near future, e.g., ELGs from eBOSS, DESI etc. Either a more flexible emulator or forward modeling is needed to achieve unbiased cosmological inference.

\bibliographystyle{yahapj}
\bibliography{emu_gc_bib,software}

\begin{thebibliography}{}
\providecommand\natexlab[1]{#1}
\providecommand\JournalTitle[1]{#1}

\bibitem[{{Abazajian} {et~al.}(2009){Abazajian}, {Adelman-McCarthy},
  {Ag{\"u}eros}, {Allam}, {Allende Prieto}, {An}, {Anderson}, {Anderson},
  {Annis}, {Bahcall}, \& et~al.}]{Abazajian_2009}
{Abazajian}, K.~N., {Adelman-McCarthy}, J.~K., {Ag{\"u}eros}, M.~A., {et~al.}
  2009,
  \href{http://dx.doi.org/10.1088/0067-0049/182/2/543}{\JournalTitle{\apjs},
  182, 543}

\bibitem[{{Abbott} {et~al.}(2022){Abbott}, {Aguena}, {Alarcon}, {Allam},
  {Alves}, {Amon}, {Andrade-Oliveira}, {Annis}, {Avila}, {Bacon}, {Baxter},
  {Bechtol}, {Becker}, {Bernstein}, {Bhargava}, {Birrer}, {Blazek},
  {Brandao-Souza}, {Bridle}, {Brooks}, {Buckley-Geer}, {Burke}, {Camacho},
  {Campos}, {Carnero Rosell}, {Carrasco Kind}, {Carretero}, {Castander},
  {Cawthon}, {Chang}, {Chen}, {Chen}, {Choi}, {Conselice}, {Cordero},
  {Costanzi}, {Crocce}, {da Costa}, {da Silva Pereira}, {Davis}, {Davis}, {De
  Vicente}, {DeRose}, {Desai}, {Di Valentino}, {Diehl}, {Dietrich}, {Dodelson},
  {Doel}, {Doux}, {Drlica-Wagner}, {Eckert}, {Eifler}, {Elsner}, {Elvin-Poole},
  {Everett}, {Evrard}, {Fang}, {Farahi}, {Fernandez}, {Ferrero}, {Fert{\'e}},
  {Fosalba}, {Friedrich}, {Frieman}, {Garc{\'\i}a-Bellido}, {Gatti},
  {Gaztanaga}, {Gerdes}, {Giannantonio}, {Giannini}, {Gruen}, {Gruendl},
  {Gschwend}, {Gutierrez}, {Harrison}, {Hartley}, {Herner}, {Hinton},
  {Hollowood}, {Honscheid}, {Hoyle}, {Huff}, {Huterer}, {Jain}, {James},
  {Jarvis}, {Jeffrey}, {Jeltema}, {Kovacs}, {Krause}, {Kron}, {Kuehn},
  {Kuropatkin}, {Lahav}, {Leget}, {Lemos}, {Liddle}, {Lidman}, {Lima}, {Lin},
  {MacCrann}, {Maia}, {Marshall}, {Martini}, {McCullough}, {Melchior},
  {Mena-Fern{\'a}ndez}, {Menanteau}, {Miquel}, {Mohr}, {Morgan}, {Muir},
  {Myles}, {Nadathur}, {Navarro-Alsina}, {Nichol}, {Ogando}, {Omori},
  {Palmese}, {Pandey}, {Park}, {Paz-Chinch{\'o}n}, {Petravick}, {Pieres},
  {Plazas Malag{\'o}n}, {Porredon}, {Prat}, {Raveri}, {Rodriguez-Monroy},
  {Rollins}, {Romer}, {Roodman}, {Rosenfeld}, {Ross}, {Rykoff}, {Samuroff},
  {S{\'a}nchez}, {Sanchez}, {Sanchez}, {Sanchez Cid}, {Scarpine}, {Schubnell},
  {Scolnic}, {Secco}, {Serrano}, {Sevilla-Noarbe}, {Sheldon}, {Shin}, {Smith},
  {Soares-Santos}, {Suchyta}, {Swanson}, {Tabbutt}, {Tarle}, {Thomas}, {To},
  {Troja}, {Troxel}, {Tucker}, {Tutusaus}, {Varga}, {Walker}, {Weaverdyck},
  {Wechsler}, {Weller}, {Yanny}, {Yin}, {Zhang}, {Zuntz}, \& {DES
  Collaboration}}]{Abbott_2022}
{Abbott}, T.~M.~C., {Aguena}, M., {Alarcon}, A., {et~al.} 2022,
  \href{http://dx.doi.org/10.1103/PhysRevD.105.023520}{\JournalTitle{\prd},
  105, 023520}

\bibitem[{{Alam} {et~al.}(2017){Alam}, {Ata}, {Bailey}, {Beutler}, {Bizyaev},
  {Blazek}, {Bolton}, {Brownstein}, {Burden}, {Chuang}, {Comparat}, {Cuesta},
  {Dawson}, {Eisenstein}, {Escoffier}, {Gil-Mar{\'\i}n}, {Grieb}, {Hand}, {Ho},
  {Kinemuchi}, {Kirkby}, {Kitaura}, {Malanushenko}, {Malanushenko}, {Maraston},
  {McBride}, {Nichol}, {Olmstead}, {Oravetz}, {Padmanabhan},
  {Palanque-Delabrouille}, {Pan}, {Pellejero-Ibanez}, {Percival}, {Petitjean},
  {Prada}, {Price-Whelan}, {Reid}, {Rodr{\'\i}guez-Torres}, {Roe}, {Ross},
  {Ross}, {Rossi}, {Rubi{\~n}o-Mart{\'\i}n}, {Saito}, {Salazar-Albornoz},
  {Samushia}, {S{\'a}nchez}, {Satpathy}, {Schlegel}, {Schneider},
  {Sc{\'o}ccola}, {Seo}, {Sheldon}, {Simmons}, {Slosar}, {Strauss}, {Swanson},
  {Thomas}, {Tinker}, {Tojeiro}, {Maga{\~n}a}, {Vazquez}, {Verde}, {Wake},
  {Wang}, {Weinberg}, {White}, {Wood-Vasey}, {Y{\`e}che}, {Zehavi}, {Zhai}, \&
  {Zhao}}]{Alam_2017}
{Alam}, S., {Ata}, M., {Bailey}, S., {et~al.} 2017,
  \href{http://dx.doi.org/10.1093/mnras/stx721}{\JournalTitle{\mnras}, 470,
  2617}

\bibitem[{{Alcock} \& {Paczynski}(1979)}]{Alcock_1979}
{Alcock}, C., \& {Paczynski}, B. 1979,
  \href{http://dx.doi.org/10.1038/281358a0}{\JournalTitle{\nat}, 281, 358}

\bibitem[{{Ambikasaran} {et~al.}(2015){Ambikasaran}, {Foreman-Mackey},
  {Greengard}, {Hogg}, \& {O'Neil}}]{george_2014}
{Ambikasaran}, S., {Foreman-Mackey}, D., {Greengard}, L., {Hogg}, D.~W., \&
  {O'Neil}, M. 2015,
  \href{http://dx.doi.org/10.1109/TPAMI.2015.2448083}{\JournalTitle{IEEE
  Transactions on Pattern Analysis and Machine Intelligence}, 38},
  \href{http://arxiv.org/abs/1403.6015}{{\sffamily arXiv:1403.6015 [math.NA]}}

\bibitem[{{Angulo} \& {Pontzen}(2016)}]{Angulo_2016}
{Angulo}, R.~E., \& {Pontzen}, A. 2016,
  \href{http://dx.doi.org/10.1093/mnrasl/slw098}{\JournalTitle{\mnras}, 462,
  L1}

\bibitem[{{Aric{\`o}} {et~al.}(2020){Aric{\`o}}, {Angulo}, {Contreras},
  {Ondaro-Mallea}, {Pellejero-Iba{\~n}ez}, \& {Zennaro}}]{Arico_2020}
{Aric{\`o}}, G., {Angulo}, R.~E., {Contreras}, S., {et~al.} 2020,
  \JournalTitle{arXiv e-prints}, arXiv:2011.15018

\bibitem[{{Asgari} {et~al.}(2020){Asgari}, {Tr{\"o}ster}, {Heymans},
  {Hildebrandt}, {van den Busch}, {Wright}, {Choi}, {Erben}, {Joachimi},
  {Joudaki}, {Kannawadi}, {Kuijken}, {Lin}, {Schneider}, \&
  {Zuntz}}]{Asgari_2020}
{Asgari}, M., {Tr{\"o}ster}, T., {Heymans}, C., {et~al.} 2020,
  \href{http://dx.doi.org/10.1051/0004-6361/201936512}{\JournalTitle{\aap},
  634, A127}

\bibitem[{{Bautista} {et~al.}(2021){Bautista}, {Paviot}, {Vargas Maga{\~n}a},
  {de la Torre}, {Fromenteau}, {Gil-Mar{\'\i}n}, {Ross}, {Burtin}, {Dawson},
  {Hou}, {Kneib}, {de Mattia}, {Percival}, {Rossi}, {Tojeiro}, {Zhao}, {Zhao},
  {Alam}, {Brownstein}, {Chapman}, {Choi}, {Chuang}, {Escoffier}, {de la
  Macorra}, {du Mas des Bourboux}, {Mohammad}, {Moon}, {M{\"u}ller},
  {Nadathur}, {Newman}, {Schneider}, {Seo}, \& {Wang}}]{Bautista_2021}
{Bautista}, J.~E., {Paviot}, R., {Vargas Maga{\~n}a}, M., {et~al.} 2021,
  \href{http://dx.doi.org/10.1093/mnras/staa2800}{\JournalTitle{\mnras}, 500,
  736}

\bibitem[{{Beutler} {et~al.}(2012){Beutler}, {Blake}, {Colless}, {Jones},
  {Staveley-Smith}, {Poole}, {Campbell}, {Parker}, {Saunders}, \&
  {Watson}}]{Beutler_2012}
{Beutler}, F., {Blake}, C., {Colless}, M., {et~al.} 2012,
  \href{http://dx.doi.org/10.1111/j.1365-2966.2012.21136.x}{\JournalTitle{\mnras},
  423, 3430}

\bibitem[{{Bird} {et~al.}(2019){Bird}, {Rogers}, {Peiris}, {Verde},
  {Font-Ribera}, \& {Pontzen}}]{Simeon_2019}
{Bird}, S., {Rogers}, K.~K., {Peiris}, H.~V., {et~al.} 2019,
  \href{http://dx.doi.org/10.1088/1475-7516/2019/02/050}{\JournalTitle{\jcap},
  2019, 050}

\bibitem[{{Blake} {et~al.}(2012){Blake}, {Brough}, {Colless}, {Contreras},
  {Couch}, {Croom}, {Croton}, {Davis}, {Drinkwater}, {Forster}, {Gilbank},
  {Gladders}, {Glazebrook}, {Jelliffe}, {Jurek}, {Li}, {Madore}, {Martin},
  {Pimbblet}, {Poole}, {Pracy}, {Sharp}, {Wisnioski}, {Woods}, {Wyder}, \&
  {Yee}}]{Blake_2012}
{Blake}, C., {Brough}, S., {Colless}, M., {et~al.} 2012,
  \href{http://dx.doi.org/10.1111/j.1365-2966.2012.21473.x}{\JournalTitle{\mnras},
  425, 405}

\bibitem[{{Blake} {et~al.}(2013){Blake}, {Baldry}, {Bland-Hawthorn},
  {Christodoulou}, {Colless}, {Conselice}, {Driver}, {Hopkins}, {Liske},
  {Loveday}, {Norberg}, {Peacock}, {Poole}, \& {Robotham}}]{Blake_2013}
{Blake}, C., {Baldry}, I.~K., {Bland-Hawthorn}, J., {et~al.} 2013,
  \href{http://dx.doi.org/10.1093/mnras/stt1791}{\JournalTitle{\mnras}, 436,
  3089}

\bibitem[{{Bolton} {et~al.}(2012){Bolton}, {Schlegel}, {Aubourg}, {Bailey},
  {Bhardwaj}, {Brownstein}, {Burles}, {Chen}, {Dawson}, {Eisenstein}, {Gunn},
  {Knapp}, {Loomis}, {Lupton}, {Maraston}, {Muna}, {Myers}, {Olmstead},
  {Padmanabhan}, {P{\^a}ris}, {Percival}, {Petitjean}, {Rockosi}, {Ross},
  {Schneider}, {Shu}, {Strauss}, {Thomas}, {Tremonti}, {Wake}, {Weaver}, \&
  {Wood-Vasey}}]{Bolton_2012}
{Bolton}, A.~S., {Schlegel}, D.~J., {Aubourg}, {\'E}., {et~al.} 2012,
  \href{http://dx.doi.org/10.1088/0004-6256/144/5/144}{\JournalTitle{\aj}, 144,
  144}

\bibitem[{{Buchner} {et~al.}(2014){Buchner}, {Georgakakis}, {Nandra}, {Hsu},
  {Rangel}, {Brightman}, {Merloni}, {Salvato}, {Donley}, \&
  {Kocevski}}]{Buchner_2014}
{Buchner}, J., {Georgakakis}, A., {Nandra}, K., {et~al.} 2014,
  \href{http://dx.doi.org/10.1051/0004-6361/201322971}{\JournalTitle{\aap},
  564, A125}

\bibitem[{{Chapman} {et~al.}(2021){Chapman}, {Mohammad}, {Zhai}, {Percival},
  {Tinker}, {Bautista}, {Brownstein}, {Burtin}, {Dawson}, {Gil-Mar{\'\i}n}, {de
  la Macorra}, {Ross}, {Rossi}, {Schneider}, \& {Zhao}}]{Chapman_2021}
{Chapman}, M.~J., {Mohammad}, F.~G., {Zhai}, Z., {et~al.} 2021,
  \JournalTitle{arXiv e-prints}, arXiv:2106.14961

\bibitem[{{Chen} {et~al.}(2022){Chen}, {Vlah}, \& {White}}]{Chen_2022}
{Chen}, S.-F., {Vlah}, Z., \& {White}, M. 2022,
  \href{http://dx.doi.org/10.1088/1475-7516/2022/02/008}{\JournalTitle{\jcap},
  2022, 008}

\bibitem[{{Chuang} {et~al.}(2019){Chuang}, {Yepes}, {Kitaura},
  {Pellejero-Ibanez}, {Rodr{\'\i}guez-Torres}, {Feng}, {Metcalf}, {Wechsler},
  {Zhao}, {To}, {Alam}, {Banerjee}, {DeRose}, {Giocoli}, {Knebe}, \&
  {Reyes}}]{Chuang_2019}
{Chuang}, C.-H., {Yepes}, G., {Kitaura}, F.-S., {et~al.} 2019,
  \href{http://dx.doi.org/10.1093/mnras/stz1233}{\JournalTitle{\mnras}, 487,
  48}

\bibitem[{{Cole} {et~al.}(2005){Cole}, {Percival}, {Peacock}, {Norberg},
  {Baugh}, {Frenk}, {Baldry}, {Bland-Hawthorn}, {Bridges}, {Cannon}, {Colless},
  {Collins}, {Couch}, {Cross}, {Dalton}, {Eke}, {De Propris}, {Driver},
  {Efstathiou}, {Ellis}, {Glazebrook}, {Jackson}, {Jenkins}, {Lahav}, {Lewis},
  {Lumsden}, {Maddox}, {Madgwick}, {Peterson}, {Sutherland}, \&
  {Taylor}}]{Cole_2005}
{Cole}, S., {Percival}, W.~J., {Peacock}, J.~A., {et~al.} 2005,
  \href{http://dx.doi.org/10.1111/j.1365-2966.2005.09318.x}{\JournalTitle{\mnras},
  362, 505}

\bibitem[{{Colless} {et~al.}(2001){Colless}, {Dalton}, {Maddox}, {Sutherland},
  {Norberg}, {Cole}, {Bland-Hawthorn}, {Bridges}, {Cannon}, {Collins}, {Couch},
  {Cross}, {Deeley}, {De Propris}, {Driver}, {Efstathiou}, {Ellis}, {Frenk},
  {Glazebrook}, {Jackson}, {Lahav}, {Lewis}, {Lumsden}, {Madgwick}, {Peacock},
  {Peterson}, {Price}, {Seaborne}, \& {Taylor}}]{Colless_2001}
{Colless}, M., {Dalton}, G., {Maddox}, S., {et~al.} 2001,
  \href{http://dx.doi.org/10.1046/j.1365-8711.2001.04902.x}{\JournalTitle{\mnras},
  328, 1039}

\bibitem[{{Conroy} {et~al.}(2006){Conroy}, {Wechsler}, \&
  {Kravtsov}}]{Conroy_2006}
{Conroy}, C., {Wechsler}, R.~H., \& {Kravtsov}, A.~V. 2006,
  \href{http://dx.doi.org/10.1086/503602}{\JournalTitle{\apj}, 647, 201}

\bibitem[{{Contreras} {et~al.}(2020){Contreras}, {Angulo}, \&
  {Zennaro}}]{Contreras_2020}
{Contreras}, S., {Angulo}, R., \& {Zennaro}, M. 2020, \JournalTitle{arXiv
  e-prints}, arXiv:2005.03672

\bibitem[{{Davis} \& {Peebles}(1983)}]{Davis_1983}
{Davis}, M., \& {Peebles}, P.~J.~E. 1983,
  \href{http://dx.doi.org/10.1086/160884}{\JournalTitle{\apj}, 267, 465}

\bibitem[{{Dawson} {et~al.}(2013){Dawson}, {Schlegel}, {Ahn}, {Anderson},
  {Aubourg}, {Bailey}, {Barkhouser}, {Bautista}, {Beifiori}, {Berlind},
  {Bhardwaj}, {Bizyaev}, {Blake}, {Blanton}, {Blomqvist}, {Bolton}, {Borde},
  {Bovy}, {Brandt}, {Brewington}, {Brinkmann}, {Brown}, {Brownstein}, {Bundy},
  {Busca}, {Carithers}, {Carnero}, {Carr}, {Chen}, {Comparat}, {Connolly},
  {Cope}, {Croft}, {Cuesta}, {da Costa}, {Davenport}, {Delubac}, {de Putter},
  {Dhital}, {Ealet}, {Ebelke}, {Eisenstein}, {Escoffier}, {Fan}, {Filiz Ak},
  {Finley}, {Font-Ribera}, {G{\'e}nova-Santos}, {Gunn}, {Guo}, {Haggard},
  {Hall}, {Hamilton}, {Harris}, {Harris}, {Ho}, {Hogg}, {Holder}, {Honscheid},
  {Huehnerhoff}, {Jordan}, {Jordan}, {Kauffmann}, {Kazin}, {Kirkby}, {Klaene},
  {Kneib}, {Le Goff}, {Lee}, {Long}, {Loomis}, {Lundgren}, {Lupton}, {Maia},
  {Makler}, {Malanushenko}, {Malanushenko}, {Mandelbaum}, {Manera}, {Maraston},
  {Margala}, {Masters}, {McBride}, {McDonald}, {McGreer}, {McMahon}, {Mena},
  {Miralda-Escud{\'e}}, {Montero-Dorta}, {Montesano}, {Muna}, {Myers},
  {Naugle}, {Nichol}, {Noterdaeme}, {Nuza}, {Olmstead}, {Oravetz}, {Oravetz},
  {Owen}, {Padmanabhan}, {Palanque-Delabrouille}, {Pan}, {Parejko},
  {P{\^a}ris}, {Percival}, {P{\'e}rez-Fournon}, {P{\'e}rez-R{\`a}fols},
  {Petitjean}, {Pfaffenberger}, {Pforr}, {Pieri}, {Prada}, {Price-Whelan},
  {Raddick}, {Rebolo}, {Rich}, {Richards}, {Rockosi}, {Roe}, {Ross}, {Ross},
  {Rossi}, {Rubi{\~n}o-Martin}, {Samushia}, {S{\'a}nchez}, {Sayres}, {Schmidt},
  {Schneider}, {Sc{\'o}ccola}, {Seo}, {Shelden}, {Sheldon}, {Shen}, {Shu},
  {Slosar}, {Smee}, {Snedden}, {Stauffer}, {Steele}, {Strauss}, {Streblyanska},
  {Suzuki}, {Swanson}, {Tal}, {Tanaka}, {Thomas}, {Tinker}, {Tojeiro},
  {Tremonti}, {Vargas Maga{\~n}a}, {Verde}, {Viel}, {Wake}, {Watson}, {Weaver},
  {Weinberg}, {Weiner}, {West}, {White}, {Wood-Vasey}, {Yeche}, {Zehavi},
  {Zhao}, \& {Zheng}}]{Dawson_BOSS}
{Dawson}, K.~S., {Schlegel}, D.~J., {Ahn}, C.~P., {et~al.} 2013,
  \href{http://dx.doi.org/10.1088/0004-6256/145/1/10}{\JournalTitle{\aj}, 145,
  10}

\bibitem[{{Dawson} {et~al.}(2016){Dawson}, {Kneib}, {Percival}, {Alam},
  {Albareti}, {Anderson}, {Armengaud}, {Aubourg}, {Bailey}, {Bautista},
  {Berlind}, {Bershady}, {Beutler}, {Bizyaev}, {Blanton}, {Blomqvist},
  {Bolton}, {Bovy}, {Brandt}, {Brinkmann}, {Brownstein}, {Burtin}, {Busca},
  {Cai}, {Chuang}, {Clerc}, {Comparat}, {Cope}, {Croft}, {Cruz-Gonzalez}, {da
  Costa}, {Cousinou}, {Darling}, {de la Macorra}, {de la Torre}, {Delubac}, {du
  Mas des Bourboux}, {Dwelly}, {Ealet}, {Eisenstein}, {Eracleous}, {Escoffier},
  {Fan}, {Finoguenov}, {Font-Ribera}, {Frinchaboy}, {Gaulme}, {Georgakakis},
  {Green}, {Guo}, {Guy}, {Ho}, {Holder}, {Huehnerhoff}, {Hutchinson}, {Jing},
  {Jullo}, {Kamble}, {Kinemuchi}, {Kirkby}, {Kitaura}, {Klaene}, {Laher},
  {Lang}, {Laurent}, {Le Goff}, {Li}, {Liang}, {Lima}, {Lin}, {Lin}, {Lin},
  {Long}, {Lundgren}, {MacDonald}, {Geimba Maia}, {Malanushenko},
  {Malanushenko}, {Mariappan}, {McBride}, {McGreer}, {M{\'e}nard}, {Merloni},
  {Meza}, {Montero-Dorta}, {Muna}, {Myers}, {Nandra}, {Naugle}, {Newman},
  {Noterdaeme}, {Nugent}, {Ogando}, {Olmstead}, {Oravetz}, {Oravetz},
  {Padmanabhan}, {Palanque-Delabrouille}, {Pan}, {Parejko}, {P{\^a}ris},
  {Peacock}, {Petitjean}, {Pieri}, {Pisani}, {Prada}, {Prakash}, {Raichoor},
  {Reid}, {Rich}, {Ridl}, {Rodriguez-Torres}, {Carnero Rosell}, {Ross},
  {Rossi}, {Ruan}, {Salvato}, {Sayres}, {Schneider}, {Schlegel}, {Seljak},
  {Seo}, {Sesar}, {Shandera}, {Shu}, {Slosar}, {Sobreira}, {Streblyanska},
  {Suzuki}, {Taylor}, {Tao}, {Tinker}, {Tojeiro}, {Vargas-Maga{\~n}a}, {Wang},
  {Weaver}, {Weinberg}, {White}, {Wood-Vasey}, {Yeche}, {Zhai}, {Zhao}, {Zhao},
  {Zheng}, {Ben Zhu}, \& {Zou}}]{eBOSS_Dawson}
{Dawson}, K.~S., {Kneib}, J.-P., {Percival}, W.~J., {et~al.} 2016,
  \href{http://dx.doi.org/10.3847/0004-6256/151/2/44}{\JournalTitle{\aj}, 151,
  44}

\bibitem[{{de Jong} {et~al.}(2016){de Jong}, {Barden}, {Bellido-Tirado},
  {Brynnel}, {Frey}, {Giannone}, {Haynes}, {Johl}, {Phillips}, {Schnurr},
  {Walcher}, {Winkler}, {Ansorge}, {Feltzing}, {McMahon}, {Baker}, {Caillier},
  {Dwelly}, {Gaessler}, {Iwert}, {Mandel}, {Piskunov}, {Pragt}, {Walton},
  {Bensby}, {Bergemann}, {Chiappini}, {Christlieb}, {Cioni}, {Driver},
  {Finoguenov}, {Helmi}, {Irwin}, {Kitaura}, {Kneib}, {Liske}, {Merloni},
  {Minchev}, {Richard}, \& {Starkenburg}}]{Jong_2016}
{de Jong}, R.~S., {Barden}, S.~C., {Bellido-Tirado}, O., {et~al.} 2016,
  \href{http://dx.doi.org/10.1117/12.2232832}{in \procspie, Vol. 9908,
  Ground-based and Airborne Instrumentation for Astronomy VI}, 99081O

\bibitem[{{de la Torre} {et~al.}(2013){de la Torre}, {Guzzo}, {Peacock},
  {Branchini}, {Iovino}, {Granett}, {Abbas}, {Adami}, {Arnouts}, {Bel},
  {Bolzonella}, {Bottini}, {Cappi}, {Coupon}, {Cucciati}, {Davidzon}, {De
  Lucia}, {Fritz}, {Franzetti}, {Fumana}, {Garilli}, {Ilbert}, {Krywult}, {Le
  Brun}, {Le F{\`e}vre}, {Maccagni}, {Ma{\l}ek}, {Marulli}, {McCracken},
  {Moscardini}, {Paioro}, {Percival}, {Polletta}, {Pollo}, {Schlagenhaufer},
  {Scodeggio}, {Tasca}, {Tojeiro}, {Vergani}, {Zanichelli}, {Burden}, {Di
  Porto}, {Marchetti}, {Marinoni}, {Mellier}, {Monaco}, {Nichol}, {Phleps},
  {Wolk}, \& {Zamorani}}]{de_la_Torre_2013}
{de la Torre}, S., {Guzzo}, L., {Peacock}, J.~A., {et~al.} 2013,
  \href{http://dx.doi.org/10.1051/0004-6361/201321463}{\JournalTitle{\aap},
  557, A54}

\bibitem[{{DeRose} {et~al.}(2019){DeRose}, {Wechsler}, {Tinker}, {Becker},
  {Mao}, {McClintock}, {McLaughlin}, {Rozo}, \& {Zhai}}]{DeRose_2018}
{DeRose}, J., {Wechsler}, R.~H., {Tinker}, J.~L., {et~al.} 2019,
  \href{http://dx.doi.org/10.3847/1538-4357/ab1085}{\JournalTitle{\apj}, 875,
  69}

\bibitem[{{DESI Collaboration} {et~al.}(2016){DESI Collaboration}, {Aghamousa},
  {Aguilar}, {Ahlen}, {Alam}, {Allen}, {Allende Prieto}, {Annis}, {Bailey},
  {Balland}, \& et~al.}]{DESI_2016}
{DESI Collaboration}, {Aghamousa}, A., {Aguilar}, J., {et~al.} 2016,
  \JournalTitle{ArXiv e-prints},
  \href{http://arxiv.org/abs/1611.00036}{{\sffamily arXiv:1611.00036
  [astro-ph.IM]}}

\bibitem[{{Dressler} {et~al.}(2012){Dressler}, {Spergel}, {Mountain},
  {Postman}, {Elliott}, {Bendek}, {Bennett}, {Dalcanton}, {Gaudi}, \&
  {Gehrels}}]{Dressler_2012}
{Dressler}, A., {Spergel}, D., {Mountain}, M., {et~al.} 2012,
  \JournalTitle{arXiv e-prints}, arXiv:1210.7809

\bibitem[{{Drinkwater} {et~al.}(2010){Drinkwater}, {Jurek}, {Blake}, {Woods},
  {Pimbblet}, {Glazebrook}, {Sharp}, {Pracy}, {Brough}, {Colless}, {Couch},
  {Croom}, {Davis}, {Forbes}, {Forster}, {Gilbank}, {Gladders}, {Jelliffe},
  {Jones}, {Li}, {Madore}, {Martin}, {Poole}, {Small}, {Wisnioski}, {Wyder}, \&
  {Yee}}]{Drinkwater_2010}
{Drinkwater}, M.~J., {Jurek}, R.~J., {Blake}, C., {et~al.} 2010,
  \href{http://dx.doi.org/10.1111/j.1365-2966.2009.15754.x}{\JournalTitle{\mnras},
  401, 1429}

\bibitem[{{Euclid Collaboration} {et~al.}(2019){Euclid Collaboration},
  {Knabenhans}, {Stadel}, {Marelli}, {Potter}, {Teyssier}, {Legrand},
  {Schneider}, {Sudret}, {Blot}, {Awan}, {Burigana}, {Carvalho},
  {Kurki-Suonio}, \& {Sirri}}]{Knabenhans_2019}
{Euclid Collaboration}, {Knabenhans}, M., {Stadel}, J., {et~al.} 2019,
  \href{http://dx.doi.org/10.1093/mnras/stz197}{\JournalTitle{\mnras}, 484,
  5509}

\bibitem[{{Feroz} {et~al.}(2009){Feroz}, {Hobson}, \& {Bridges}}]{Feroz_2009}
{Feroz}, F., {Hobson}, M.~P., \& {Bridges}, M. 2009,
  \href{http://dx.doi.org/10.1111/j.1365-2966.2009.14548.x}{\JournalTitle{\mnras},
  398, 1601}

\bibitem[{{Fisher} {et~al.}(1994){Fisher}, {Davis}, {Strauss}, {Yahil}, \&
  {Huchra}}]{Fisher_1994}
{Fisher}, K.~B., {Davis}, M., {Strauss}, M.~A., {Yahil}, A., \& {Huchra}, J.
  1994, \href{http://dx.doi.org/10.1093/mnras/266.1.50}{\JournalTitle{\mnras},
  266, 50}

\bibitem[{{Gao} {et~al.}(2005){Gao}, {Springel}, \& {White}}]{Gao_2005}
{Gao}, L., {Springel}, V., \& {White}, S.~D.~M. 2005,
  \href{http://dx.doi.org/10.1111/j.1745-3933.2005.00084.x}{\JournalTitle{\mnras},
  363, L66}

\bibitem[{{Giblin} {et~al.}(2019){Giblin}, {Cataneo}, {Moews}, \&
  {Heymans}}]{Giblin_2019}
{Giblin}, B., {Cataneo}, M., {Moews}, B., \& {Heymans}, C. 2019,
  \href{http://dx.doi.org/10.1093/mnras/stz2659}{\JournalTitle{\mnras}, 490,
  4826}

\bibitem[{{Green} {et~al.}(2012){Green}, {Schechter}, {Baltay}, {Bean},
  {Bennett}, {Brown}, {Conselice}, {Donahue}, {Fan}, \& {Gaudi}}]{Green_2012}
{Green}, J., {Schechter}, P., {Baltay}, C., {et~al.} 2012, \JournalTitle{arXiv
  e-prints}, arXiv:1208.4012

\bibitem[{{Guo} {et~al.}(2012){Guo}, {Zehavi}, \& {Zheng}}]{Guo_2012}
{Guo}, H., {Zehavi}, I., \& {Zheng}, Z. 2012,
  \href{http://dx.doi.org/10.1088/0004-637X/756/2/127}{\JournalTitle{\apj},
  756, 127}

\bibitem[{{Guo} {et~al.}(2015){Guo}, {Zheng}, {Zehavi}, {Dawson}, {Skibba},
  {Tinker}, {Weinberg}, {White}, \& {Schneider}}]{Guo_2015}
{Guo}, H., {Zheng}, Z., {Zehavi}, I., {et~al.} 2015,
  \href{http://dx.doi.org/10.1093/mnras/stu2120}{\JournalTitle{\mnras}, 446,
  578}

\bibitem[{{Han} {et~al.}(2019){Han}, {Li}, {Jing}, {Nishimichi}, {Wang}, \&
  {Jiang}}]{Han_2019}
{Han}, J., {Li}, Y., {Jing}, Y., {et~al.} 2019,
  \href{http://dx.doi.org/10.1093/mnras/sty2822}{\JournalTitle{\mnras}, 482,
  1900}

\bibitem[{{Harker} {et~al.}(2006){Harker}, {Cole}, {Helly}, {Frenk}, \&
  {Jenkins}}]{Harker_2006}
{Harker}, G., {Cole}, S., {Helly}, J., {Frenk}, C., \& {Jenkins}, A. 2006,
  \href{http://dx.doi.org/10.1111/j.1365-2966.2006.10022.x}{\JournalTitle{\mnras},
  367, 1039}

\bibitem[{{Hearin} {et~al.}(2021){Hearin}, {Chaves-Montero}, {Becker}, \&
  {Alarcon}}]{Hearin_2021}
{Hearin}, A.~P., {Chaves-Montero}, J., {Becker}, M.~R., \& {Alarcon}, A. 2021,
  \href{http://dx.doi.org/10.21105/astro.2105.05859}{\JournalTitle{The Open
  Journal of Astrophysics}, 4, 7}

\bibitem[{{Hearin} {et~al.}(2016){Hearin}, {Zentner}, {van den Bosch},
  {Campbell}, \& {Tollerud}}]{Hearin_2016}
{Hearin}, A.~P., {Zentner}, A.~R., {van den Bosch}, F.~C., {Campbell}, D., \&
  {Tollerud}, E. 2016,
  \href{http://dx.doi.org/10.1093/mnras/stw840}{\JournalTitle{\mnras}, 460,
  2552}

\bibitem[{{Heitmann} {et~al.}(2009){Heitmann}, {Higdon}, {White}, {Habib},
  {Williams}, {Lawrence}, \& {Wagner}}]{Heitmann_2009}
{Heitmann}, K., {Higdon}, D., {White}, M., {et~al.} 2009,
  \href{http://dx.doi.org/10.1088/0004-637X/705/1/156}{\JournalTitle{\apj},
  705, 156}

\bibitem[{{Heitmann} {et~al.}(2014){Heitmann}, {Lawrence}, {Kwan}, {Habib}, \&
  {Higdon}}]{Heitmann_2014}
{Heitmann}, K., {Lawrence}, E., {Kwan}, J., {Habib}, S., \& {Higdon}, D. 2014,
  \href{http://dx.doi.org/10.1088/0004-637X/780/1/111}{\JournalTitle{\apj},
  780, 111}

\bibitem[{{Heitmann} {et~al.}(2010){Heitmann}, {White}, {Wagner}, {Habib}, \&
  {Higdon}}]{Heitmann_2010}
{Heitmann}, K., {White}, M., {Wagner}, C., {Habib}, S., \& {Higdon}, D. 2010,
  \href{http://dx.doi.org/10.1088/0004-637X/715/1/104}{\JournalTitle{\apj},
  715, 104}

\bibitem[{{Hoshino} {et~al.}(2015){Hoshino}, {Leauthaud}, {Lackner}, {Hikage},
  {Rozo}, {Rykoff}, {Mandelbaum}, {More}, {More}, {Saito}, \&
  {Vulcani}}]{Hoshino_2015}
{Hoshino}, H., {Leauthaud}, A., {Lackner}, C., {et~al.} 2015,
  \href{http://dx.doi.org/10.1093/mnras/stv1271}{\JournalTitle{\mnras}, 452,
  998}

\bibitem[{{Howlett} {et~al.}(2015){Howlett}, {Ross}, {Samushia}, {Percival}, \&
  {Manera}}]{Howlett_2015}
{Howlett}, C., {Ross}, A.~J., {Samushia}, L., {Percival}, W.~J., \& {Manera},
  M. 2015,
  \href{http://dx.doi.org/10.1093/mnras/stu2693}{\JournalTitle{\mnras}, 449,
  848}

\bibitem[{Hunter(2007)}]{matplotlib}
Hunter, J.~D. 2007,
  \href{http://dx.doi.org/10.1109/MCSE.2007.55}{\JournalTitle{Computing in
  Science Engineering}, 9, 90}

\bibitem[{{Ishiyama} {et~al.}(2020){Ishiyama}, {Prada}, {Klypin}, {Sinha},
  {Metcalf}, {Jullo}, {Altieri}, {Cora}, {Croton}, {de la Torre},
  {Mill{\'a}n-Calero}, {Oogi}, {Ruedas}, \&
  {Vega-Mart{\'\i}nez}}]{Ishiyama_2020}
{Ishiyama}, T., {Prada}, F., {Klypin}, A.~A., {et~al.} 2020,
  \JournalTitle{arXiv e-prints}, arXiv:2007.14720

\bibitem[{{Ivanov}(2021)}]{Ivanov_2021}
{Ivanov}, M.~M. 2021,
  \href{http://dx.doi.org/10.1103/PhysRevD.104.103514}{\JournalTitle{\prd},
  104, 103514}

\bibitem[{Jones {et~al.}(2001--)Jones, Oliphant, Peterson, {et~al.}}]{scipy}
Jones, E., Oliphant, T., Peterson, P., {et~al.} 2001--, {SciPy}: Open source
  scientific tools for {Python}, [Online;
  \href{http://www.scipy.org/}{scipy.org}]

\bibitem[{{Klypin} \& {Prada}(2018)}]{Klypin_2018}
{Klypin}, A., \& {Prada}, F. 2018,
  \href{http://dx.doi.org/10.1093/mnras/sty1340}{\JournalTitle{\mnras}, 478,
  4602}

\bibitem[{{Klypin} {et~al.}(2016){Klypin}, {Yepes}, {Gottl{\"o}ber}, {Prada},
  \& {He{\ss}}}]{Klypin_2016}
{Klypin}, A., {Yepes}, G., {Gottl{\"o}ber}, S., {Prada}, F., \& {He{\ss}}, S.
  2016, \href{http://dx.doi.org/10.1093/mnras/stw248}{\JournalTitle{\mnras},
  457, 4340}

\bibitem[{{Klypin} {et~al.}(2011){Klypin}, {Trujillo-Gomez}, \&
  {Primack}}]{Klypin_2011}
{Klypin}, A.~A., {Trujillo-Gomez}, S., \& {Primack}, J. 2011,
  \href{http://dx.doi.org/10.1088/0004-637X/740/2/102}{\JournalTitle{\apj},
  740, 102}

\bibitem[{{Kobayashi} {et~al.}(2021){Kobayashi}, {Nishimichi}, {Takada}, \&
  {Miyatake}}]{Kobayashi_2021}
{Kobayashi}, Y., {Nishimichi}, T., {Takada}, M., \& {Miyatake}, H. 2021,
  \JournalTitle{arXiv e-prints}, arXiv:2110.06969

\bibitem[{{Kobayashi} {et~al.}(2020){Kobayashi}, {Nishimichi}, {Takada},
  {Takahashi}, \& {Osato}}]{Kobayashi_2020}
{Kobayashi}, Y., {Nishimichi}, T., {Takada}, M., {Takahashi}, R., \& {Osato},
  K. 2020,
  \href{http://dx.doi.org/10.1103/PhysRevD.102.063504}{\JournalTitle{\prd},
  102, 063504}

\bibitem[{{Kokron} {et~al.}(2021){Kokron}, {DeRose}, {Chen}, {White}, \&
  {Wechsler}}]{Kokron_2021}
{Kokron}, N., {DeRose}, J., {Chen}, S.-F., {White}, M., \& {Wechsler}, R.~H.
  2021, \href{http://dx.doi.org/10.1093/mnras/stab1358}{\JournalTitle{\mnras},
  505, 1422}

\bibitem[{{Kravtsov} {et~al.}(2004){Kravtsov}, {Berlind}, {Wechsler}, {Klypin},
  {Gottl{\"o}ber}, {Allgood}, \& {Primack}}]{Kravtsov_2004}
{Kravtsov}, A.~V., {Berlind}, A.~A., {Wechsler}, R.~H., {et~al.} 2004,
  \href{http://dx.doi.org/10.1086/420959}{\JournalTitle{\apj}, 609, 35}

\bibitem[{{Krolewski} {et~al.}(2021){Krolewski}, {Ferraro}, \&
  {White}}]{Krolewski_2021}
{Krolewski}, A., {Ferraro}, S., \& {White}, M. 2021,
  \href{http://dx.doi.org/10.1088/1475-7516/2021/12/028}{\JournalTitle{\jcap},
  2021, 028}

\bibitem[{{Landy} \& {Szalay}(1993)}]{LS_1993}
{Landy}, S.~D., \& {Szalay}, A.~S. 1993,
  \href{http://dx.doi.org/10.1086/172900}{\JournalTitle{\apj}, 412, 64}

\bibitem[{{Lange} {et~al.}(2021){Lange}, {Hearin}, {Leauthaud}, {van den
  Bosch}, {Guo}, \& {DeRose}}]{Lange_2021}
{Lange}, J.~U., {Hearin}, A.~P., {Leauthaud}, A., {et~al.} 2021,
  \JournalTitle{arXiv e-prints}, arXiv:2101.12261

\bibitem[{{Laureijs} {et~al.}(2011){Laureijs}, {Amiaux}, {Arduini},
  {Augu{\`e}res}, {Brinchmann}, {Cole}, {Cropper}, {Dabin}, {Duvet}, \&
  {Ealet}}]{Laureijs_2011}
{Laureijs}, R., {Amiaux}, J., {Arduini}, S., {et~al.} 2011, \JournalTitle{arXiv
  e-prints}, arXiv:1110.3193

\bibitem[{{Laureijs} {et~al.}(2012){Laureijs}, {Gondoin}, {Duvet}, {Saavedra
  Criado}, {Hoar}, {Amiaux}, {Augu{\`e}res}, {Cole}, {Cropper}, {Ealet},
  {Ferruit}, {Escudero Sanz}, {Jahnke}, {Kohley}, {Maciaszek}, {Mellier},
  {Oosterbroek}, {Pasian}, {Sauvage}, {Scaramella}, {Sirianni}, \&
  {Valenziano}}]{Laureijs_2012}
{Laureijs}, R., {Gondoin}, P., {Duvet}, L., {et~al.} 2012,
  \href{http://dx.doi.org/10.1117/12.926496}{in \procspie, Vol. 8442, Space
  Telescopes and Instrumentation 2012: Optical, Infrared, and Millimeter Wave},
  84420T

\bibitem[{{Lawrence} {et~al.}(2010){Lawrence}, {Heitmann}, {White}, {Higdon},
  {Wagner}, {Habib}, \& {Williams}}]{Lawrence_2010}
{Lawrence}, E., {Heitmann}, K., {White}, M., {et~al.} 2010,
  \href{http://dx.doi.org/10.1088/0004-637X/713/2/1322}{\JournalTitle{\apj},
  713, 1322}

\bibitem[{{Leauthaud} {et~al.}(2016){Leauthaud}, {Bundy}, {Saito}, {Tinker},
  {Maraston}, {Tojeiro}, {Huang}, {Brownstein}, {Schneider}, \&
  {Thomas}}]{leauthaud_etal:16}
{Leauthaud}, A., {Bundy}, K., {Saito}, S., {et~al.} 2016,
  \href{http://dx.doi.org/10.1093/mnras/stw117}{\JournalTitle{\mnras}, 457,
  4021}

\bibitem[{{Leauthaud} {et~al.}(2017){Leauthaud}, {Saito}, {Hilbert},
  {Barreira}, {More}, {White}, {Alam}, {Behroozi}, {Bundy}, {Coupon}, {Erben},
  {Heymans}, {Hildebrandt}, {Mandelbaum}, {Miller}, {Moraes}, {Pereira},
  {Rodr{\'\i}guez-Torres}, {Schmidt}, {Shan}, {Viel}, \&
  {Villaescusa-Navarro}}]{Leauthaud_2017}
{Leauthaud}, A., {Saito}, S., {Hilbert}, S., {et~al.} 2017,
  \href{http://dx.doi.org/10.1093/mnras/stx258}{\JournalTitle{\mnras}, 467,
  3024}

\bibitem[{{Lehmann} {et~al.}(2017){Lehmann}, {Mao}, {Becker}, {Skillman}, \&
  {Wechsler}}]{Lehmann_2017}
{Lehmann}, B.~V., {Mao}, Y.-Y., {Becker}, M.~R., {Skillman}, S.~W., \&
  {Wechsler}, R.~H. 2017,
  \href{http://dx.doi.org/10.3847/1538-4357/834/1/37}{\JournalTitle{\apj}, 834,
  37}

\bibitem[{Lue {et~al.}(2004)Lue, Scoccimarro, \& Starkman}]{Lue_04}
Lue, A., Scoccimarro, R., \& Starkman, G. 2004,
  \href{http://dx.doi.org/10.1103/physrevd.69.044005}{\JournalTitle{Physical
  Review D}, 69}

\bibitem[{{Mao} {et~al.}(2015){Mao}, {Williamson}, \& {Wechsler}}]{Mao_2015}
{Mao}, Y.-Y., {Williamson}, M., \& {Wechsler}, R.~H. 2015,
  \href{http://dx.doi.org/10.1088/0004-637X/810/1/21}{\JournalTitle{\apj}, 810,
  21}

\bibitem[{{Mao} {et~al.}(2018){Mao}, {Zentner}, \& {Wechsler}}]{Mao_2018}
{Mao}, Y.-Y., {Zentner}, A.~R., \& {Wechsler}, R.~H. 2018,
  \href{http://dx.doi.org/10.1093/mnras/stx3111}{\JournalTitle{\mnras}, 474,
  5143}

\bibitem[{{McClintock} {et~al.}(2019{\natexlab{a}}){McClintock}, {Rozo},
  {Becker}, {DeRose}, {Mao}, {McLaughlin}, {Tinker}, {Wechsler}, \&
  {Zhai}}]{McClintock_2018}
{McClintock}, T., {Rozo}, E., {Becker}, M.~R., {et~al.} 2019{\natexlab{a}},
  \href{http://dx.doi.org/10.3847/1538-4357/aaf568}{\JournalTitle{\apj}, 872,
  53}

\bibitem[{{McClintock} {et~al.}(2019{\natexlab{b}}){McClintock}, {Rozo},
  {Banerjee}, {Becker}, {DeRose}, {McLaughlin}, {Tinker}, {Wechsler}, \&
  {Zhai}}]{McClintock_2019}
{McClintock}, T., {Rozo}, E., {Banerjee}, A., {et~al.} 2019{\natexlab{b}},
  \JournalTitle{arXiv e-prints}, arXiv:1907.13167

\bibitem[{{Miyatake} {et~al.}(2020){Miyatake}, {Kobayashi}, {Takada},
  {Nishimichi}, {Shirasaki}, {Sugiyama}, {Takahashi}, {Osato}, {More}, \&
  {Park}}]{Miyatake_2021}
{Miyatake}, H., {Kobayashi}, Y., {Takada}, M., {et~al.} 2020,
  \JournalTitle{arXiv e-prints}, arXiv:2101.00113

\bibitem[{{Nishimichi} {et~al.}(2019){Nishimichi}, {Takada}, {Takahashi},
  {Osato}, {Shirasaki}, {Oogi}, {Miyatake}, {Oguri}, {Murata}, {Kobayashi}, \&
  {Yoshida}}]{Nishimichi_2019}
{Nishimichi}, T., {Takada}, M., {Takahashi}, R., {et~al.} 2019,
  \href{http://dx.doi.org/10.3847/1538-4357/ab3719}{\JournalTitle{\apj}, 884,
  29}

\bibitem[{{Pedersen} {et~al.}(2020){Pedersen}, {Font-Ribera}, {Rogers},
  {McDonald}, {Peiris}, {Pontzen}, \& {Slosar}}]{Pedersen_2020}
{Pedersen}, C., {Font-Ribera}, A., {Rogers}, K.~K., {et~al.} 2020,
  \JournalTitle{arXiv e-prints}, arXiv:2011.15127

\bibitem[{{Planck Collaboration} {et~al.}(2015){Planck Collaboration}, {Ade},
  {Aghanim}, {Arnaud}, {Ashdown}, {Aumont}, {Baccigalupi}, {Banday},
  {Barreiro}, {Bartlett}, \& et~al.}]{Planck_2015}
{Planck Collaboration}, {Ade}, P.~A.~R., {Aghanim}, N., {et~al.} 2015,
  \JournalTitle{ArXiv e-prints},
  \href{http://arxiv.org/abs/1502.01589}{{\sffamily arXiv:1502.01589}}

\bibitem[{{Planck Collaboration} {et~al.}(2020){Planck Collaboration},
  {Aghanim}, {Akrami}, {Ashdown}, {Aumont}, {Baccigalupi}, {Ballardini},
  {Banday}, {Barreiro}, {Bartolo}, {Basak}, {Battye}, {Benabed}, {Bernard},
  {Bersanelli}, {Bielewicz}, {Bock}, {Bond}, {Borrill}, {Bouchet}, {Boulanger},
  {Bucher}, {Burigana}, {Butler}, {Calabrese}, {Cardoso}, {Carron},
  {Challinor}, {Chiang}, {Chluba}, {Colombo}, {Combet}, {Contreras}, {Crill},
  {Cuttaia}, {de Bernardis}, {de Zotti}, {Delabrouille}, {Delouis}, {Di
  Valentino}, {Diego}, {Dor{\'e}}, {Douspis}, {Ducout}, {Dupac}, {Dusini},
  {Efstathiou}, {Elsner}, {En{\ss}lin}, {Eriksen}, {Fantaye}, {Farhang},
  {Fergusson}, {Fernandez-Cobos}, {Finelli}, {Forastieri}, {Frailis},
  {Fraisse}, {Franceschi}, {Frolov}, {Galeotta}, {Galli}, {Ganga},
  {G{\'e}nova-Santos}, {Gerbino}, {Ghosh}, {Gonz{\'a}lez-Nuevo}, {G{\'o}rski},
  {Gratton}, {Gruppuso}, {Gudmundsson}, {Hamann}, {Handley}, {Hansen},
  {Herranz}, {Hildebrandt}, {Hivon}, {Huang}, {Jaffe}, {Jones}, {Karakci},
  {Keih{\"a}nen}, {Keskitalo}, {Kiiveri}, {Kim}, {Kisner}, {Knox},
  {Krachmalnicoff}, {Kunz}, {Kurki-Suonio}, {Lagache}, {Lamarre}, {Lasenby},
  {Lattanzi}, {Lawrence}, {Le Jeune}, {Lemos}, {Lesgourgues}, {Levrier},
  {Lewis}, {Liguori}, {Lilje}, {Lilley}, {Lindholm}, {L{\'o}pez-Caniego},
  {Lubin}, {Ma}, {Mac{\'\i}as-P{\'e}rez}, {Maggio}, {Maino}, {Mandolesi},
  {Mangilli}, {Marcos-Caballero}, {Maris}, {Martin}, {Martinelli},
  {Mart{\'\i}nez-Gonz{\'a}lez}, {Matarrese}, {Mauri}, {McEwen}, {Meinhold},
  {Melchiorri}, {Mennella}, {Migliaccio}, {Millea}, {Mitra},
  {Miville-Desch{\^e}nes}, {Molinari}, {Montier}, {Morgante}, {Moss}, {Natoli},
  {N{\o}rgaard-Nielsen}, {Pagano}, {Paoletti}, {Partridge}, {Patanchon},
  {Peiris}, {Perrotta}, {Pettorino}, {Piacentini}, {Polastri}, {Polenta},
  {Puget}, {Rachen}, {Reinecke}, {Remazeilles}, {Renzi}, {Rocha}, {Rosset},
  {Roudier}, {Rubi{\~n}o-Mart{\'\i}n}, {Ruiz-Granados}, {Salvati}, {Sandri},
  {Savelainen}, {Scott}, {Shellard}, {Sirignano}, {Sirri}, {Spencer},
  {Sunyaev}, {Suur-Uski}, {Tauber}, {Tavagnacco}, {Tenti}, {Toffolatti},
  {Tomasi}, {Trombetti}, {Valenziano}, {Valiviita}, {Van Tent}, {Vibert},
  {Vielva}, {Villa}, {Vittorio}, {Wandelt}, {Wehus}, {White}, {White},
  {Zacchei}, \& {Zonca}}]{Planck_2020}
{Planck Collaboration}, {Aghanim}, N., {Akrami}, Y., {et~al.} 2020,
  \href{http://dx.doi.org/10.1051/0004-6361/201833910}{\JournalTitle{\aap},
  641, A6}

\bibitem[{{Ramachandra} {et~al.}(2020){Ramachandra}, {Valogiannis}, {Ishak}, \&
  {Heitmann}}]{Ramachandra_2020}
{Ramachandra}, N., {Valogiannis}, G., {Ishak}, M., \& {Heitmann}, K. 2020,
  \JournalTitle{arXiv e-prints}, arXiv:2010.00596

\bibitem[{{Reid} {et~al.}(2016){Reid}, {Ho}, {Padmanabhan}, {Percival},
  {Tinker}, {Tojeiro}, {White}, {Eisenstein}, {Maraston}, {Ross},
  {S{\'a}nchez}, {Schlegel}, {Sheldon}, {Strauss}, {Thomas}, {Wake}, {Beutler},
  {Bizyaev}, {Bolton}, {Brownstein}, {Chuang}, {Dawson}, {Harding}, {Kitaura},
  {Leauthaud}, {Masters}, {McBride}, {More}, {Olmstead}, {Oravetz}, {Nuza},
  {Pan}, {Parejko}, {Pforr}, {Prada}, {Rodr{\'\i}guez-Torres},
  {Salazar-Albornoz}, {Samushia}, {Schneider}, {Sc{\'o}ccola}, {Simmons}, \&
  {Vargas-Magana}}]{Reid_2016}
{Reid}, B., {Ho}, S., {Padmanabhan}, N., {et~al.} 2016,
  \href{http://dx.doi.org/10.1093/mnras/stv2382}{\JournalTitle{\mnras}, 455,
  1553}

\bibitem[{{Reid} {et~al.}(2014){Reid}, {Seo}, {Leauthaud}, {Tinker}, \&
  {White}}]{Reid_2014}
{Reid}, B.~A., {Seo}, H.-J., {Leauthaud}, A., {Tinker}, J.~L., \& {White}, M.
  2014, \href{http://dx.doi.org/10.1093/mnras/stu1391}{\JournalTitle{\mnras},
  444, 476}

\bibitem[{{Rogers} {et~al.}(2019){Rogers}, {Peiris}, {Pontzen}, {Bird},
  {Verde}, \& {Font-Ribera}}]{Rogers_2019}
{Rogers}, K.~K., {Peiris}, H.~V., {Pontzen}, A., {et~al.} 2019,
  \href{http://dx.doi.org/10.1088/1475-7516/2019/02/031}{\JournalTitle{\jcap},
  2019, 031}

\bibitem[{{Salcedo} {et~al.}(2018){Salcedo}, {Maller}, {Berlind}, {Sinha},
  {McBride}, {Behroozi}, {Wechsler}, \& {Weinberg}}]{Salcedo_2018}
{Salcedo}, A.~N., {Maller}, A.~H., {Berlind}, A.~A., {et~al.} 2018,
  \href{http://dx.doi.org/10.1093/mnras/sty109}{\JournalTitle{\mnras}, 475,
  4411}

\bibitem[{{Salcedo} {et~al.}(2020){Salcedo}, {Zu}, {Zhang}, {Wang}, {Yang},
  {Wu}, {Jing}, {Mo}, \& {Weinberg}}]{Salcedo_2020}
{Salcedo}, A.~N., {Zu}, Y., {Zhang}, Y., {et~al.} 2020, \JournalTitle{arXiv
  e-prints}, arXiv:2010.04176

\bibitem[{{Samushia} {et~al.}(2014){Samushia}, {Reid}, {White}, {Percival},
  {Cuesta}, {Zhao}, {Ross}, {Manera}, {Aubourg}, {Beutler}, {Brinkmann},
  {Brownstein}, {Dawson}, {Eisenstein}, {Ho}, {Honscheid}, {Maraston},
  {Montesano}, {Nichol}, {Roe}, {Ross}, {S{\'a}nchez}, {Schlegel}, {Schneider},
  {Streblyanska}, {Thomas}, {Tinker}, {Wake}, {Weaver}, \&
  {Zehavi}}]{Samushia_2014}
{Samushia}, L., {Reid}, B.~A., {White}, M., {et~al.} 2014,
  \href{http://dx.doi.org/10.1093/mnras/stu197}{\JournalTitle{\mnras}, 439,
  3504}

\bibitem[{{Sheth} \& {Tormen}(2004)}]{Sheth_2004}
{Sheth}, R.~K., \& {Tormen}, G. 2004,
  \href{http://dx.doi.org/10.1111/j.1365-2966.2004.07733.x}{\JournalTitle{\mnras},
  350, 1385}

\bibitem[{{Shi} \& {Sheth}(2018)}]{Shi_2018}
{Shi}, J., \& {Sheth}, R.~K. 2018,
  \href{http://dx.doi.org/10.1093/mnras/stx2277}{\JournalTitle{\mnras}, 473,
  2486}

\bibitem[{{Singh} {et~al.}(2020){Singh}, {Mandelbaum}, {Seljak},
  {Rodr{\'\i}guez-Torres}, \& {Slosar}}]{Singh_2020b}
{Singh}, S., {Mandelbaum}, R., {Seljak}, U., {Rodr{\'\i}guez-Torres}, S., \&
  {Slosar}, A. 2020,
  \href{http://dx.doi.org/10.1093/mnras/stz2922}{\JournalTitle{\mnras}, 491,
  51}

\bibitem[{{Sinha} \& {Garrison}(2020)}]{Sinha_2020}
{Sinha}, M., \& {Garrison}, L.~H. 2020,
  \href{http://dx.doi.org/10.1093/mnras/stz3157}{\JournalTitle{\mnras}, 491,
  3022}

\bibitem[{{Skilling}(2004)}]{Skilling_2004}
{Skilling}, J. 2004, \href{http://dx.doi.org/10.1063/1.1835238}{in American
  Institute of Physics Conference Series, Vol. 735, Bayesian Inference and
  Maximum Entropy Methods in Science and Engineering: 24th International
  Workshop on Bayesian Inference and Maximum Entropy Methods in Science and
  Engineering, ed. R.~{Fischer}, R.~{Preuss}, \& U.~V. {Toussaint}}, 395

\bibitem[{{Spergel} {et~al.}(2015){Spergel}, {Gehrels}, {Baltay}, {Bennett},
  {Breckinridge}, {Donahue}, {Dressler}, {Gaudi}, {Greene}, \&
  {Guyon}}]{Spergel_2015}
{Spergel}, D., {Gehrels}, N., {Baltay}, C., {et~al.} 2015, \JournalTitle{arXiv
  e-prints}, arXiv:1503.03757

\bibitem[{{Szewciw} {et~al.}(2021){Szewciw}, {Beltz-Mohrmann}, {Berlind}, \&
  {Sinha}}]{Szewciw_2021}
{Szewciw}, A.~O., {Beltz-Mohrmann}, G.~D., {Berlind}, A.~A., \& {Sinha}, M.
  2021, \JournalTitle{arXiv e-prints}, arXiv:2110.03701

\bibitem[{{Takada} {et~al.}(2014){Takada}, {Ellis}, {Chiba}, {Greene},
  {Aihara}, {Arimoto}, {Bundy}, {Cohen}, {Dor{\'e}}, {Graves}, {Gunn},
  {Heckman}, {Hirata}, {Ho}, {Kneib}, {Le F{\`e}vre}, {Lin}, {More},
  {Murayama}, {Nagao}, {Ouchi}, {Seiffert}, {Silverman}, {Sodr{\'e}},
  {Spergel}, {Strauss}, {Sugai}, {Suto}, {Takami}, \& {Wyse}}]{Takada_2014}
{Takada}, M., {Ellis}, R.~S., {Chiba}, M., {et~al.} 2014,
  \href{http://dx.doi.org/10.1093/pasj/pst019}{\JournalTitle{\pasj}, 66, R1}

\bibitem[{{Tinker} {et~al.}(2008){Tinker}, {Conroy}, {Norberg}, {Patiri},
  {Weinberg}, \& {Warren}}]{Tinker_void_2008}
{Tinker}, J.~L., {Conroy}, C., {Norberg}, P., {et~al.} 2008,
  \href{http://dx.doi.org/10.1086/589983}{\JournalTitle{\apj}, 686, 53}

\bibitem[{Tinker {et~al.}(2005)Tinker, Weinberg, Zheng, \&
  Zehavi}]{Tinker_analytical}
Tinker, J.~L., Weinberg, D.~H., Zheng, Z., \& Zehavi, I. 2005,
  \href{http://stacks.iop.org/0004-637X/631/i=1/a=41}{\JournalTitle{\apj}, 631,
  41}

\bibitem[{{Tinker} {et~al.}(2012){Tinker}, {Sheldon}, {Wechsler}, {Becker},
  {Rozo}, {Zu}, {Weinberg}, {Zehavi}, {Blanton}, {Busha}, \&
  {Koester}}]{Tinker_2012}
{Tinker}, J.~L., {Sheldon}, E.~S., {Wechsler}, R.~H., {et~al.} 2012,
  \href{http://dx.doi.org/10.1088/0004-637X/745/1/16}{\JournalTitle{\apj}, 745,
  16}

\bibitem[{{Tinker} {et~al.}(2017){Tinker}, {Brownstein}, {Guo}, {Leauthaud},
  {Maraston}, {Masters}, {Montero-Dorta}, {Thomas}, {Tojeiro}, {Weiner},
  {Zehavi}, \& {Olmstead}}]{Tinker_2017}
{Tinker}, J.~L., {Brownstein}, J.~R., {Guo}, H., {et~al.} 2017,
  \href{http://dx.doi.org/10.3847/1538-4357/aa6845}{\JournalTitle{\apj}, 839,
  121}

\bibitem[{{Vakili} \& {Hahn}(2019)}]{Vakili_2019}
{Vakili}, M., \& {Hahn}, C. 2019,
  \href{http://dx.doi.org/10.3847/1538-4357/aaf1a1}{\JournalTitle{\apj}, 872,
  115}

\bibitem[{{Vale} \& {Ostriker}(2004)}]{Vale_2004}
{Vale}, A., \& {Ostriker}, J.~P. 2004,
  \href{http://dx.doi.org/10.1111/j.1365-2966.2004.08059.x}{\JournalTitle{\mnras},
  353, 189}

\bibitem[{van~der Walt {et~al.}(2011)van~der Walt, Colbert, \&
  Varoquaux}]{numpy}
van~der Walt, S., Colbert, S.~C., \& Varoquaux, G. 2011,
  \href{http://dx.doi.org/10.1109/MCSE.2011.37}{\JournalTitle{Computing in
  Science Engineering}, 13, 22}

\bibitem[{{Villarreal} {et~al.}(2017){Villarreal}, {Zentner}, {Mao}, {Purcell},
  {van den Bosch}, {Diemer}, {Lange}, {Wang}, \& {Campbell}}]{Villareal_2017}
{Villarreal}, A.~S., {Zentner}, A.~R., {Mao}, Y.-Y., {et~al.} 2017,
  \href{http://dx.doi.org/10.1093/mnras/stx2045}{\JournalTitle{\mnras}, 472,
  1088}

\bibitem[{{Walsh} \& {Tinker}(2019)}]{Walsh_2019}
{Walsh}, K., \& {Tinker}, J. 2019,
  \href{http://dx.doi.org/10.1093/mnras/stz1351}{\JournalTitle{\mnras}, 488,
  470}

\bibitem[{{Walther} {et~al.}(2020){Walther}, {Armengaud}, {Ravoux},
  {Palanque-Delabrouille}, {Y{\`e}che}, \& {Luki{\'c}}}]{Walther_2020}
{Walther}, M., {Armengaud}, E., {Ravoux}, C., {et~al.} 2020,
  \JournalTitle{arXiv e-prints}, arXiv:2012.04008

\bibitem[{{Wang} {et~al.}(2019){Wang}, {Mao}, {Zentner}, {van den Bosch},
  {Lange}, {Schafer}, {Villarreal}, {Hearin}, \& {Campbell}}]{Wang_2019}
{Wang}, K., {Mao}, Y.-Y., {Zentner}, A.~R., {et~al.} 2019,
  \href{http://dx.doi.org/10.1093/mnras/stz1733}{\JournalTitle{\mnras}, 488,
  3541}

\bibitem[{{Wang} {et~al.}(2021){Wang}, {Zhai}, {Alavi}, {Massara}, {Pisani},
  {Benson}, {Hirata}, {Samushia}, {Weinberg}, {Colbert}, {Dor{\'e}}, {Eifler},
  {Heinrich}, {Ho}, {Krause}, {Padmanabhan}, {Spergel}, \&
  {Teplitz}}]{Wang_2021}
{Wang}, Y., {Zhai}, Z., {Alavi}, A., {et~al.} 2021, \JournalTitle{arXiv
  e-prints}, arXiv:2110.01829

\bibitem[{{Wechsler} {et~al.}(2002){Wechsler}, {Bullock}, {Primack},
  {Kravtsov}, \& {Dekel}}]{Wechsler2001}
{Wechsler}, R.~H., {Bullock}, J.~S., {Primack}, J.~R., {Kravtsov}, A.~V., \&
  {Dekel}, A. 2002,
  \href{http://dx.doi.org/10.1086/338765}{\JournalTitle{\apj}, 568, 52}

\bibitem[{{Wechsler} \& {Tinker}(2018)}]{Wechsler_2018}
{Wechsler}, R.~H., \& {Tinker}, J.~L. 2018,
  \href{http://dx.doi.org/10.1146/annurev-astro-081817-051756}{\JournalTitle{\araa},
  56, 435}

\bibitem[{{Wechsler} {et~al.}(2006){Wechsler}, {Zentner}, {Bullock},
  {Kravtsov}, \& {Allgood}}]{Wechsler2006}
{Wechsler}, R.~H., {Zentner}, A.~R., {Bullock}, J.~S., {Kravtsov}, A.~V., \&
  {Allgood}, B. 2006,
  \href{http://dx.doi.org/10.1086/507120}{\JournalTitle{\apj}, 652, 71}

\bibitem[{{White} {et~al.}(2022){White}, {Zhou}, {DeRose}, {Ferraro}, {Chen},
  {Kokron}, {Bailey}, {Brooks}, {Garc{\'\i}a-Bellido}, {Guy}, {Honscheid},
  {Kehoe}, {Kremin}, {Levi}, {Palanque-Delabrouille}, {Poppett}, {Schlegel}, \&
  {Tarle}}]{White_2022}
{White}, M., {Zhou}, R., {DeRose}, J., {et~al.} 2022,
  \href{http://dx.doi.org/10.1088/1475-7516/2022/02/007}{\JournalTitle{\jcap},
  2022, 007}

\bibitem[{{Wibking} {et~al.}(2020){Wibking}, {Weinberg}, {Salcedo}, {Wu},
  {Singh}, {Rodr{\'\i}guez-Torres}, {Garrison}, \& {Eisenstein}}]{Wibking_2020}
{Wibking}, B.~D., {Weinberg}, D.~H., {Salcedo}, A.~N., {et~al.} 2020,
  \href{http://dx.doi.org/10.1093/mnras/stz3423}{\JournalTitle{\mnras}, 492,
  2872}

\bibitem[{{Wibking} {et~al.}(2017){Wibking}, {Salcedo}, {Weinberg}, {Garrison},
  {Ferrer}, {Tinker}, {Eisenstein}, {Metchnik}, \& {Pinto}}]{Wibking_2017}
{Wibking}, B.~D., {Salcedo}, A.~N., {Weinberg}, D.~H., {et~al.} 2017,
  \JournalTitle{ArXiv e-prints},
  \href{http://arxiv.org/abs/1709.07099}{{\sffamily arXiv:1709.07099}}

\bibitem[{{Xu} {et~al.}(2020){Xu}, {Zehavi}, \& {Contreras}}]{Xu_2020}
{Xu}, X., {Zehavi}, I., \& {Contreras}, S. 2020, \JournalTitle{arXiv e-prints},
  arXiv:2007.05545

\bibitem[{{York} {et~al.}(2000){York}, {Adelman}, {Anderson}, {Anderson},
  {Annis}, {Bahcall}, {Bakken}, {Barkhouser}, {Bastian}, {Berman}, {Boroski},
  {Bracker}, {Briegel}, {Briggs}, {Brinkmann}, {Brunner}, {Burles}, {Carey},
  {Carr}, {Castander}, {Chen}, {Colestock}, {Connolly}, {Crocker}, {Csabai},
  {Czarapata}, {Davis}, {Doi}, {Dombeck}, {Eisenstein}, {Ellman}, {Elms},
  {Evans}, {Fan}, {Federwitz}, {Fiscelli}, {Friedman}, {Frieman}, {Fukugita},
  {Gillespie}, {Gunn}, {Gurbani}, {de Haas}, {Haldeman}, {Harris}, {Hayes},
  {Heckman}, {Hennessy}, {Hindsley}, {Holm}, {Holmgren}, {Huang}, {Hull},
  {Husby}, {Ichikawa}, {Ichikawa}, {Ivezi{\'c}}, {Kent}, {Kim}, {Kinney},
  {Klaene}, {Kleinman}, {Kleinman}, {Knapp}, {Korienek}, {Kron}, {Kunszt},
  {Lamb}, {Lee}, {Leger}, {Limmongkol}, {Lindenmeyer}, {Long}, {Loomis},
  {Loveday}, {Lucinio}, {Lupton}, {MacKinnon}, {Mannery}, {Mantsch}, {Margon},
  {McGehee}, {McKay}, {Meiksin}, {Merelli}, {Monet}, {Munn}, {Narayanan},
  {Nash}, {Neilsen}, {Neswold}, {Newberg}, {Nichol}, {Nicinski}, {Nonino},
  {Okada}, {Okamura}, {Ostriker}, {Owen}, {Pauls}, {Peoples}, {Peterson},
  {Petravick}, {Pier}, {Pope}, {Pordes}, {Prosapio}, {Rechenmacher}, {Quinn},
  {Richards}, {Richmond}, {Rivetta}, {Rockosi}, {Ruthmansdorfer}, {Sandford},
  {Schlegel}, {Schneider}, {Sekiguchi}, {Sergey}, {Shimasaku}, {Siegmund},
  {Smee}, {Smith}, {Snedden}, {Stone}, {Stoughton}, {Strauss}, {Stubbs},
  {SubbaRao}, {Szalay}, {Szapudi}, {Szokoly}, {Thakar}, {Tremonti}, {Tucker},
  {Uomoto}, {Vanden Berk}, {Vogeley}, {Waddell}, {Wang}, {Watanabe},
  {Weinberg}, {Yanny}, {Yasuda}, \& {SDSS Collaboration}}]{SDSS_York}
{York}, D.~G., {Adelman}, J., {Anderson}, Jr., J.~E., {et~al.} 2000,
  \href{http://dx.doi.org/10.1086/301513}{\JournalTitle{\aj}, 120, 1579}

\bibitem[{{Yuan} {et~al.}(2021){Yuan}, {Garrison}, {Hadzhiyska}, {Bose}, \&
  {Eisenstein}}]{Yuan_2021}
{Yuan}, S., {Garrison}, L.~H., {Hadzhiyska}, B., {Bose}, S., \& {Eisenstein},
  D.~J. 2021, \JournalTitle{arXiv e-prints}, arXiv:2110.11412

\bibitem[{{Yuan} {et~al.}(2020){Yuan}, {Hadzhiyska}, {Bose}, {Eisenstein}, \&
  {Guo}}]{Yuan_2020}
{Yuan}, S., {Hadzhiyska}, B., {Bose}, S., {Eisenstein}, D.~J., \& {Guo}, H.
  2020, \JournalTitle{arXiv e-prints}, arXiv:2010.04182

\bibitem[{{Zentner} {et~al.}(2019){Zentner}, {Hearin}, {van den Bosch},
  {Lange}, \& {Villarreal}}]{Zentner_2019}
{Zentner}, A.~R., {Hearin}, A., {van den Bosch}, F.~C., {Lange}, J.~U., \&
  {Villarreal}, A. 2019,
  \href{http://dx.doi.org/10.1093/mnras/stz470}{\JournalTitle{\mnras}, 485,
  1196}

\bibitem[{{Zhai} {et~al.}(2017){Zhai}, {Tinker}, {Hahn}, {Seo}, {Blanton},
  {Tojeiro}, {Camacho}, {Lima}, {Carnero Rosell}, {Sobreira}, {da Costa},
  {Bautista}, {Brownstein}, {Comparat}, {Dawson}, {Newman}, {Prakash},
  {Roman-Lopes}, \& {Schneider}}]{Zhai_2017}
{Zhai}, Z., {Tinker}, J.~L., {Hahn}, C., {et~al.} 2017,
  \href{http://dx.doi.org/10.3847/1538-4357/aa8eee}{\JournalTitle{\apj}, 848,
  76}

\bibitem[{{Zhai} {et~al.}(2019){Zhai}, {Tinker}, {Becker}, {DeRose}, {Mao},
  {McClintock}, {McLaughlin}, {Rozo}, \& {Wechsler}}]{Zhai_2019}
{Zhai}, Z., {Tinker}, J.~L., {Becker}, M.~R., {et~al.} 2019,
  \href{http://dx.doi.org/10.3847/1538-4357/ab0d7b}{\JournalTitle{\apj}, 874,
  95}

\bibitem[{{Zheng} {et~al.}(2005){Zheng}, {Berlind}, {Weinberg}, {Benson},
  {Baugh}, {Cole}, {Dav{\'e}}, {Frenk}, {Katz}, \& {Lacey}}]{Zheng_2005}
{Zheng}, Z., {Berlind}, A.~A., {Weinberg}, D.~H., {et~al.} 2005,
  \href{http://dx.doi.org/10.1086/466510}{\JournalTitle{\apj}, 633, 791}

\bibitem[{{Zu}(2020)}]{Zu_2020}
{Zu}, Y. 2020, \JournalTitle{arXiv e-prints}, arXiv:2010.01143

\end{thebibliography}

\end{document}